\documentclass[amsmath,amssymb,amsfonts,twocolumn,english,showpacs,superscriptaddress,prb]{revtex4-1}
\pdfoutput=1

\usepackage[T1]{fontenc}
\usepackage[latin9]{inputenc}
\usepackage{times}
\usepackage{graphicx}
\usepackage{color}
\usepackage{mathrsfs}
\usepackage{float}
\usepackage{indentfirst}
\usepackage{babel}
\usepackage{wasysym}
\usepackage{tikz}
\usepackage{amsthm}
\usetikzlibrary{calc}
\usepackage{verbatim}
\usepackage{tabularx}
\usepackage{tabulary}
\usepackage{mwe}
\newcommand{\ket} [1] {\vert #1 \rangle}
\newcommand{\bra} [1] {\langle #1 \vert}

\usepackage{booktabs}
\usepackage{array, multirow}
\usepackage{subfigure}

\usepackage[colorlinks,bookmarks=false,citecolor=blue,linkcolor=blue,urlcolor=blue]{hyperref}
\setcitestyle{numbers,square}
\usepackage{array}

\graphicspath{{.}}

\begin{document}

\title{Commuting-projector Hamiltonians for 2D topological insulators: edge physics and many-body invariants}

\author{Jun Ho Son}
\affiliation{Department of Physics, Stanford University, Stanford, CA 94305, USA}

\author{Jason Alicea}
\affiliation{Department of Physics and Institute for Quantum Information and Matter, California Institute of Technology, Pasadena, CA 91125, USA}
\affiliation{Walter Burke Institute for Theoretical Physics, California Institute of Technology, Pasadena, CA 91125, USA}

\begin{abstract}

Inspired by a recently constructed commuting-projector Hamiltonian for a two-dimensional (2D) time-reversal-invariant topological superconductor [Wang et al., Phys.~Rev.~B {\bf 98}, 094502 (2018)], we introduce a commuting-projector model that describes an interacting yet exactly solvable 2D topological insulator. We explicitly show that both the gapped and gapless boundaries of our model are consistent with those of band-theoretic, weakly interacting topological insulators. Interestingly, on certain lattices our time-reversal-symmetric models also enjoy $\mathcal{CP}$ symmetry, leading to intuitive interpretations of the bulk invariant for a $\mathcal{CP}$-symmetric topological insulator upon putting the system on a Klein bottle. We also briefly discuss how these many-body invariants may be able to characterize models with only time-reversal symmetry. 

\end{abstract}

\maketitle

\section{Introduction} \label{sec:Intro}
 
 Commuting-projector Hamiltonians consist of sums of local terms that commute with each other. In these models, both ground states and excited states can be obtained simply by finding simultaneous eigenstates of all local terms---typically implying exact solvability despite strong interactions among microscopic degrees of freedom.   
Commuting-projector models have yielded great insights into interacting, gapped topological phases of matter. 
 Historically, the two now-canonical commuting-projector models---Kitaev quantum double \cite{Kitaev2003} and Levin-Wen string-net models \cite{Levin2006}---laid important cornerstones for the study of bosonic topologically ordered phases. More recently, various commuting-projector models for fermionic topologically ordered states and symmetry-protected topological phases \cite{Gu2014,Chen2014,Tarantino2016,Ware2016,Bhardwaj2017,Tantivasadakarn2018,Ellison2019,Aasen2017,Walker_unpub} have been constructed, establishing concrete lattice models of topological phases predicted by more abstract formalisms. 
 
 Despite the remarkable recent progress, commuting projector models for fermionic topological phases with both antiunitary symmetry \emph{and} continuous on-site symmetry have, to our knowledge, so far eluded construction. The first goal of the paper is to build a commuting-projector model for the most well-known example of such phases: a two-dimensional (2D) topological insulator with time-reversal symmetry $\mathcal{T}$ and U(1) particle conservation, also known as a quantum spin Hall system (we use both names interchangeably in this paper). Topological insulators are one of the first-discovered symmetry-protected topological phases and have been extensively explored via band theory.  Nevertheless, it has remained unclear how such states can emerge in a local lattice model outside of band-theoretic frameworks. 
 
 We will show that decorating Ising-spin domain walls with two Kitaev chains \citep{Kitaev2001} in a time-reversal symmetric and particle-conserving manner (instead of a single Kitaev chain as done in Ref.~\onlinecite{Wang2017}) allows one to construct commuting-projector models for $\mathcal{T}$-symmetric topological insulators. Our model, though strongly-interacting, possesses the same symmetry and edge properties of band-theoretic quantum spin Hall insulators. In particular, to study gapless edge states of our model, we derive a \emph{strictly} one-dimensional (1D) Hamiltonian whose low-energy physics is identical to that of fully symmetric quantum-spin-Hall edge states.  This Hamiltonian constitutes a generalization of a 1D model that recently appeared in Ref.~\onlinecite{jones2019}. 

 
 We further point out that the exactly solvable models for  $\mathcal{T}$-invariant topological superconductors and insulators also possess $\mathcal{CP}$ symmetry when defined on certain lattices (here $\mathcal{C}$ and $\mathcal{P}$ respectively denote charge conjugation and spatial reflection symmetries). Recent studies have explored $\mathcal{CP}$-protected topological phases \citep{Hsieh2014,Hsieh2014b,Witten2016,Shiozaki2018}.  While $\mathcal{CP}$ symmetry is rather unnatural in realistic condensed-matter setups, many topological phases are described by Lorentz-symmetric field theories in the infrared even though they emerge from non-relativistic settings; moreover, in relativistic theories $\mathcal{CP}$ is equivalent to $\mathcal{T}$ due to the $\mathcal{CPT}$ theorem. Hence, studying $\mathcal{CP}$-invariant topological phases may shed light on how to understand $\mathcal{T}$-invariant topological phases as well. We will show that our $\mathcal{CP}$-symmetric models can be defined on a Klein bottle, upon which many-body invariants that characterize $\mathcal{CP}$-protected topological phases obtain intuitive interpretations. We also comment on possible applications of these ideas to $\mathcal{T}$-symmetric topological insulators without any exact $\mathcal{CP}$ symmetry built in.
 
The rest of the paper is organized as follows.  Section~\ref{sec:model} briefly reviews the commuting-projector model for $\mathcal{T}$-invariant topological superconductors \cite{Wang2017} and then generalizes the construction to 2D topological insulators. Sections~\ref{sec:edgegapped} and \ref{sec:edgegapless} explore gapped and gapless edge phases in our commuting-projector topological-insulator Hamiltonians, demonstrating consistency with band-theoretic phenomenology for weakly interacting topological insulators.  We explain how to define our models on a Klein bottle and how to compute topological invariants in Sec.~\ref{sec:kleinCP}.  Finally, concluding remarks appear in Sec.~\ref{sec:conc}.
 
\section{The Model}
\label{sec:model}

\subsection{Symmetry actions in a Majorana representation}
\label{sec:symmetry}

 
 In this paper it is convenient to express models in terms of Majorana fermions rather than complex fermions.  Hence we review how U$(1)$, $\mathcal{T}$, and $\mathcal{CP}$ symmetries act on Majorana operators.
We start with the first two symmetries, which are relevant to time-reversal-symmetric topological insulators. 
Let $f_{i,s}$ denote fermion operators for spatial index $i$ and spin $s = \uparrow,\downarrow$.  
Symmetries $U_{\alpha} \in \text{U}(1)$ and $\mathcal{T}$ act on these operators as follows:
\begin{equation} \label{eq:cconserve}
U_{\alpha} \text{ : } f^{\dagger}_{i,s} \rightarrow e^{i\alpha}f^{\dagger}_{i,s}, \quad f_{i,s} \rightarrow e^{-i\alpha}f_{i,s} 
\end{equation}
\begin{equation}
\begin{split}
\mathcal{T}\text{ : } & f^{\dagger}_{i,\uparrow} \rightarrow f^{\dagger}_{i,\downarrow}, \quad f_{i,\downarrow}^{\dagger} \rightarrow -f_{i,\uparrow}^{\dagger} \\
& f_{i,\uparrow} \rightarrow f_{i,\downarrow}, \quad f_{i,\downarrow} \rightarrow -f_{i,\uparrow}^{\dagger}
\end{split}.
\end{equation}
When acting on individual fermion operators, time-reversal and charge conservation symmetries satisfy
\begin{equation}
\label{eq:basicrelT}
\begin{split} 
&\mathcal{T}^{2} = -1, \quad U_{\alpha}\mathcal{T} = \mathcal{T}U_{-\alpha}.
\end{split}
\end{equation}
Non-commutation between $\mathcal{T}$ and U(1) encoded in the second relation reflects antiunitarity of $\mathcal{T}$.   
Hence, a topological insulator is often denoted as protected by U$(1)\rtimes \mathcal{T}$, where the semidirect product emphasizes the above relation. 
 
 Now we define two Majorana operators associated with each $f_{i,s}$ operator via
\begin{equation} \label{eq:Majrep}
\gamma_{1,i,s} = \frac{f_{i,s}+f_{i,s}^{\dagger}}{2}, \qquad  \gamma_{2,i,s} = \frac{f_{i,s}^{\dagger}-f_{i,s}}{2i}.
\end{equation}
One can then show that $U_{\alpha}$ and $\mathcal{T}$ transform Majorana operators according to 
\begin{equation} \label{eq:symactT1}
\begin{split}
&U_{\alpha} \text{ : }  \begin{pmatrix}
\gamma_{1,i,\uparrow}\\
\gamma_{2,i,\uparrow}\\
\gamma_{1,i,\downarrow}\\
\gamma_{2,i,\downarrow}
\end{pmatrix} \rightarrow   M_{\alpha}\begin{pmatrix}
\gamma_{1,i,\uparrow}\\
\gamma_{2,i,\uparrow}\\
\gamma_{1,i,\downarrow}\\
\gamma_{2,i,\downarrow}
\end{pmatrix}\\
&\mathcal{T} \text{ : }  \begin{pmatrix}
\gamma_{1,i,\uparrow}\\
\gamma_{2,i,\uparrow}\\
\gamma_{1,i,\downarrow}\\
\gamma_{2,i,\downarrow}
\end{pmatrix} \rightarrow M_{\mathcal{T}} \begin{pmatrix}
\gamma_{1,i,\uparrow}\\
\gamma_{2,i,\uparrow}\\
\gamma_{1,i,\downarrow}\\
\gamma_{2,i,\downarrow}
\end{pmatrix}
\end{split}
\end{equation} 
where
\begin{equation} \label{eq:symactT2}
\begin{split}
&M_{\alpha} = \begin{pmatrix}
\cos{\alpha} & -\sin{\alpha} & 0 & 0 \\
\sin{\alpha} & \cos{\alpha} & 0 & 0 \\
0 & 0 & \cos{\alpha} & -\sin{\alpha} \\
0 & 0 & \sin{\alpha} & \cos{\alpha} 
\end{pmatrix} \\
&M_{\mathcal{T}} = \begin{pmatrix}
0 & 0 & 1 & 0 \\
0 & 0 & 0 & -1 \\
-1 & 0 & 0 & 0 \\
0 & 1 & 0 & 0 
\end{pmatrix}.
\end{split}
\end{equation}
The matrices $M_{\alpha}$ and $M_{\mathcal{T}}$ satisfy relations analogous to Eq.~\eqref{eq:basicrelT}, i.e.,
\begin{equation}\label{eq:Tmatrixrel}
M_{\mathcal{T}}^2 = -1,~~~M_{\alpha} M_{\mathcal{T}} = M_{\mathcal{T}} M_{-\alpha}.
\end{equation}
Thus, one may take the above relations as  \textit{defining properties}---that is, the SO$(2)$ transformation matrix on Majorana operators $M_{\alpha}$ and time-reversal transformation matrix $M_{\mathcal{T}}$ encode the correct symmetries of a 2D topological insulator provided Eq.~\eqref{eq:Tmatrixrel} holds.

 
 One can similarly encapsulate symmetries of a U$(1) \rtimes\mathcal{CP}$ topological insulator in real matrices representing the symmetry action on Majorana operators. Complex fermion operators transform under $\mathcal{CP}$ as
\begin{equation}
\mathcal{CP}\text{ : } f^{\dagger}_{i,s} \rightarrow f_{-i,s}, \quad f_{i,s} \rightarrow f_{-i,s}^{\dagger};
\end{equation}
here $-i$ in the subscript denotes the spatial index obtained from reflecting site $i$ with respect to some axis of our choice.  
The $U_\alpha$ and $\mathcal{CP}$ symmetries satisfy
\begin{equation}
(\mathcal{CP})^{2} = 1, \quad U_{\alpha}\mathcal{CP} = \mathcal{CP}U_{-\alpha}.
\label{CPrels}
\end{equation}
Here non-commutativity between $\mathcal{CP}$ and $U_{\alpha}$ comes from the fact that $\mathcal{CP}$ exchanges particles and holes, which acquire opposite U(1) phases.  

In the Majorana representation $\mathcal{CP}$ sends
\begin{equation} \label{eq:symactCP}
\begin{split}
&\mathcal{CP} \text{ : }  \begin{pmatrix}
\gamma_{1,i,\uparrow}\\
\gamma_{2,i,\uparrow}\\
\gamma_{1,i,\downarrow}\\
\gamma_{2,i,\downarrow}
\end{pmatrix} \rightarrow   M_{\mathcal{CP}}\begin{pmatrix}
\gamma_{1,-i,\uparrow}\\
\gamma_{2,-i,\uparrow}\\
\gamma_{1,-i,\downarrow}\\
\gamma_{2,-i,\downarrow}
\end{pmatrix}\end{split}
\end{equation} 
with
\begin{equation}
M_{\mathcal{CP}} = \begin{pmatrix}
0 & 0 & 1 & 0 \\
0 & 0 & 0 & -1 \\
1 & 0 & 0 & 0 \\
0 & -1 & 0 & 0 
\end{pmatrix}.
\end{equation}
As expected, $M_{\alpha}$ and $M_{\mathcal{CP}}$ satisfy relations akin to Eq.~\eqref{CPrels},
\begin{equation}
\label{eq:cprel}
M_{\mathcal{CP}}^{2} = 1, \quad M_{\alpha}M_{\mathcal{CP}} =  M_{\mathcal{CP}} M_{-\alpha},
\end{equation}
which may be taken as the defining property of the symmetry transformations.


\subsection{Review of the commuting-projector Hamiltonian for $\mathcal{T}$-invariant topological superconductors}

 We briefly review the exactly solvable model for $\mathcal{T}$-invariant 2D topological superconductors introduced in Ref.~\onlinecite{Wang2017}; our commuting-projector Hamiltonian for 2D topological insulators naturally extends this model as we will see in Sec.~\ref{2DTImodel}.  
For brevity and ease of generalization to $\mathcal{CP}$-protected topological phases, throughout the main text we focus on models constructed on the honeycomb lattice. We emphasize, however, that the Hamiltonians can be easily generalized into arbitrary trivalent lattices.
 
\begin{figure}
\includegraphics[width=0.8\linewidth]{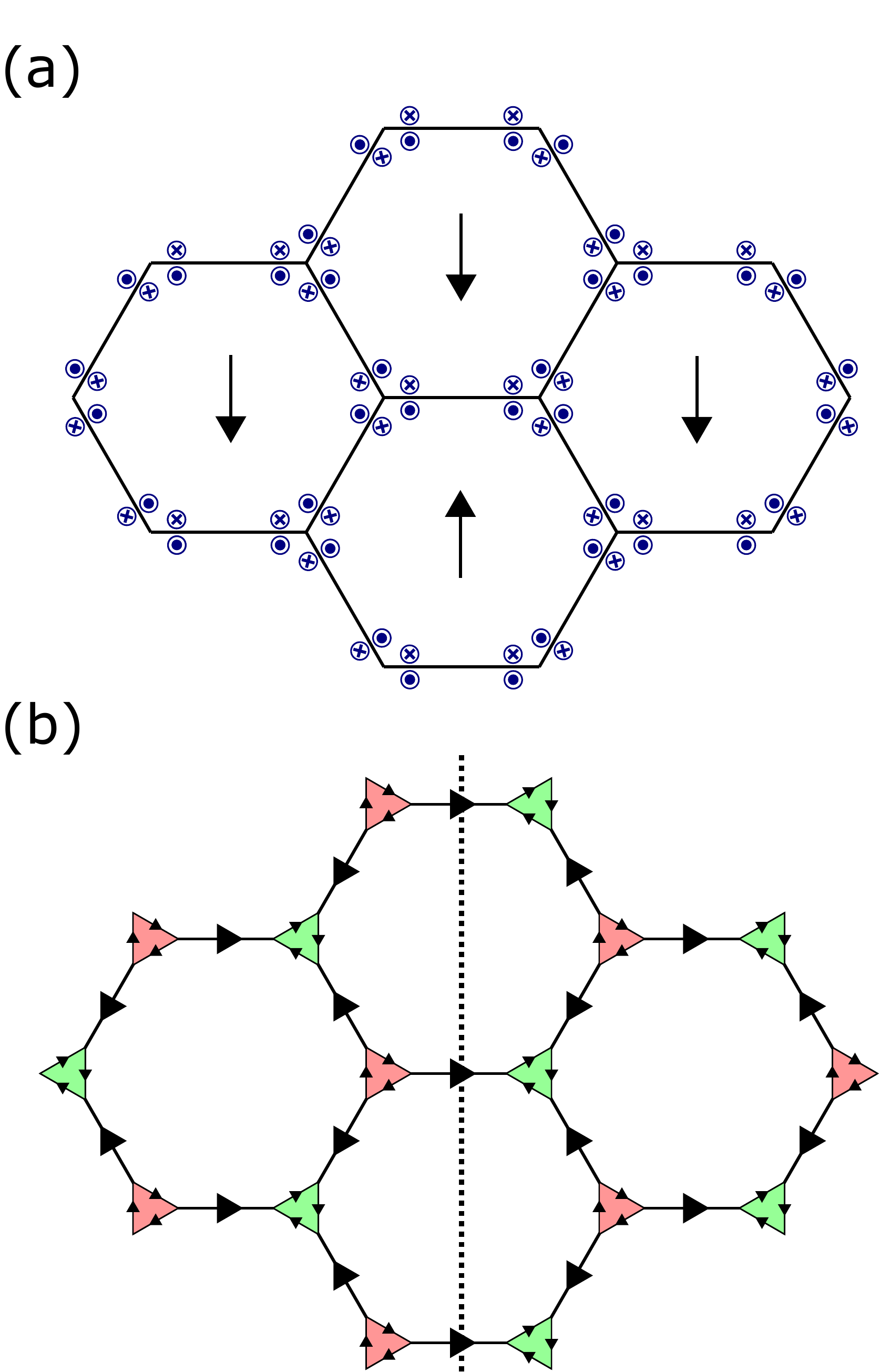}
\caption{(a) Basic ingredients in the topological-superconductor model. Each honeycomb plaquette hosts a bosonic spin-$\frac{1}{2}$ degree of freedom, while each honeycomb edge harbors four Majorana fermions.  Alternatively, each vertex of the Fisher lattice shown in (b) harbors two Majorana fermions.  Spin indices associated with the Majorana fermions are illustrated as out-of-page (for spin up) or into-the-page (for spin down) symbols familiar from freshman electromagnetism. (b) Choice of Kasteleyn orientation employed in the main text. Small triangles corresponding to $A$ and $B$ sublattices of the honeycomb lattice are respectively colored red and green.  }
\label{fig:1_dofandkast}
\end{figure}

The degrees of freedom in the 2D topological superconductor model are two spinful Majorana fermions per honeycomb-lattice edge and one spin-$\frac{1}{2}$ per plaquette.  See Fig.~\ref{fig:1_dofandkast}(a) for an illustration.  In an equivalent picture that we will frequently exploit, one can modify the honeycomb lattice by replacing each vertex with a small triangle, generating the Fisher lattice sketched in Fig.~\ref{fig:1_dofandkast}(b); the two Majorana fermions at each honeycomb-lattice edge can then be viewed as living on Fisher-lattice vertices.  We label Pauli operators for the spin at plaquette $p$ by $\sigma^{x}_{p}$ and $\sigma_{p}^{z}$, and denote the Majorana operators at Fisher-lattice vertex $v$ by $\gamma_{v,s}$, where $s = \uparrow,\downarrow$ labels spin. 
 
Due to subtleties with global fermion parity, defining the model consistently requires specifying a Kasteleyn orientation on the the Fisher lattice. Kasteleyn orientations are defined as a choice of arrows on the lattice that satisfies the following condition, often denoted as the `clockwise-odd rule': Around any closed clockwise cycle, there are an odd number of clockwise-oriented arrows. 
(See Refs.~\onlinecite{Tarantino2016} and \onlinecite{Ware2016} for a detailed discussion.)  We adopt the following convention:
\begin{itemize}
\item As in Ref.~\onlinecite{Wang2017}, let `long edges' denote Fisher-lattice edges derived from the original honeycomb lattice.  Moreover, label the two honeycomb sublattices by $A$ and $B$ [colored red and green, respectively, in Fig.~\ref{fig:1_dofandkast}(b)].  
Arrows on all long edges point from the $A$ sublattice to $B$ sublattice.  
\item Let `short edges' denote the edges of the small triangles in the Fisher lattice.  Arrows on short edges orient clockwise on going around any small triangle. 
\end{itemize}
Figure~\ref{fig:1_dofandkast}(b) illustrates the resulting Kasteleyn orientation. 

The Hamiltonian,
\begin{equation}
  H_{\text{TSC}} = -\sum_{t} A_{t} -\sum_{p} B_{p},
\end{equation}
consists of vertex terms $A_{t}$ defined for each vertex $t$ of the \textit{honeycomb} lattice (corresponding to a small triangle on the Fisher lattice, hence the subscript $t$) along with plaquette terms $B_{p}$. Before explaining these terms in detail, we define a sense of `pairing' of Majorana fermions belonging to Fisher-lattice vertices $v$ and $v'$ linked by either a long or short edge: $\gamma_{v,s}$ and $\gamma_{v',s'}$ are paired in a state $\ket{\psi}$ if $i\gamma_{v,s}\gamma_{v',s'}\ket{\psi} = g_{vv'}\ket{\psi}$.  Here $g_{vv'} = 1$ if the Kasteleyn arrow points from $v'$ to $v$; otherwise $g_{vv'} = -1$. 

\begin{figure}
\includegraphics[width=\linewidth]{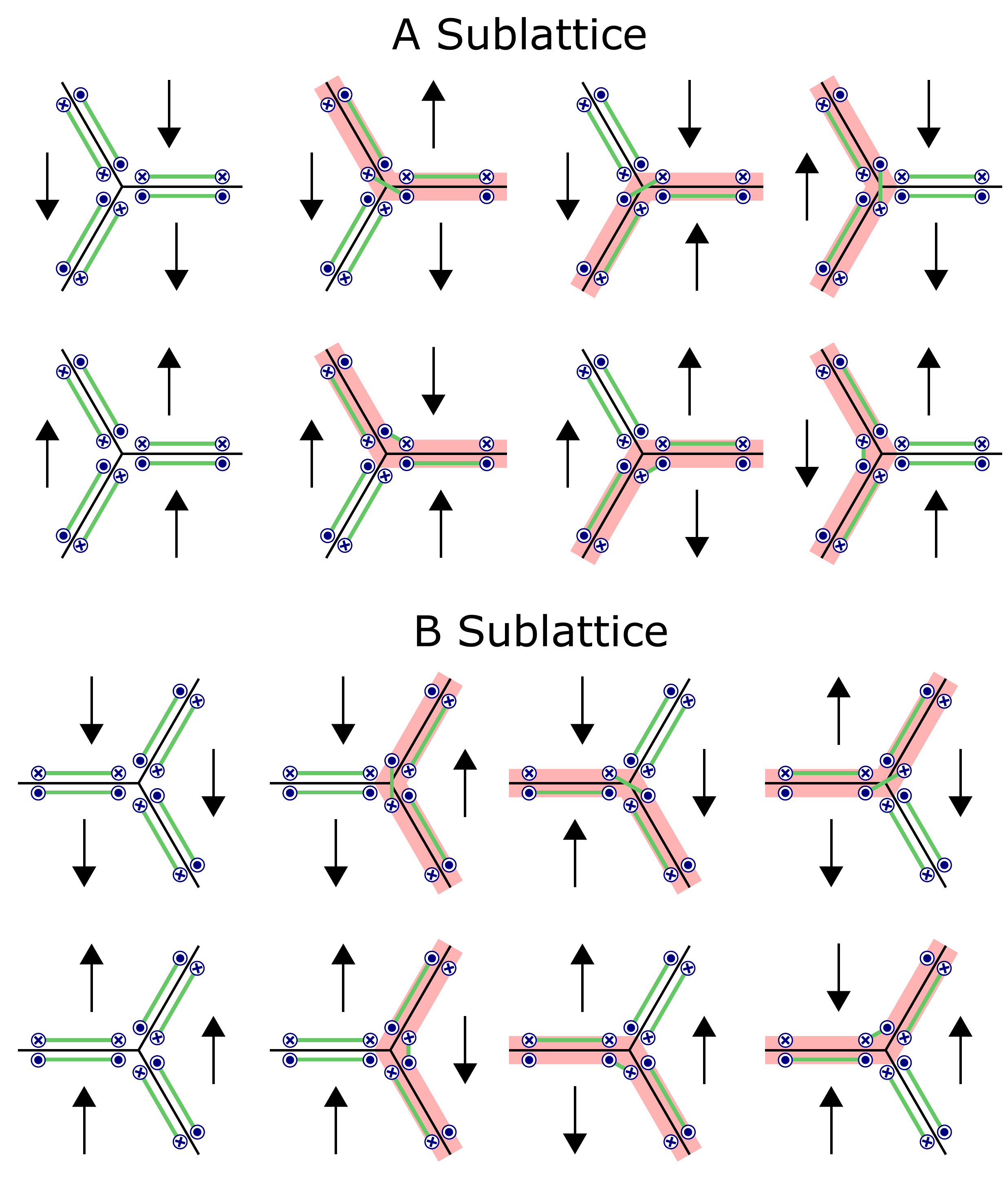}
\caption{Constraints energetically enforced by the $A_{t}$ vertex terms in the topological-superconductor commuting-projector model. Green lines connect Majorana fermions that `pair' in a given bosonic spin configuration, while shaded bonds indicate bosonic-spin domain walls.  }
\label{fig:2_vertexrule}
\end{figure}
 
 The role of $A_{t}$ is to give an energy advantage whenever Ising spin configurations around the vertex (i.e., $\sigma_{p}^{z}$ values around the vertex) and Majorana pairings are consistent with the local rules illustrated in Fig.~\ref{fig:2_vertexrule}. Some key features of these local rules are the following:
\begin{itemize} 
\item The local rules enforce Majorana pairing along short edges \textit{only when there is a domain wall of Ising spin configuration along it}. Away from such domain walls, Majoranas always pair along long edges. One can alternatively understand short-edge pairings as decorating Kitaev chains along domain walls. 
\item Majorana fermions with the same spin pair along long edges.  Meanwhile, short-edge Majorana pairings are always between opposite spins.
\item The Majorana spins involved in short-edge pairings also depend on the adjacent Ising spins.  In particular, the upper and lower panels for each sublattice in Fig.~\ref{fig:2_vertexrule} are related by an Ising-spin flip, and have Majoranas with opposite spin indices paired.
\end{itemize} 
After defining the projector 
\begin{equation}\label{eq:proj}
P_{vs,v's'} = \frac{1+ig_{vv'}\gamma_{v,s}\gamma_{v',s'}}{2},
\end{equation}
which projects onto the state where $\gamma_{v,s}$ and $\gamma_{v',s'}$ are paired, one can explicitly express $A_{t}$ as
\begin{equation} \label{eq:tscav}
\begin{split}
&A_{t} = \sum_{u_{t}} \mathcal{A}_{u_{t}} \mathcal{S}_{u_{t}} \\
&\mathcal{A}_{u_{t}} = \left( \prod_{(vs,v's') \in \mathcal{P}_{u_{t}}} P_{vs,v's'} \right) .
\end{split}
\end{equation}
Here $u_{t}$ denotes an Ising spin configuration around the honeycomb vertex $t$; the $u_{t}$ sum runs over all eight possible Ising spin configurations. The factor $\mathcal{S}_{u_{t}}$ projects onto states with Ising spin configuration $u_{t}$. And $\mathcal{P}_{u_{t}}$ is a set whose elements are pairings $(vs,v's')$ enforced by the Ising spin configuration $u_{t}$ according to the local rules. Simply stated, $A_{t} =1$ if the state is consistent with the local rules illustrated in Fig.~\ref{fig:2_vertexrule}, while $A_{t}=0$ on all other states.

\begin{figure}
\includegraphics[width=\linewidth]{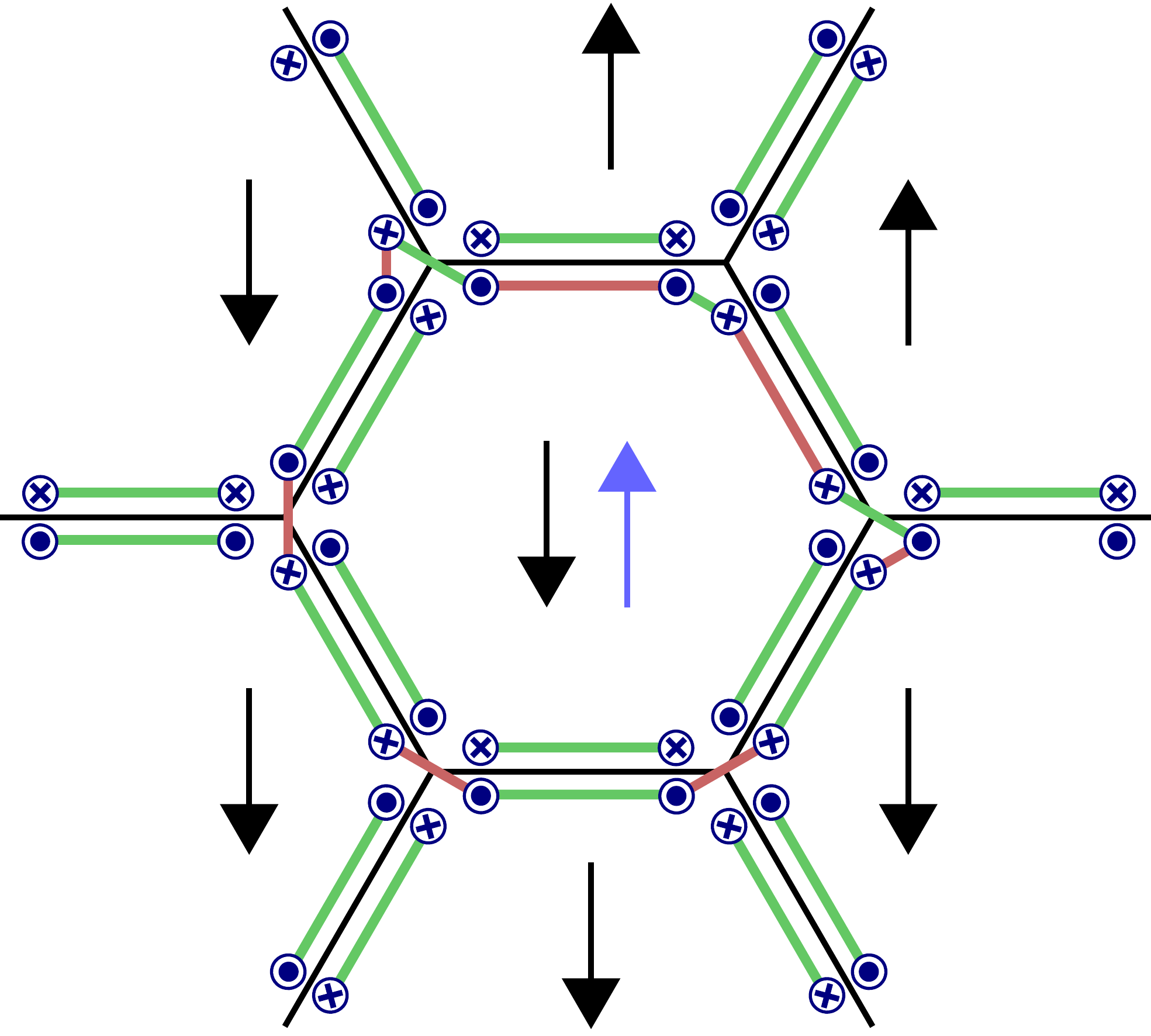}
\caption{Action of the $B_{p}$ plaquette terms in the topological-superconductor commuting-projector model. Black arrows denote the bosonic spin configuration before applying $B_{p}$, while the central blue arrow denotes the spin flipped by $B_{p}$. Green lines correspond to Majorana pairings enforced by the starting spin configuration [i.e., the set $\mathcal{P}_{u_{p}}$ associated with the string of projectors in the second line of Eq.~\eqref{eq:tscbp}]. Red lines, obtained by taking a subset of green lines and shifting by one Majorana unit, correspond to pairings enforced by the new spin configuration obtained by applying $B_p$ [i.e., the set $\overline{\mathcal{P}}_{u_{p}}$ in the string of projectors in the first line of Eq.~\eqref{eq:tscbp}.]}
\label{fig:3_bp}
\end{figure}

Plaquette terms $B_{p}$ allow Majorana pairings and the plaquette spin to fluctuate in a way that preserves the local rules if they are satisfied initially.  These terms read 
\begin{equation} 
\label{eq:tscbp}
\begin{split}
&B_{p} = \sum_{u_{p}} \mathcal{B}_{u_{p}}\sigma_{p}^{x}\mathcal{S}_{u_{p}} \\
&\mathcal{B}_{u_{p}} = \frac{1}{\sqrt{2}} \left( \prod_{(v_{1}s_{1},v_{2}s_{2}) \in \overline{\mathcal{P}}_{u_{p}} } \sqrt{2} P_{v_{1}s_{1},v_{2}s_{2}} \right) \\
&\left( \prod_{(v_{3}s_{3},v_{4}s_{4}) \in \mathcal{P}_{u_{p}} } P_{v_{3}s_{3},v_{4}s_{4}} \right) .
\end{split}
\end{equation}
In the above equations $u_{p}$ denotes an Ising spin configuration for plaquette $p$ together with the six neighboring plaquettes. Moreover, $\mathcal{S}_{u_{p}}$ projects onto Ising spin configuration $u_{p}$; $\mathcal{P}_{u_{p}}$ denotes a set of Majorana pairings consistent with $u_{p}$; and $\overline{\mathcal{P}}_{u_{p}}$ is a modified set of Majorana pairings consistent with the spin configuration in which $\sigma^z_p$ is flipped.  
Thus, the bosonic part $\sigma_{p}^{x}\mathcal{S}_{u_{p}}$ flips the plaquette spin, and the two strings of Majorana projectors in $\mathcal{B}_{u_p}$ project onto the state with Majorana pairings consistent with the new Ising spin configuration. Finally, some integer powers of $\sqrt{2}$ are added to normalize $B_{p}$ to have an eigenvalue $\pm 1$ and $0$ on entire Hilbert space. See Fig.~\ref{fig:3_bp} for an illustration of the action of $B_{p}$.

 One can explicitly show that all $A_{t}$ and $B_{p}$ operators commute with each other. Also, $B_{p}$ is \emph{Hermitian and unitary} on the subspace in which $A_{t}=1$. Hence, on this subspace, $B_{p}^{2} = 1$. The ground state can be easily constructed by diagonalizing each term, and corresponds to an equal superposition of each Ising spin configuration accompanied with correct Majorana pairings.  

 Finally, let us discuss symmetry of the Hamiltonian.  Denote a vertex of the Fisher lattice adjacent to sublattice $A$ and $B$ of the honeycomb lattice as $v_{A}$ and $v_B$, respectively.  We can define an antiunitary symmetry $\mathcal{T}_{\text{TSC}}$ that acts on Majorana operators and Ising spins according to  
\begin{equation}
\label{eq:TTSCunitary}
\begin{split}
\mathcal{T}_{\text{TSC}} \text{ : }  &\sigma_{p}^{z} \rightarrow - \sigma_{p}^{z}, \quad \sigma_{p}^{x} \rightarrow  \sigma_{p}^{x},
\\ &\begin{pmatrix}
\gamma_{v_{A},\uparrow} \\
\gamma_{v_{B},\uparrow} \\
\gamma_{v_{A},\downarrow} \\
\gamma_{v_{B},\downarrow} 
\end{pmatrix} \rightarrow \begin{pmatrix}
0 & 0 & 1 & 0 \\
0 & 0 & 0 & -1 \\
-1 & 0 & 0 & 0 \\
0 & 1 & 0 & 0
\end{pmatrix} \begin{pmatrix}
\gamma_{v_{A},\uparrow} \\
\gamma_{v_{B},\uparrow} \\
\gamma_{v_{A},\downarrow} \\
\gamma_{v_{B},\downarrow} 
\end{pmatrix}
\end{split}
\end{equation}
Notice that the action on Majorana fermions is identical to how $\mathcal{T}$ is defined in Eqs.~\eqref{eq:symactT1} and \eqref{eq:symactT2}. One can prove that the Hamiltonian preserves $\mathcal{T}_{\text{TSC}}$. The following observation comes in handy for the proof: for all possible Majorana pairs $\gamma_{v,s}$ and $\gamma_{v's'}$ that can be paired due to the local rules, \textit{only one acquires a minus sign under time-reversal}. An additional minus sign from complex conjugation results in a transformation of $i\gamma_{v,s}\gamma_{v',s'}$ to $i\gamma_{v,\overline{s}}\gamma_{v',\overline{s'}}$, where the overline denotes an opposite spin index.

\subsection{Extension to $\mathcal{T}$-invariant topological insulators}
\label{2DTImodel}

Next we will generalize the model reviewed in the previous subsection to construct a commuting-projector Hamiltonian for the celebrated quantum spin Hall insulator. As a first step we double the number of Majorana fermions per Fisher-lattice vertex from two to four.  The corresponding Majorana operators on vertex $v$ are denoted by $\gamma_{v,s,a}$, the subscript $a=1,2$ being a `layer index'.  Essentially, we will just construct a Hamiltonian in which the ground state is decorated with \textit{two layers of Kitaev chains} instead of one as in the topological superconductor case. 
 
The Hamiltonian once again takes the form 
\begin{equation}
  H_{\rm TI} = -\sum_{t} A_{t} - \sum_{p} B_{p}.
\end{equation}
Vertex and plaquette terms are straightforwardly modified to reflect that there are now two layers of Majorana fermions that pair identically according to the same local rules summarized in Fig.~\ref{fig:2_vertexrule}.
To this end, define the projector
\begin{equation}
\label{eq:tiproj}
P_{vs,v's'}^{\text{TI}} = \frac{1+ig_{vv'}\gamma_{v,s,1}\gamma_{v',s',1}}{2}\frac{1+ig_{vv'}\gamma_{v,s,2}\gamma_{v',s',2}}{2},
\end{equation}
which is a product of two projectors in Eq.~\eqref{eq:proj}, one for layer 1 and another for layer 2.
We then have 
\begin{equation} \label{eq:tiav}
\begin{split}
&A_{t} = \sum_{u_{t}} \mathcal{A}_{u_{t}} \mathcal{S}_{u_{t}} \\
&\mathcal{A}_{u_{t}} = \sum_{u_{t}} \left( \prod_{(vs,v's') \in \mathcal{P}_{u_{t}} } P_{vs,v's'}^{\text{TI}} \right) \mathcal{S}_{u_{t}}.
\end{split}
\end{equation}
and
\begin{equation} 
\label{eq:tibp}
\begin{split}
&B_{p} = \sum_{u_{p}} \mathcal{B}_{u_{p}}\sigma_{p}^{x}\mathcal{S}_{u_{p}} \\
&\mathcal{B}_{u_{p}} = \frac{1}{2} \sum_{u_{p}}  \left( \prod_{(v_{1}s_{1},v_{2}s_{2}) \in \overline{\mathcal{P}}_{u_{p}} } 2 P_{v_{1}s_{1},v_{2}s_{2}}^{\text{TI}} \right) \\
& \left( \prod_{(v_{3}s_{3},v_{4}s_{4}) \in \mathcal{P}_{u_{p}} } P_{v_{3}s_{3},v_{4}s_{4}}^{\text{TI}} \right) .
\end{split}
\end{equation}
Aside from normalization factors in $\mathcal{B}_p$, the vertex and plaquette terms are identical to those in the topological superconductor model, but with $P_{vs,v's'} \rightarrow P_{vs,v's'}^{\text{TI}}$.  


The Hamiltonian $H_{\rm TI}$ exhibits the following three symmetries:
\begin{enumerate}
\item The time-reversal transformation $\mathcal{T}_{\text{TSC}}$ inherited from the original topological superconductor model applied to plaquette spins and Majorana fermions on both layers.  We will soon find out that this operation does \emph{not} coincide with physical time-reversal symmetry of an electronic topological insulator---which we hereafter denote by $\mathcal{T}_{\text{TI}}$.  
\item A $\mathbb{Z}_{2}$ layer interchange symmetry $\mathcal{L}$, which transforms $\gamma_{v,s,1} \leftrightarrow \gamma_{v,s,2}$.  The Hamiltonian obviously preserves this symmetry since the layers are treated identically.  
\item A U$(1)$ symmetry with elements $U_\alpha$ that acts on Majorana operators as
\begin{equation}
\begin{pmatrix}
\gamma_{v,s,1} \\
\gamma_{v,s,2}
\end{pmatrix} \rightarrow \begin{pmatrix}
\cos{\alpha} & -\sin{\alpha} \\
\sin{\alpha} & \cos{\alpha}
\end{pmatrix} \begin{pmatrix}
\gamma_{v,s,1} \\
\gamma_{v,s,2}
\end{pmatrix}.
\end{equation}
To see that this is actually a symmetry of the Hamiltonian, we observe that $P_{vs,v's'}^{\text{TI}}$ is invariant under the above U$(1)$ transformation due to a simple corollary of a lemma presented in Appendix~\ref{app:lemma}.  Since all fermionic terms in the Hamiltonian appear in the form of $P_{vs,v's'}^{\text{TI}}$, invariance of $P_{vs,v's'}^{\text{TI}}$ implies U$(1)$ invariance of the whole Hamiltonian.
\end{enumerate}

The transformations specified above imply that $U_{\alpha}\mathcal{L} = \mathcal{L}U_{-\alpha}$, which is reminiscent of the second relation satisfied by time-reversal symmetry in Eq.~\eqref{eq:basicrelT}.  Nevertheless, $\mathcal{L}$ clearly is not the time-reversal symmetry appropriate for a quantum spin Hall insulator since it is unitary and obeys $\mathcal{L}^2 = 1$.  
Furthermore, although $\mathcal{T}_{\text{TSC}}$ is antiunitary and obeys $\mathcal{T}_{\text{TSC}}^2 =-1$, this operation can not be the desired time-reversal symmetry either since it commutes with U$(1)$. Instead, the correct time-reversal symmetry is 
\begin{equation}
  \mathcal{T}_{\text{TI}} = \mathcal{T}_{\text{TSC}}\mathcal{L}. 
  \label{TTIdef}
\end{equation}  
One can explicitly show that $\mathcal{T}_{\text{TI}}$ and $U_{\alpha}$ satisfy the relations given in Eq.~\eqref{eq:basicrelT}, and equivalently in Eq.~\eqref{eq:Tmatrixrel}.


We close this subsection with two remarks.  First, defining time-reversal symmetry as layer interchange followed by a symmetry inherited from the topological superconductor may appear somewhat artificial. However, we will see in Secs.~\ref{sec:edgegapped} and \ref{sec:kleinCP} that this choice allows one to see topological properties transparently, indicating that our choice is indeed physically correct. Second, our model has an extra $\mathbb{Z}_{2}$ layer-interchange symmetry (or equivalently, $\mathcal{T}_{\text{TSC}}$ symmetry) that one may suspect plays a crucial role in topological properties.  It does not.  This `accidental' symmetry can be broken without altering the topological properties of the model as discussed later on.  


\subsection{$\mathcal{CP}$ symmetry}


Both the $\mathcal{T}$-symmetric topological insulator and topological superconductor models additionally possess $\mathcal{CP}$ symmetry.  The dashed line in Fig.~\ref{fig:1_dofandkast}(b) shows our choice of the reflection axis associated with $\mathcal{P}$.  We note two important features of this reflection: First, it maps maps sublattice $A$ onto sublattice $B$ and vice versa. 
Second, in the preceding sections we adopted a `special Kasteleyn orientation' for which \emph{all arrows flip under this reflection}.  
 
 For simplicity, let us first consider the topological superconductor model. The second point above suggests that naive parity symmetry is absent. However, one can consider the following operation:
\begin{equation}
\label{eq:cpact1}
\begin{split}
\mathcal{CP}_{\text{TSC}} \text{ : } &\sigma_{p}^{z} \rightarrow -\sigma_{-p}^{z}\\
&\gamma_{v_{A},\uparrow} \rightarrow \gamma_{-v_{A}, \downarrow} \\
&\gamma_{v_{A},\downarrow} \rightarrow -\gamma_{-v_{A}, \uparrow} \\
&\gamma_{v_{B},\uparrow} \rightarrow -\gamma_{-v_{B}, \downarrow}\\
&\gamma_{v_{B},\downarrow} \rightarrow \gamma_{-v_{B}, \uparrow}.
\end{split}
\end{equation}
As in Sec.~\ref{sec:symmetry}, the minus sign on the vertex and plaquette indices refers to the reflection operation. Note that since $-v_{A}$ and $-v_{B}$ respectively belong to sublattice $B$ and $A$, $(\mathcal{CP}_{\text{TSC}})^2 = 1$ despite some minus signs picked up by Majorana fermions in the transformation. Equation~\eqref{eq:cpact1} essentially flips the plaquette spins and Majorana spins, spatially reflects, and adds some minus signs. 

 To see why $\mathcal{CP}_{\text{TSC}}$ is a symmetry of the Hamiltonian, first observe that the vertex rules illustrated in Fig.~\ref{fig:2_vertexrule} are symmetric under spatial reflection and simultaneous flipping of plaquette spins and Majorana spins. This property is not sufficient, however, because we earlier observed that the spatial reflection flips the Kasteleyn-orientation arrows. The minus signs added in the transformation rule for Majorana fermions in Eq.~\eqref{eq:cpact1} remedy the issue, which can be seen as follows.  Long-edge pairing always pairs Majoranas with the same spin but opposite sublattice. The transformation rule dictates that precisely one of those Majorana fermions acquires an extra minus sign, thereby correcting the Kasteleyn-orientation flip from the spatial reflection. Similarly, short-edge pairing always pairs Majoranas on the same sublattice but with opposite spin; here too only one such Majorana fermion acquires a minus sign, again correcting the reversed Kasteleyn orientation.  
 
The topological-insulator model, which has doubled Majorana degrees of freedom and exhibits U$(1)$ and $\mathcal{L}$ layer-interchange symmetry, straightforwardly inherits $\mathcal{CP}_{\text{TSC}}$ symmetry.  As we already encountered for time-reversal symmetry, however, $\mathcal{CP}_{\text{TSC}}$ is not the symmetry appropriate for the quantum-spin-Hall setting---this operation satisfies neither Eq.~\eqref{CPrels} nor the equivalent Eq.~\eqref{eq:cprel}.  We can nevertheless construct $\mathcal{CP}_{\text{TI}}$, the $\mathcal{CP}$ symmetry relevant to the $\mathcal{CP}$-protected topological insulator, by combining $\mathcal{L}$ with $\mathcal{CP}_{\text{TSC}}$:
\begin{equation}
  \mathcal{CP}_{\text{TI}} = \mathcal{CP}_{\text{TSC}} \mathcal{L}.
\end{equation}
Note the similarity to Eq.~\eqref{TTIdef}.
One can explicitly show that $\mathcal{CP}_{\text{TI}}^2 =1$, and that the U$(1)$ and $\mathcal{CP}_{\text{TI}}$ action on Majorana degrees of freedoms satisfies Eq.~\eqref{eq:cprel}.

 
Let us briefly comment on the fate of $\mathcal{CP}$ symmetry \emph{beyond} the honeycomb-lattice setting. Recall that the existence of $\mathcal{CP}$ symmetry above relied on the bipartite and reflection-symmetric nature of the lattice. We detail in Apppendix~\ref{app:moregeneral} how $\mathcal{CP}$ symmetry may survive in models defined on any reflection-symmetric lattice (not necessarily bipartite) as well.  Conversely, if $\mathcal{T}$-symmetric topological phases are defined on a non-parity-symmetric lattice, there cannot be any strict $\mathcal{CP}$ symmetry. We will, however, see in Sec.~\ref{sec:kleinCP} that some bulk diagnostic of non-trivial topology of $\mathcal{CP}$-symmetric phases still remains in such models despite the lack of $\mathcal{CP}$ symmetry. 
 

\section{Gapped, symmetry-breaking edge properties of the $\mathcal{T}$-invariant topological insulator model} \label{sec:edgegapped}

 Quantum spin Hall insulators are protected by the interplay of U$(1)$ and $\mathcal{T}$ symmetries.  Hence, breaking either symmetry (e.g., via introduction of superconductivity or magnetism) suffices to gap out the edge, in turn enabling multiple inequivalent gapped edge phases.  Domain walls separating incompatibly gapped regions of the edges bind interesting zero-dimensional modes that have been widely studied, typically by assuming that the edge is governed by an effective model for helical 1D Dirac fermions.  In this section we show that our commuting-projector model reproduces precisely the same physics.  Interestingly, due to the exact solvability, these zero modes can be observed explicitly on the lattice level without resorting to an analysis of an effective low-energy edge description.


\subsection{Construction of broken-symmetry gapped edges}
\label{GappedSec}

\begin{figure}
\includegraphics[width=0.9\linewidth]{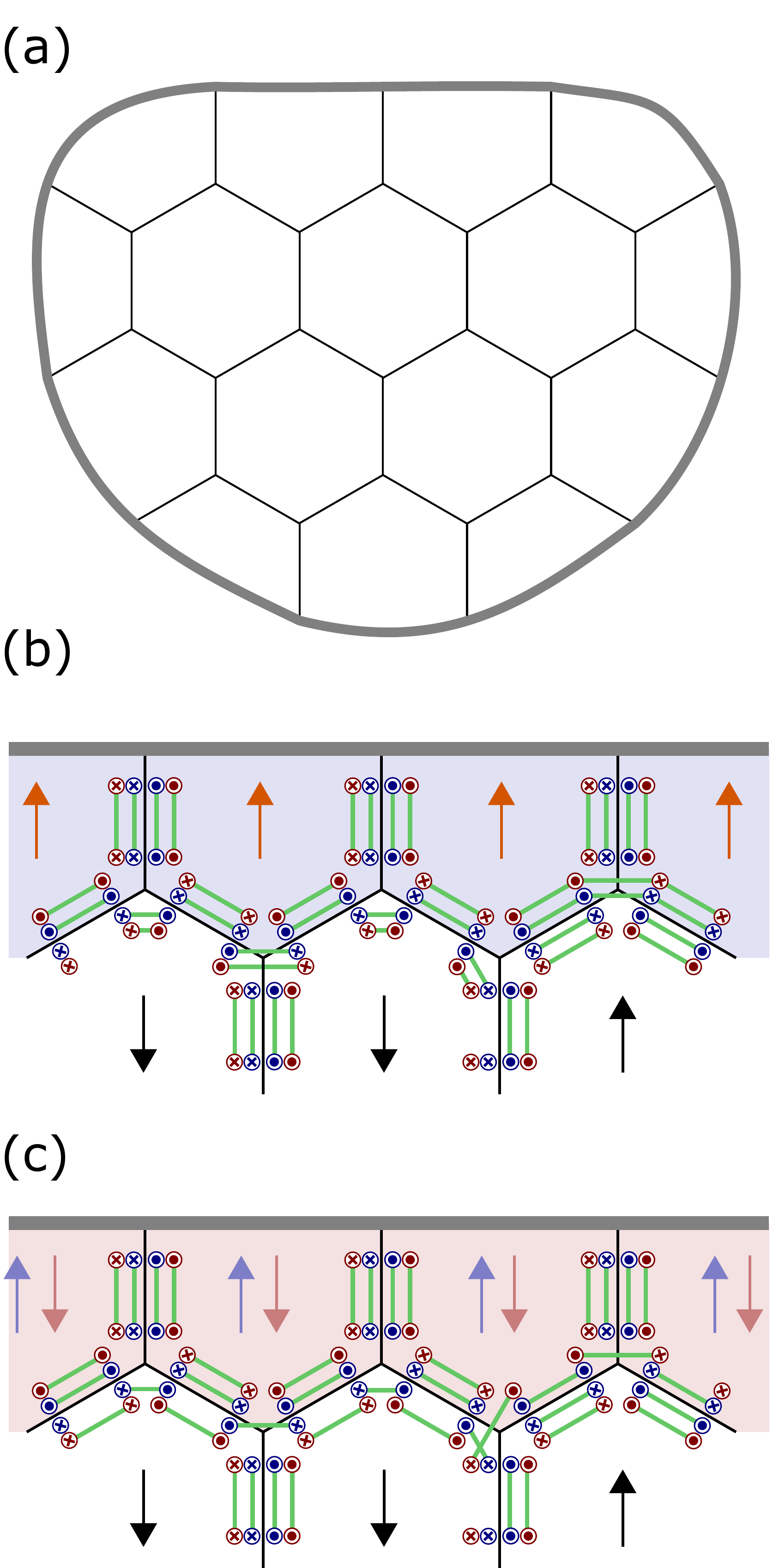}
\caption{(a) `Dangling-bond edge termination' where all vertices are trivalent but plaquettes at the interface are incomplete. (b) Snapshot of a plaquette-spin configuration and corresponding Majorana pairings for a $\mathcal{T}$-breaking interface. The orange plaquette spins are polarized by a Zeeman field. (c) Similar snapshot for a U$(1)$-breaking interface.  Low-saturation blue and red arrows at the interface denote \emph{fictitious} spins that are not actual degrees of freedom in the model; Majorana fermions on layers 1 and 2 pair as if these fictitious spins point up and down, respectively.  }
\label{fig:5_gapped}
\end{figure}

First, we discuss how one can break $\mathcal{T}$ symmetry [but preserve U$(1)$] at the boundary without losing exact solvability or generating spurious gapless degrees of freedom.  Consider the `dangling-bond edge termination' shown in Fig.~\ref{fig:5_gapped}(a) in which boundary plaquettes are incomplete. In such a geometry, one can define vertex terms $A_{t}$ precisely as before since all vertices remain trivalent.  Likewise, one can define $B_{p}$ as before for all complete `bulk' plaquettes. For incomplete edge plaquettes indexed by $\tilde p$, we violate $\mathcal{T}$ by polarizing edge spins with a Zeeman term 
\begin{equation}
  H_{\rm Zeeman} = h \sum_{\tilde{p}} \sigma_{\tilde{p}}^{z}.  
  \label{HZeeman}
\end{equation}
Since $H_{\rm Zeeman}$ clearly commutes with all other Hamiltonian terms, exact solvability is retained and the spectrum remains gapped. The ground state is readily constructed by freezing the edge spins to minimize the Zeeman term, then allowing bulk spins to fluctuate and Majorana pairings to follow the constraint given by $A_{t}$'s. See Fig.~\ref{fig:5_gapped}(b) for an illustration.

We can also break U$(1)$ at the edge while preserving time-reversal symmetry.  To do so we employ the same lattice termination as above, but now \emph{we remove the spins from all incomplete plaquettes instead of polarizing them via a Zeeman field}.  Vertex and plaquette terms that invoked these eliminated spins in their definition must then be replaced.  We do so by imposing the following Majorana-pairing rule at the boundary: 
\emph{Majorana fermions on layer 1 pair as if the eliminated spins point down, whereas Majorana fermions on layer 2 pair as if the eliminated spins point up}. Figure~\ref{fig:5_gapped}(c) shows an example of Majorana pairings consistent with this rule.
 
Since Majorana fermions in layers 1 and 2 couple differently, the Hamiltonian explicitly breaks both the layer-mixing U$(1)$ symmetry and the layer-interchange $\mathcal{L}$ symmetry. The Hamiltonian also breaks $\mathcal{T}_{\rm TSC}$ since $(i)$ this symmetry acts separately on Majorana fermions within each layer and $(ii)$ the boundary Majorana fermions behave as if the eliminated spins were polarized.  The modified rule above does, however, preserve $\mathcal{T}_{\text{TI}} = \mathcal{T}_{\text{TSC}}\mathcal{L}$, emphasizing that this somewhat-arbitrary-looking composite operation is indeed the correct antiunitary symmetry that protects the topological insulator!

Note that we can alternatively construct a time-reversal-invariant, gapped boundary by pairing Majorana fermions as if the eliminated edge spins point up for layer 2 and down for layer 1 (opposite to what we described in the previous paragraph). In fact, a further generalization is possible.  Define $\gamma_{v,s,1}(\theta)$ and $\gamma_{v,s,2}(\theta)$ as 
\begin{equation}
\begin{pmatrix}
\gamma_{v,s,1}(\theta) \\
\gamma_{v,s,2}(\theta)
\end{pmatrix} = \begin{pmatrix}
\cos{\frac{\theta}{2}} & \sin{\frac{\theta}{2}} \\
-\sin{\frac{\theta}{2}} & \cos{\frac{\theta}{2}}
\end{pmatrix} \begin{pmatrix}
\gamma_{v,s,1} \\
\gamma_{v,s,2}
\end{pmatrix}.
\label{gammathetas}
\end{equation}
Essentially, $\left( \gamma_{v,s,1}(\theta), \gamma_{v,s,2}(\theta) \right)$ correspond to a U$(1)$-twisted version of $\left( \gamma_{v,s,1}, \gamma_{v,s,2} \right)$.  We can construct a $\mathcal{T}$-symmetric gapped edge for any $\theta$ by pairing $\gamma_{v,s,1}(\theta)$'s as if the eliminated edge spins orient up and pairing $\gamma_{v,s,2}(\theta)$'s as if the eliminated edge spins point down. The example at the beginning of the paragraph corresponds to $\theta = \pi$.  Physically, $\theta$ corresponds to the phase of the superconducting order parameter that gaps the boundary.  The factor of 2 in the trigonometric functions in Eq.~\eqref{gammathetas} is important.  It follows that $\left( \gamma_{v,s,1}(2\pi), \gamma_{v,s,2}(2\pi) \right) = \left( -\gamma_{v,s,1}, -\gamma_{v,s,2} \right)$, though fermion-parity preservation ensures that the edge states corresponding to $\theta =0$ and $\theta = 2\pi$ are identical.  The minus signs on the right side are in fact consistent with our identification of $\theta$ as the order-parameter phase.  Winding the phase by $2\pi$ is equivalent to dragging an $h/(2e)$ superconducting vortex around the edge---which does not affect Cooper pairs but generates a minus sign for individual fermions.

\subsection{Domain walls between incompatibly gapped edge phases}
\label{DomainWallSec}

\begin{figure}
\includegraphics[width=\linewidth]{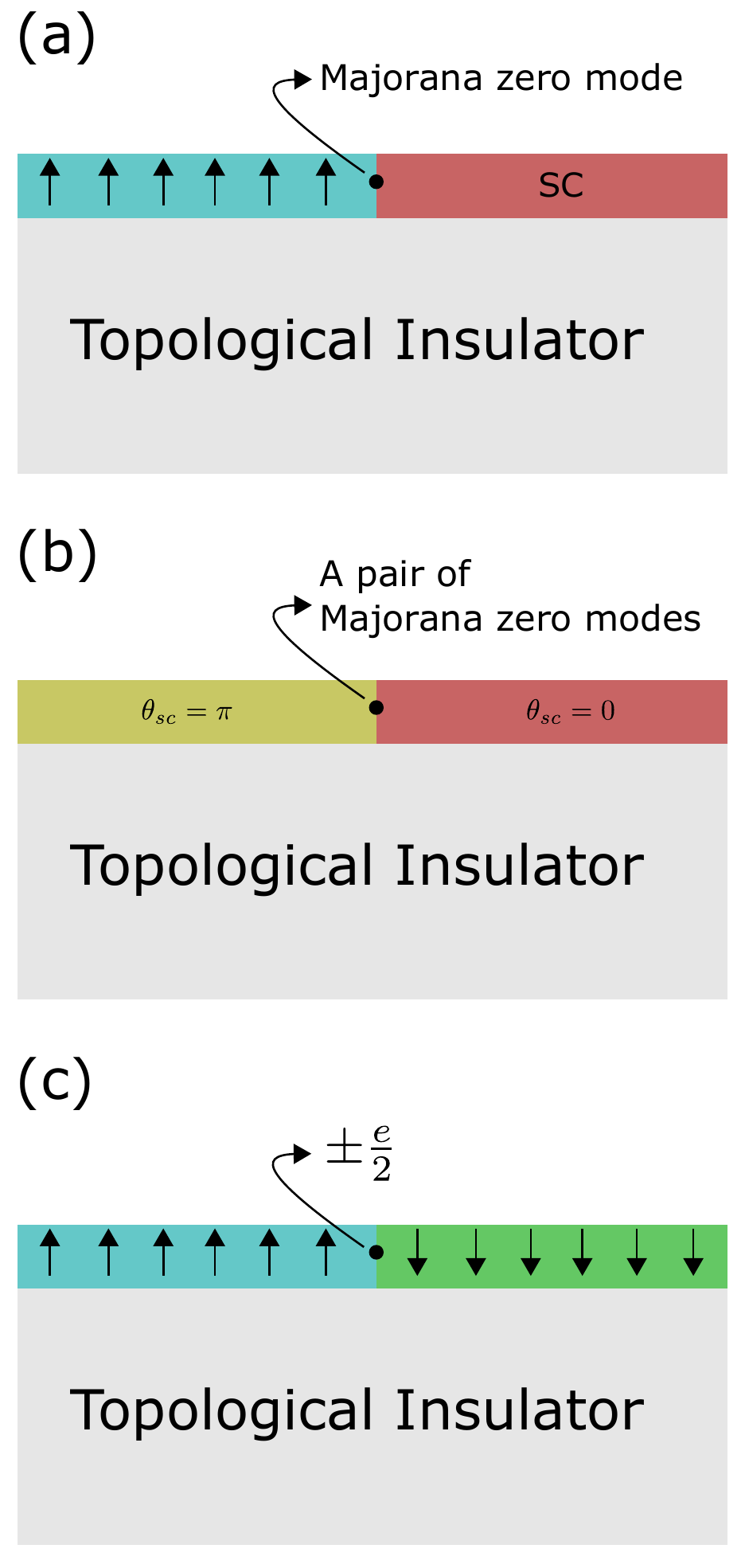}
\caption{Summary of zero modes hosted between incompatibly gapped regions of a quantu-spin-Hall edge.  }
\label{fig:6_freefdomain}
\end{figure}

We now briefly review the physics of domain walls separating inequivalent gapped, broken-symmetry regions of the edge.  Perhaps most famously, the interface between a superconducting domain that breaks U$(1)$ and a magnetic domain that breaks $\mathcal{T}$ binds a single unpaired Majorana zero mode \citep{Fu2009}.  Nontrivial domain walls are also possible even if either $\mathcal{T}$ or U$(1)$ is globally conserved.  Suppose that $\mathcal{T}$ is globally conserved.  Here one can consider a $\mathcal{T}$-invariant domain wall between two U$(1)$-breaking edge regions, one with a superconducting-order-parameter phase $\theta_{sc} = 0$ and another with $\theta_{sc} = \pi$. This $\pi$-junction domain wall hosts a Kramers pair of Majorana zero modes---which is intimately related to the fractional Josephson effect \citep{Kitaev2001}.  Now imagine that U$(1)$ is globally conserved.  Since $\mathcal{T}$ is a discrete symmetry, $\mathcal{T}$-breaking edge phases come in pairs that are time-reversed partners of each other.  In an experimental context these time-reversed partners correspond to helical edge modes gapped via a proximate ferromagnet that magnetizes either up or down; in our construction they simply reflect the two possible choices for the sign of the Zeeman field $h$ in Eq.~\eqref{HZeeman} that polarizes the edge spins. A domain wall between oppositely magnetized regions of the edge hosts a (complex) fermion zero mode with fractional charge $\pm e/2$ \citep{Jackiw1976,Qi2008}.  Figure~\ref{fig:6_freefdomain} summarizes the structure of the three types of domain walls highlighted above.  

\begin{figure}
\includegraphics[width=\linewidth]{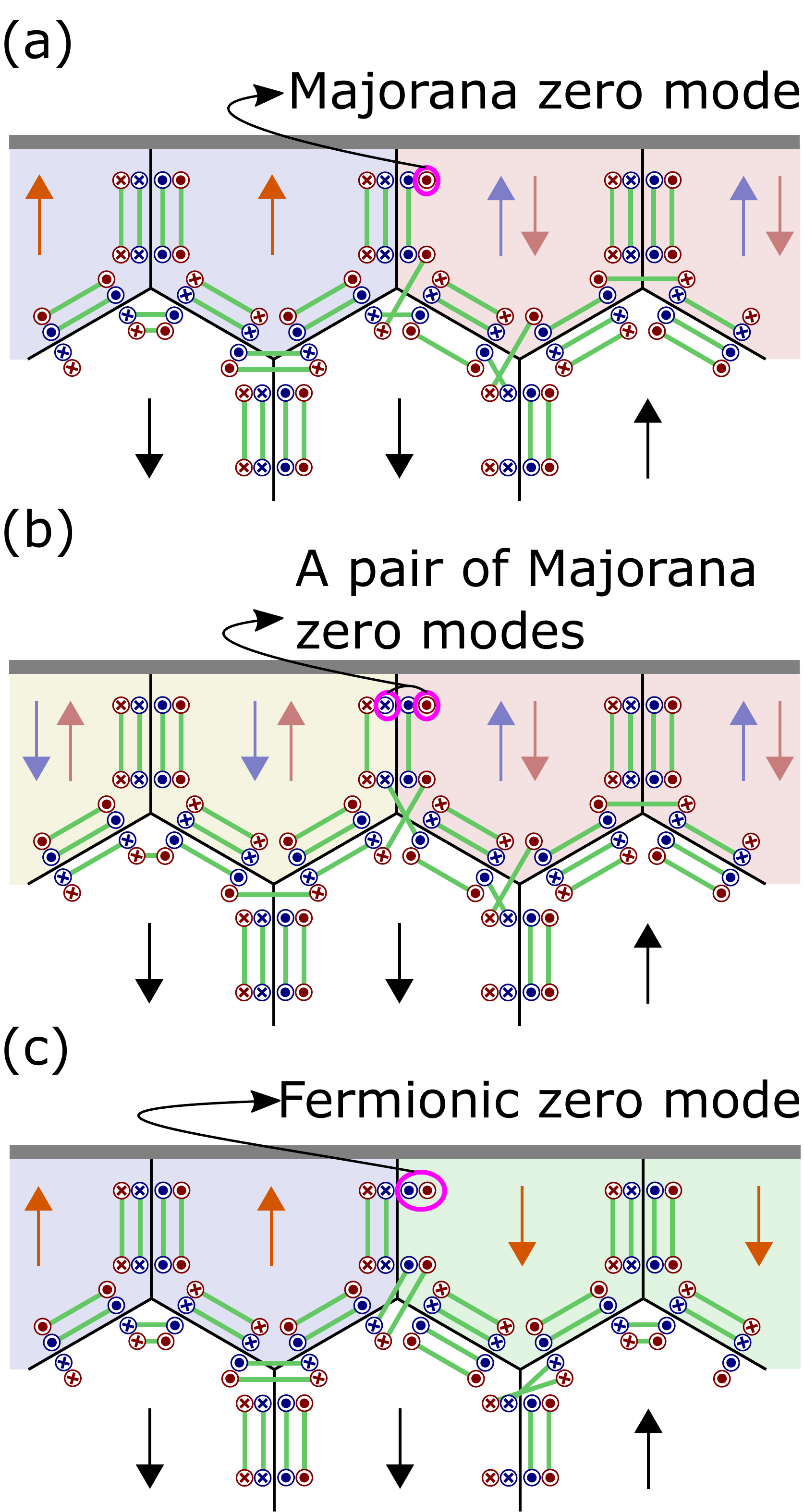}
\caption{Snapshots of microscopic configurations in the topological-insulator commuting-projector model that give rise to the zero modes sketched in Fig.~\ref{fig:6_freefdomain}.  
}
\label{fig:7_zeromodeexact}
\end{figure}

Within our framework, we can most readily capture the Majorana zero mode at a domain wall between superconducting and ferromagnetic edge phases.  To implement such a domain wall we break U$(1)$ in one region of the edge and time-reversal symmetry in another, precisely as described in the preceding subsection.  For concreteness we arbitrarily polarize the spins up in the magnetized domain.  Recalling that Majorana fermions in layer 1 pair as if the eliminated boundary spins pointed up, one sees that the domain wall does not affect Majorana pairings in that layer.  On the contrary, Majorana fermions in layer 2 pair as if the eliminated spins pointed down---yielding an `effective magnetic domain wall' for layer 2 that does influence Majorana pairings, and in particular indeed traps the expected unpaired Majorana zero mode as sketched in Fig.~\ref{fig:7_zeromodeexact}(a). 

Consider next a $\mathcal{T}$-invariant domain wall at which the superconducting-order-parameter phase jumps by $\pi$. In the previous subsection we established that the effective orientations of eliminated edge spins for layers 1 and 2 at $\theta = \pi$ are exactly opposite those at $\theta = 0$.  Hence, there is now an effective magnetic domain wall for \emph{both} layers. As Fig.~\ref{fig:7_zeromodeexact}(b) illustrates, we thereby obtain one Majorana zero mode from each layer that together form the expected Kramer's pair.


We can also readily describe U$(1)$-invariant magnetic domain walls in our setup by simply polarizing the boundary spins up in one region and down in another via a position-dependent Zeeman field.  As seen in Fig.~\ref{fig:7_zeromodeexact}(c), our Majorana-pairing rules dictate that a magnetic domain wall binds two uncoupled Majorana zero modes $\gamma_{v,\uparrow,1}$ and $\gamma_{v,\uparrow,2}$---or equivalently one complex fermion zero mode $f_{0}^{\dagger} = \gamma_{v,\uparrow,1} + i \gamma_{v,\uparrow,2}$.  Under an element U$_\alpha$ of U$(1)$, the complex fermion zero mode transforms as $f_{0}^{\dagger} \rightarrow e^{i\alpha}f_{0}^{\dagger}$, signaling that $f_{0}$ carries a single unit of electric charge $e$.  Hence, if the magnetic domain wall binds charge $q$ in a state with $f^\dagger_0 f_0 = 0$, then the charge with $f^\dagger_0 f_0 = 1$ must be $q + e$.  We will now argue that $q = -e/2$.  
 
 
As a primer, suppose that we completely gap out the boundary with a \emph{uniform} Zeeman field (i.e., no magnetic domain walls).  To fix a notion of charge neutrality, it is useful to work with complex fermions $f_{v,s} = \gamma_{v,s,1} + i \gamma_{v,s,2}$ that carry charge $e$ and define number operators $n_{v,s} = f_{v,s}^\dagger f_{v,s}$.  When Majoranas with spin $s$ at vertex $v$ pair with Majoranas with spin $s'$ at vertex $v'$, the state satisfies the constraint $i\gamma_{v,s,1}\gamma_{v,s',1} =  i\gamma_{v,s,2}\gamma_{v,s',2} = \pm 1$. Translating into complex fermion language, one finds that configurations satisfying this constraint involve an equal superposition of a state with $n_{v,s} = 1, n_{v',s'} = 0$ and another state with $n_{v,s} = 0, n_{v',s'} = 1$.  This property indicates that when the edge is gapped out with a uniform Zeeman field (or when the system is defined without a boundary), the ground state of our model always resides exactly at \emph{half-filling}.

  
Let us now examine a topological insulator defined on a manifold with a gapped boundary hosting a pair of magnetic domain walls, one hosting a complex fermion zero mode $f_0$, the other hosting $f_0'$.  The key observation is that \emph{exactly one of these zero modes must be occupied in the charge neutrality ground state}.  The state with both zero modes vacant thus has charge $-e$ relative to the ground state, while the state with both zero modes occupied has charge $+e$ relative to the ground state.  Assuming symmetric charge assignments, we conclude that each magnetic domain wall binds the expected fractional charge of $+e/2$ or $-e/2$, i.e, $q = -e/2$ as claimed above.  


 
\section{Gapless Edge State of the $\mathcal{T}$-invariant topological insulator}
\label{sec:edgegapless}

 In this section, we examine the dangling-bond edge-termination geometry in the limit where both time-reversal \emph{and} U$(1)$ symmetries are preserved everywhere.  The boundary is necessary gapless in this case, implying that the Hamiltonian can no longer consist solely of commuting projectors (which would imply a fully gapped spectrum).  Starting from our 2D model, we will nevertheless `peel off' a \emph{strictly 1D microscopic Hamiltonian} whose low-energy physics exactly reproduces that of the familiar 2D topological insulator edge. Our effective 1D model and its physics can be viewed as a natural generalization of 1D models that recently appeared in Ref.~\onlinecite{jones2019} for symmetry-protected gapless edges of a $\mathcal{T}$-symmetric topological superconductor and the Tarantino-Fidkowski model \citep{Tarantino2016}.  However, our derivations and presentation differ from those of Ref.~\onlinecite{jones2019}; we hope that these distinctions give readers complementary viewpoints on gapless edge states of fermionic symmetry-protected topological phases.  
 


\subsection{Symmetric edge termination}

The degrees of freedom in the dangling-bond edge termination are identical to those of the $\mathcal{T}$-breaking gapped boundary examined in Sec.~\ref{GappedSec}: eight Majorana fermions per edge together with Ising spins for every plaquette, including incomplete ones at the edge.  Just as for the $\mathcal{T}$-breaking case, vertex terms $A_t$ require no modification.  The key addition here involves incomplete plaquette terms.  In the $\mathcal{T}$-breaking construction, we polarized spins in incomplete plaquettes with Zeeman terms.  Instead, now we assemble $B_p$-like plaquette terms, denoted $C_{I}$ for each incomplete plaquette $I$, that allow spins and Majorana pairings to fluctuate at the edge.

 
\begin{figure}
\includegraphics[width=\linewidth]{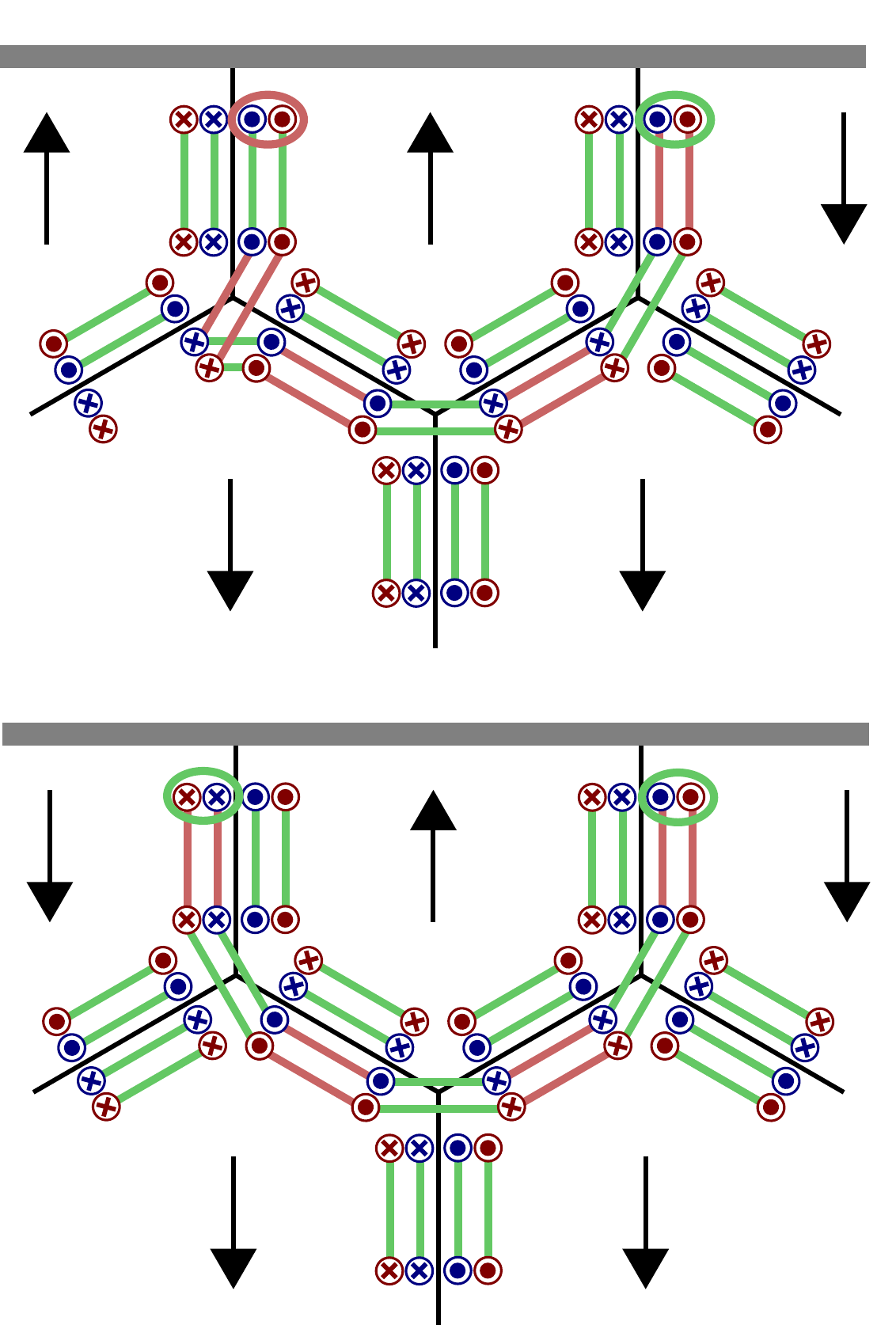}
\caption{Two examples of the action of $\mathcal{C}_{u_{I}}$, which reconfigures Majorana pairings near `incomplete plaquettes' according to Eq.~\eqref{eq:cI}. As in Fig.~\ref{fig:3_bp}, green lines correspond to Majorana pairings consistent with the original spin configuration $u_{I}$, and red lines correspond to Majorana pairings that are enforced upon flipping the spin at plaquette $I$. The major difference from the usual plaquette term $B_{p}$ is the potential existence of unpaired Majorana fermions.  In the example shown here, Majoranas in green circles are unpaired in the starting spin configuration, while those in the red circle are unpaired in the modified spin configuration.  }
\label{fig:cipic}
\end{figure}
 
Each incomplete-plaquette term is given by
\begin{equation}
\label{eq:cI}
\begin{split}
&C_{I}  = \sum_{u_{I}} \mathcal{C}_{u_{I}} \sigma_{I}^{x}\mathcal{S}_{u_{I}} \\
&\mathcal{C}_{u_{I}} = N_{u_{I}} \left( \prod_{(v_{1}s_{1},v_{2}s_{2}) \in \overline{\mathcal{P}}_{u_{I}}} 2 P_{v_{1}s_{1},v_{2}s_{2}}^{\text{TI}} \right) \\
& \left( \prod_{(v_{3}s_{3},v_{4}s_{4}) \in \mathcal{P}_{u_{I}}}  P_{v_{3}s_{3},v_{4}s_{4}}^{\text{TI}} \right) .
\end{split}
\end{equation} 
Above, $u_{I}$ refers to possible configurations of the five spins residing at incomplete plaquette $I$ and the four surrounding plaquettes; and $\mathcal{S}_{u_{I}}$ projects onto spin configuration $u_{I}$; and  $N_{u_{I}}$ is a normalization factor that we will specify shortly.  The set $\mathcal{P}_{u_{I}}$ consists of Majorana pairings enforced by spin configuration $u_{I}$ according to the usual local rules, while $\overline{\mathcal{P}}_{u_{I}}$ consists of pairings consistent with $u_I$ but with $\sigma^z_I$ flipped.  Hence, the product of projectors in $\mathcal{C}_{u_I}$ merely reconfigure the Majorana fermions, subject to the caveat below, such that they pair appropriately given the flipped incomplete-plaquette spin (the factor $2$ is added for normalization). Figure~\ref{fig:cipic} illustrates the action of these incomplete-plaquette terms. 

The expression for $C_{I}$ resembles the bulk plaquette term $B_{p}$ from Eq.~\eqref{eq:tibp}. Two key differences do, however, arise:
\begin{itemize}
\item As seen in Fig.~\ref{fig:cipic}, certain spin configurations before/after acting $C_{I}$ keep some Majorana fermions unpaired;  the action of the usual bulk plaquette terms, by contrast, always keep all Majorana fermions paired.
\item Consequently, the normalization $N_{u_{I}}$ must be more carefully chosen for $C_{I}$ to be Hermitian, as opposed to the uniform normalization $1/2$ for $B_{p}$ in Eq.~\eqref{eq:tibp}. While more than one choice of $N_{u_{I}}$ renders $C_{I}$ properly Hermitian, we take
\begin{equation}
\label{eq:normf1}
N_{u_{I}} = \begin{cases}
\frac{1}{2} & \text{If $u_{I}$ enforce four Majoranas to be unpaired}\\
1 &\text{else}
\end{cases}.
\end{equation}
\end{itemize}


Due to the structure built into the incomplete-plaquette terms, $C_{I}$ somewhat obviously commutes with all $A_{t}$ vertex terms. A slightly less obvious fact is that $C_{I}$ also commutes with all bulk plaquette terms $B_{p}$, which we investigate in detail in Appendix~\ref{app:edge}. We emphasize, however, that $C_{I}$'s do \emph{not} commute with themselves---precisely this non-commutativity underlies interesting gapless edge physics.

\subsection{Derivation of the effective 1D model for the edge}

 The fact that the low-energy physics is solely controlled by $C_{I}$ terms that commute with all bulk terms suggests the possibility of `stripping away' the bulk degrees of freedom to obtain a 1D Hamiltonian describing the gapless boundary.  However, some difficulties arise in this procedure as we now briefly discuss.  
 
\begin{figure}
\includegraphics[width=\linewidth]{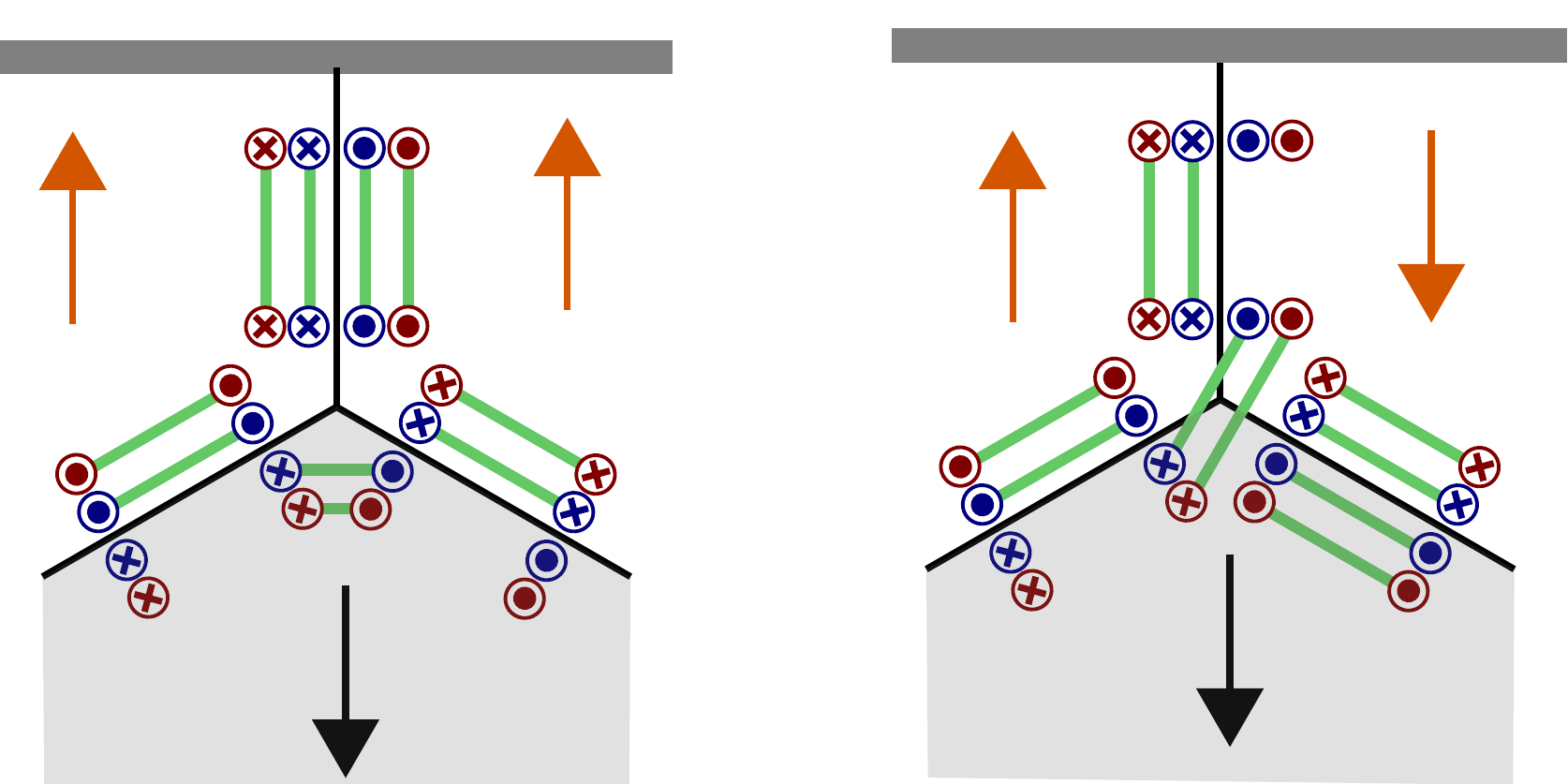}
\caption{Configurations illustrating the difficult with naively discarding bulk disagrees of freedom to obtain a 1D Hamiltonian governing the edge without breaking $\mathcal{T}$ symmetry.  With edge spins aligned as on the left, Majorana fermions in the grey region decouple from those in the white region.  With antiparallel edge spins as on the right, however, Majorana fermions pairs pair between the white and grey regions.  Thus edge and bulk degrees of freedom necessarily entangle when the edge spins fluctuate.}
\label{fig:paradox1D}
\end{figure}

 The low-energy physics occurs in the subspace that satisfies the constraint $A_{t} = B_{p} =1$ for all vertices and all bulk plaquettes.  Consider states that strictly obey this constraint. The spin operator $\sigma_{I}^{z}$ at an incomplete plaquette commutes with bulk terms, so we can further imagine states with a fixed boundary spin configuration. In the left panel of Fig.~\ref{fig:paradox1D}, two edge spins point up; there, bulk Majorana fermions in the grey area exhibit fluctuating pairings due to the $B_{p}=1$ constraint but never pair with edge Majoranas in the white area. One might therefore be tempted to derive an effective 1D edge models by simply throwing out Majoranas and spins in the grey region. However, the right panel of Fig.~\ref{fig:paradox1D} shows a different edge spin configuration which imposes pairing between Majorana fermions in the grey and white regions.  Thus one can not naively discard degrees of freedom in the grey area to derive a 1D edge model.  In fact, any choice of a `grey area'  specifying bulk degrees of freedom to be discarded similarly yields some states that entangle the `bulk' and `edge' degrees of freedom.  This conclusion reflects the familiar fact that the symmetric edge states of a quantum-spin-Hall insulator are anomalous and can not be sliced away from the accompanying bulk.    
 
 
 
Progress can nevertheless proceed if one sets out to construct a 1D lattice model \emph{whose physics at all scales may not be identical to the 2D Hamiltonian we consider, but whose physics in a certain low-energy sector certainly is}. Here we only establish the basic ideas, relegating a more detailed discussion to Appendix~\ref{app:edge}. Consider the following family of Hamiltonians parametrized by $x$:
\begin{equation}
H(x) = -a\sum_{t} A_{t} -b\sum_{p} \left( \sqrt{1-x^{2}}B_{p} + x \sigma_{p}^{z} \right) - c\sum_{I} C_{I}
\end{equation}
We keep $a,b$ sufficiently larger than $c$ so that the low-energy physics is controlled by the $C_{I}$ terms.  The limit $H(0)$ corresponds to our fully symmetric topological-insulator model. Increasing $x$ from zero to one tunes the bulk to a trivial insulator at the cost of breaking $\mathcal{T}$ symmetry. Note that $\sigma_{p}^{z}$ also commutes with $C_{I}$, so here as well all bulk and edge terms commute.

\begin{figure}
\includegraphics[width=\linewidth]{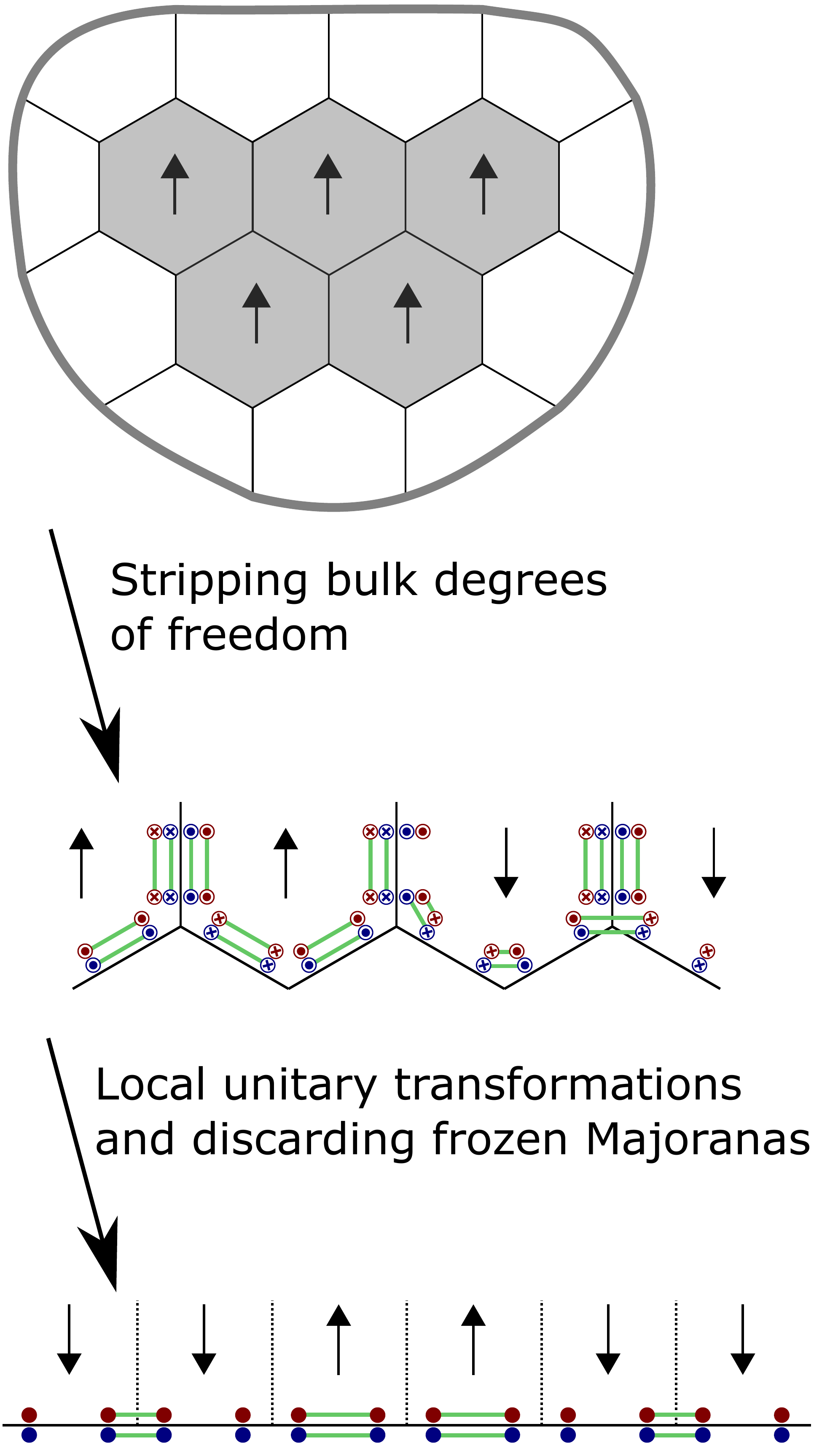}
\caption{Outline of our derivation of the strictly 1D Hamiltonian that captures the physics of a quantum-spin-Hall edge.  We first polarize the bulk spins, which allows us to discard bulk degrees of freedom.  We then perform a local unitary transformation and discard unimportant degrees of freedom to arrive at a 1D model with four Majorana fermions and one bosonic spin per unit cell.  Although polarizing the bulk spins clearly breaks time-reversal symmetry, a remnant of time-reversal symmetry persists in the 1D Hamiltonian as detailed in the main text.}
\label{fig:roadmap1D}
\end{figure}

 One may naively expect that since $H(x \neq 0)$ does not possess exact time-reversal symmetry, the edge loses its nontrivial properties and becomes similar to the magnetically gapped edge discussed in Sec.~\ref{GappedSec}. However, \emph{such a bulk transformation does not affect edge dynamics, so any $H(x)$ exhibits identical low-energy physics to $H(0)$}. This conclusion follows from the fact that the bulk Zeeman terms commute with $C_{I}$'s that give non-trivial edge dynamics. Hence, we are free to analyze $H(1)$ instead, in which $B_{p}$'s are completely replaced by $\sigma_{p}^{z}$'s. In the low-energy subspace, all bulk spins freeze in one direction, and the bulk Majorana pairings become frozen as well.  One may then discard these degrees of freedom to study the edge physics. Making a further unitary transformations and discarding unimportant degrees of freedom (as also detailed in Appendix~\ref{app:edge}), we arrive at a simplified, translationally invariant 1D model with four Majorana fermions and one bosonic spin per unit cell. Figure~\ref{fig:roadmap1D} pictorially summarizes this procedure. 
 
\begin{figure}
\includegraphics[width=0.8\linewidth]{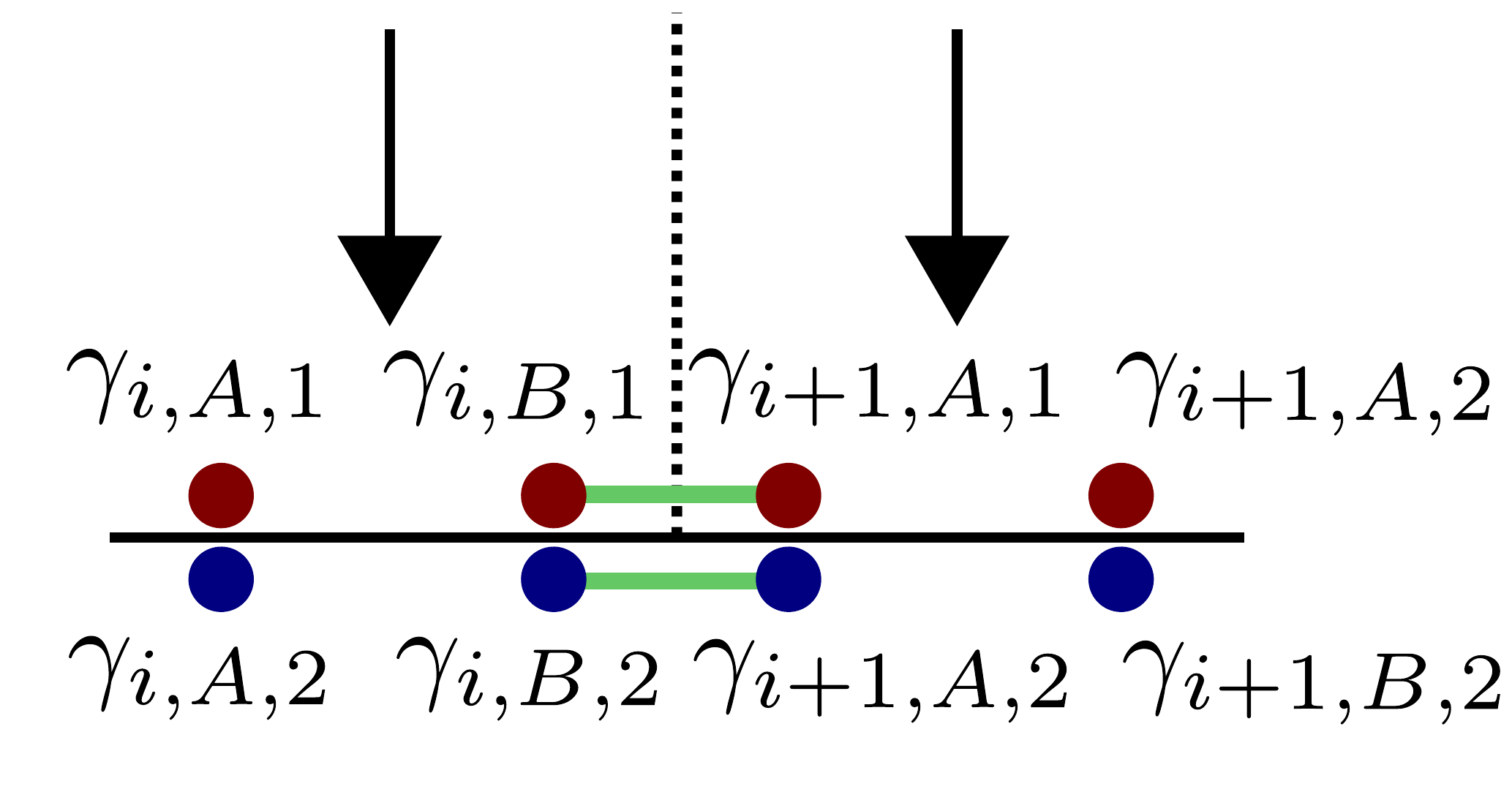}
\caption{Labeling convention for Majorana fermions in unit cells $i$ and $i+1$.}
\label{fig:label1D}
\end{figure}

Let $i$ denote 1D unit cells, and label the four Majorana fermions per unit cell as in Fig.~\ref{fig:label1D}.
Our 1D edge Hamiltonian can be written as 
\begin{equation}
\label{eq:eff1DH}
H^{L} = -\sum_{i} A_{i}^{L} - \sum_{i} C_{i}^{L}.
\end{equation}
(The meaning of the `$L$' superscripts will become apparent in the next subsection.)
The first terms are given by 
\begin{equation}
\label{eq:ail}
\begin{split}
&A_{i}^{L} = \sum_{s,t= \pm} \mathcal{A}_{i, st}^{L} \frac{1+s\sigma_{i-1}^{z}}{2}\frac{1+t\sigma_{i}^{z}}{2} \\
&\mathcal{A}_{i, --}^{L} = \frac{1+i\gamma_{i-1,B,1}^{(1)}\gamma_{i,A}^{(1)}}{2}\frac{1+i\gamma_{i-1,B}^{(2)}\gamma_{i,A}^{(2)}}{2}\\
&\mathcal{A}_{i, +-}^{L} = \frac{1+i\gamma_{i-1,A}^{(1)}\gamma_{i-1,B}^{(1)}}{2} \frac{1+i\gamma_{i-1,A}^{(2)}\gamma_{i-1,B}^{(2)}}{2} \\
&\mathcal{A}_{i, ++}^{L} = \frac{1+i\gamma_{i-1,A}^{(1)}\gamma_{i-1,B}^{(1)}}{2} \frac{1+i\gamma_{i-1,A}^{(2)}\gamma_{i-1,B}^{(2)}}{2} \\
&\mathcal{A}_{i, -+}^{L} = 1.
\end{split}
\end{equation}
The second terms read
\begin{equation}
\begin{split}
&C_{i}^{L} = \sum_{s,t,u = \pm } {\mathcal{C}}^L_{i,stu} \sigma_{i}^{x}  \frac{1+s\sigma_{i-1}^{z}}{2}\frac{1+t\sigma_{i}^{z}}{2} \frac{1+u\sigma_{i+1}^{z}}{2}  \\
& \mathcal{C}_{i,stu}^{L} = N_{stu}\mathcal{A}_{i,s\overline{t}} \mathcal{A}_{i+1,\overline{t}u} \mathcal{A}_{i,st} \mathcal{A}_{i+1,tu},
\end{split}
\end{equation}
where $N_{stu}$ takes an analogous role to $N_{u_{I}}$ in Eq.~\eqref{eq:normf1} and is defined as
\begin{equation}
N_{stu} = \begin{cases}
1 & \text{if $s=+$ and $u=+$}\\
2 & \text{else}.
\end{cases}
\end{equation}

Similar to the vertex term of the original 2D model, $A_{i}^L$ projects onto Majorana pairings consistent with bosonic spins.  The upper row of Fig.~\ref{fig:dualityT} illustrates the relevant pairing rules.  Note that here we show the pairings imposed by $A^L_i$ at a particular unit cell $i$ corresponding to the rightmost spin in each column. 
In the space where $A_{i}^L=1$ is strictly enforced for all $i$'s, Majorana fermions in each layer pair as in a non-trivial topological phase along spin-up domains, and pair as in a trivial phase in spin-down domains. The bottom of Figure~\ref{fig:roadmap1D} illustrates an example of such pairings.  (In the upper row of Fig.~\ref{fig:dualityT}, there are no green lines in unit cell $i$ when its spin points up; these `missing' green lines are filled in by $A^L_{i+1}$ to produce a pattern like that in the last row of Fig.~\ref{fig:roadmap1D}.)   
Domain walls therefore host two Majorana zero modes, reminiscent of the magnetic domain-wall structure discussed in Sec.~\ref{DomainWallSec}.  The $C_{i}$ term, derived from $C_{I}$ of the original 2D model, flips a bosonic spin and modifies Majorana pairings by projection. Note that $[A_{i},A_{j}]= [C_{i},A_{j}]=0$, though different $C_{j}$'s need not commute with each other. The low-energy physics is fully captured by the subspace in which $A_{i}=1$ is strictly observed.

\subsection{$\mathcal{T}$-symmetry in the effective 1D edge model}

\begin{figure*}
\includegraphics[width=\linewidth]{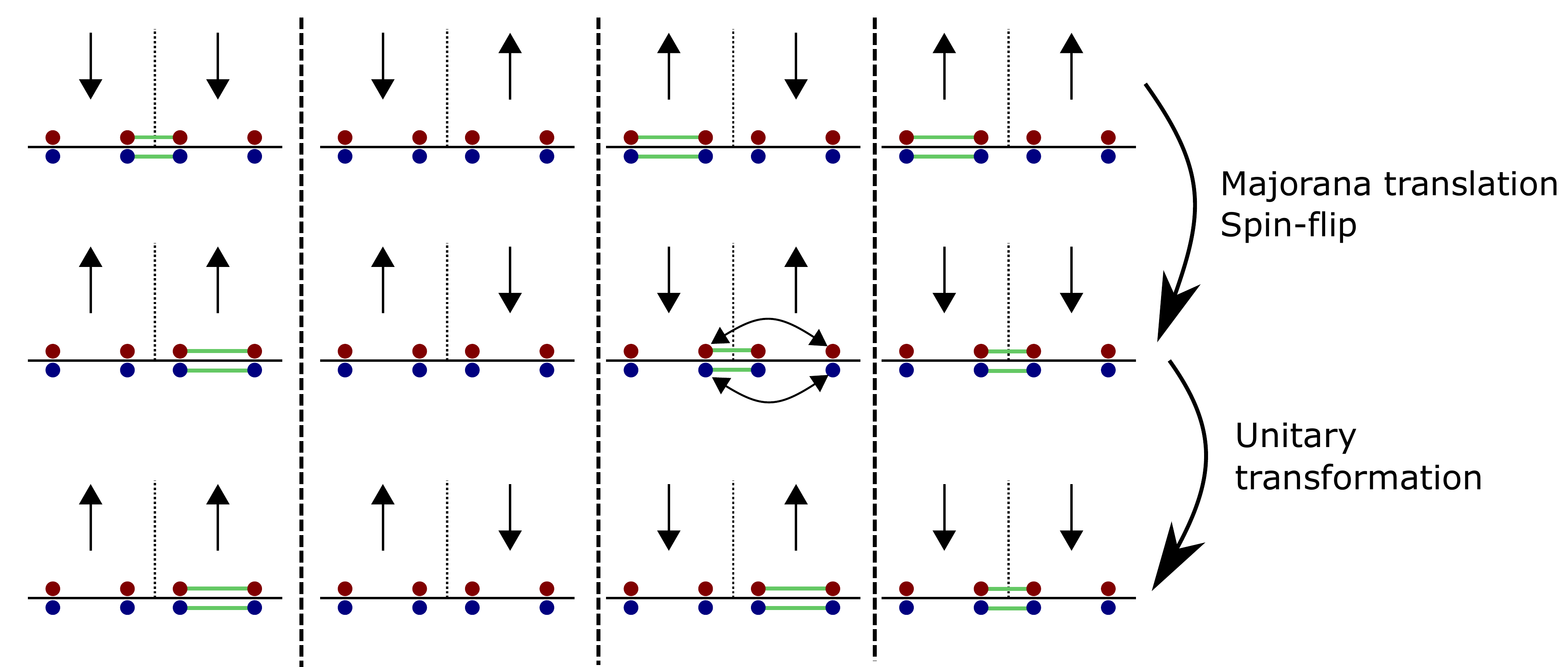}
\caption{Illustration of the `left-handed' vertex term $A_{i}^{L}$ (uppermost row) that imposes pairings Majorana pairings in our strictly 1D Hamiltonian for the quantum-spin-Hall edge, its `right handed' counterpart $A_{i}^{R}$ (lowermost row), and the duality transformation between them.}
\label{fig:dualityT}
\end{figure*}
 
Our 1D Hamiltonian $H^{L}$ still exactly preserves $U(1)$ symmetry, but is not invariant under $\mathcal{T}$.  In fact, in our derivation we explicitly broke $\mathcal{T}$-symmetry by polarizing bulk spins and threw out Kramers partners of Majoranas in the stripped-down 1D Hamiltonian.  A remnant of $\mathcal{T}$ symmetry nevertheless persists in this model as nicely explained in Ref.~\onlinecite{jones2019}; here, we recast their findings in a slightly different language.  

Define $H^{R}$, the `right-handed version' of our original `left-handed' $H^{L}$, such that $\mathcal{A}_{i,ab}^{L}$ is modified to $\mathcal{A}_{i,ab}^{R}$, defined as follows:
\begin{equation}
\label{eq:air}
\begin{split}
&\mathcal{A}_{i, --}^{R} = \frac{1+i\gamma_{i-1,B,1}^{(1)}\gamma_{i,A}^{(1)}}{2}\frac{1+i\gamma_{i-1,B}^{(2)}\gamma_{i,A}^{(2)}}{2}\\
&\mathcal{A}_{i, +-}^{R} = 1\\
&\mathcal{A}_{i, ++}^{R} = \frac{1+i\gamma_{i,A}^{(1)}\gamma_{i,B}^{(1)}}{2} \frac{1+i\gamma_{i,A}^{(2)}\gamma_{i,B}^{(2)}}{2}\\
&\mathcal{A}_{i, -+}^{R} = \frac{1+i\gamma_{i,A}^{(1)}\gamma_{i,B}^{(1)}}{2} \frac{1+i\gamma_{i,A}^{(2)}\gamma_{i,B}^{(2)}}{2}.
\end{split}
\end{equation}
We illustrate this modification in Fig.~\ref{fig:dualityT} as well. 
Interestingly, the following series of transformations map $H_{L}$ to $H_{R}$:
\begin{enumerate}
\item Global bosonic spin flip
\item Kramers-Wannier-like half-unit-cell Majorana translation combined with layer exchange, formally written as
\begin{equation}
\begin{split}
\gamma_{i,A}^{(1)} \rightarrow -\gamma_{i,B}^{(2)}, \quad \gamma_{i,A}^{(2)} \rightarrow -\gamma_{i,B}^{(1)} \\
\gamma_{i,B}^{(1)} \rightarrow \gamma_{i+1,A}^{(2)},  \quad \gamma_{i,B}^{(2)} \rightarrow \gamma_{i+1,A}^{(1)}
\end{split}
\end{equation}
\item Unitary transformation $\prod_i U_{i}$, where
\begin{equation}
\begin{split}
&U_{i} = \sum_{ab = \pm } \mathcal{U}_{i,ab}\frac{1+a\sigma_{i-1}^{z}}{2}\frac{1+b\sigma_{i}^{z}}{2} \\
&\mathcal{U}_{i,ab} = \begin{cases}
\frac{1+\gamma_{i-1,B}^{(1)}\gamma_{i,B}^{(1)}}{\sqrt{2}}
\frac{1+\gamma_{i-1,B}^{(2)}\gamma_{i,B}^{(2)}}{\sqrt{2}} &  a=-, b=+ \\
1 & \text{else}
\end{cases}
\end{split}
\end{equation}
This unitary operator transforms $\gamma_{i-1,B}^{(1)} \rightarrow -\gamma_{i,B}^{(1)}$, $\gamma_{i,B}^{(1)} \rightarrow \gamma_{i-1,B}^{(1)}$ and similarly for Majoranas on the second layer, but \emph{only when} $\sigma_{i-1}^{z}=-1$ and $\sigma_{i}^{z} =+1$.
\item Complex conjugation.
\end{enumerate}
One can check term-by-term to prove our claim. 
We argue that the above operations correspond to the implementation of $\mathcal{T}$ symmetry in our 1D model.

In the low-energy sector in which $A_{i}^{R/L}=1$ is strictly observed, $H^{L}$ and $H^{R}$ are completely identical: they exhibit identical many-body spectra, and each energy level is described by the same wavefunction.  Thus \emph{within this subspace} the 1D model preserves $\mathcal{T}$.  Due to differences in details of $A_i$'s, however, this correspondence does not hold outside of this subspace.  The situation is reminiscent of particle-hole symmetry in the lowest-Landau level of a 2D electron gas; such a symmetry clearly breaks down in the full unprojected Landau-level space. 

Although the $\mathcal{T}$-symmetry implementation is highly non-local due to Majorana-translation operations, it is explicitly anti-unitary, and also possesses correct commutation relations with U$(1)$ symmetry due to the fact that the Majorana translation additionally swaps the layer index. Hence, this nonlocal symmetry is valid to be viewed as an incarnation of physical time-reversal symmetry for the quantum-spin-Hall edge. Since our 1D model is exactly derived from a 2D quantum-spin-Hall Hamiltonian, $\mathcal{T}$ is expected to be realized only in an anomalous fashion; in our case, this `anomaly' is manifest via the non-local nature of the symmetry.  Note that \emph{either} $\mathcal{T}$ or U(1) must be realized anomalously, but not both.  Indeed, U(1) admits a simple local implementation in our 1D model similar to the action in the full 2D Hamiltonian.  
 
 Finally, we have so far discussed $\mathcal{T}$-symmetry only in the context of an infinitely long 1D chain. It is often more convenient to study a finite-size 1D model  with periodic or anti-periodic boundary conditions; more precisely, for an $L$-site chain, the projector enforcing pairing between $\gamma_{L,B}$ and $\gamma_{1,A}$ is taken to be $(1\pm i \gamma_{L,B}\gamma_{1,A})/2$, where the $+$ and $-$ signs respectively correspond to periodic and anti-periodic boundary conditions. With periodic boundary conditions, a straightforward generalization of $\mathcal{T}$-symmetry we wrote down, whereby one identifies sites $L+1$ and $1$, suffices. In the case of anti-periodic boundary conditions, however, $\mathcal{T}$ should be slightly modified as follows: In the half-unit-cell Majorana translations, $\gamma_{L,B}^{(1)}$ and $\gamma_{L,B}^{(2)}$ at the very end of the chain respectively transform to $-\gamma_{1,A}^{(2)}$ and $-\gamma_{1,A}^{(1)}$, i.e., \emph{with a minus sign}. Also, to account for the minus sign, at the last step, the unitary transformation $U_{L}^{\dagger}\prod_{i=1}^{L-1}U_{i}$ should be applied instead---in particular, the unitary at the last site $L$ is modified to its inverse. 
 


\subsection{Numerical results}

We are unaware of any exact analytical solution to $H^{L}$ or $H^{R}$.  Focusing on the restricted subspace that satisfies $A_i = 1$ for all $i$, we thus instead employ exact diagonalization to study our 1D model defined on an $L$-site chain with either periodic or anti-periodic boundary conditions.

 
\begin{figure}
\centering
\begin{minipage}{1.0\linewidth}
\includegraphics[width = 1.0\linewidth]{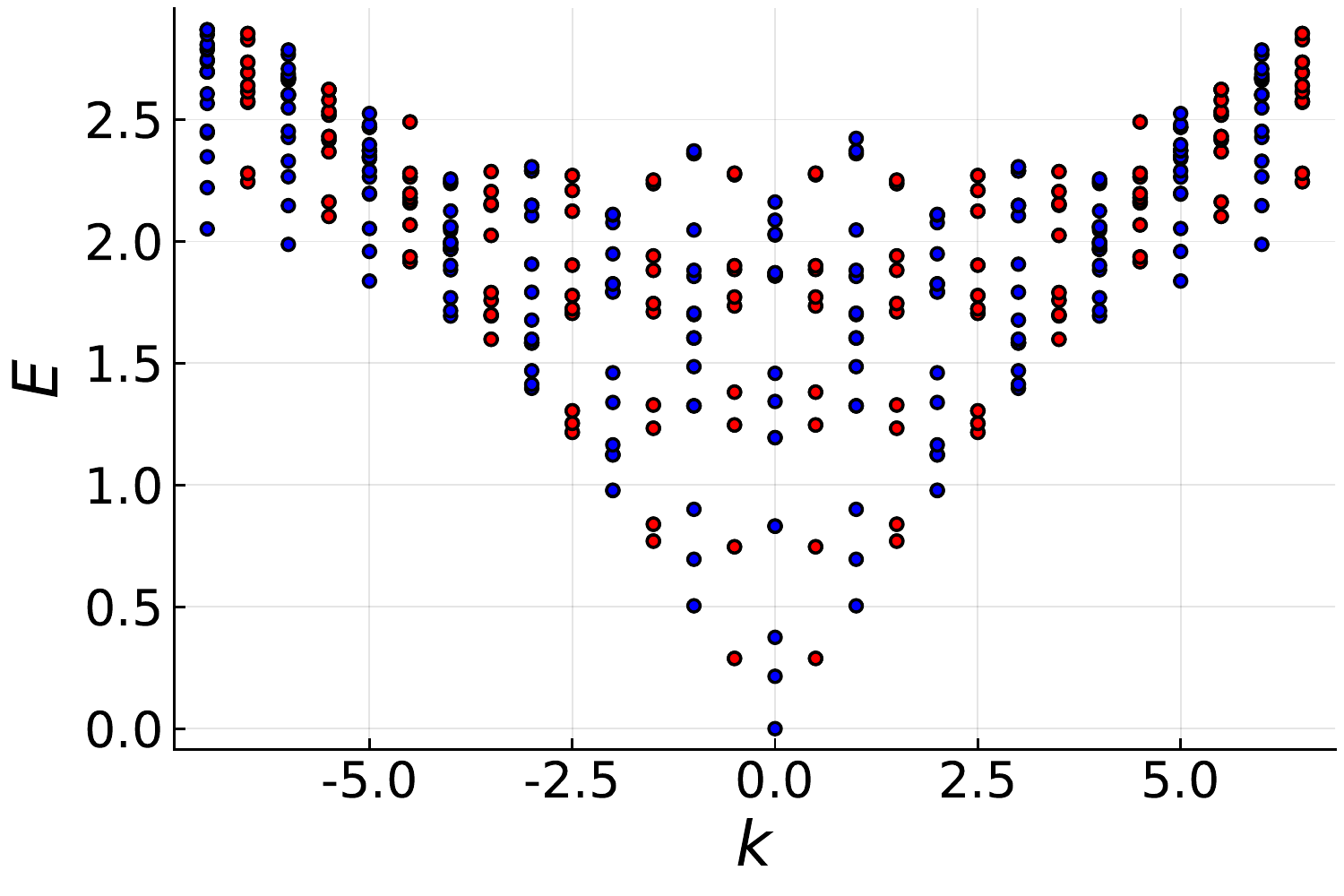}
\end{minipage}
\begin{minipage}{1.0\linewidth}
\includegraphics[width = 1.0\linewidth]{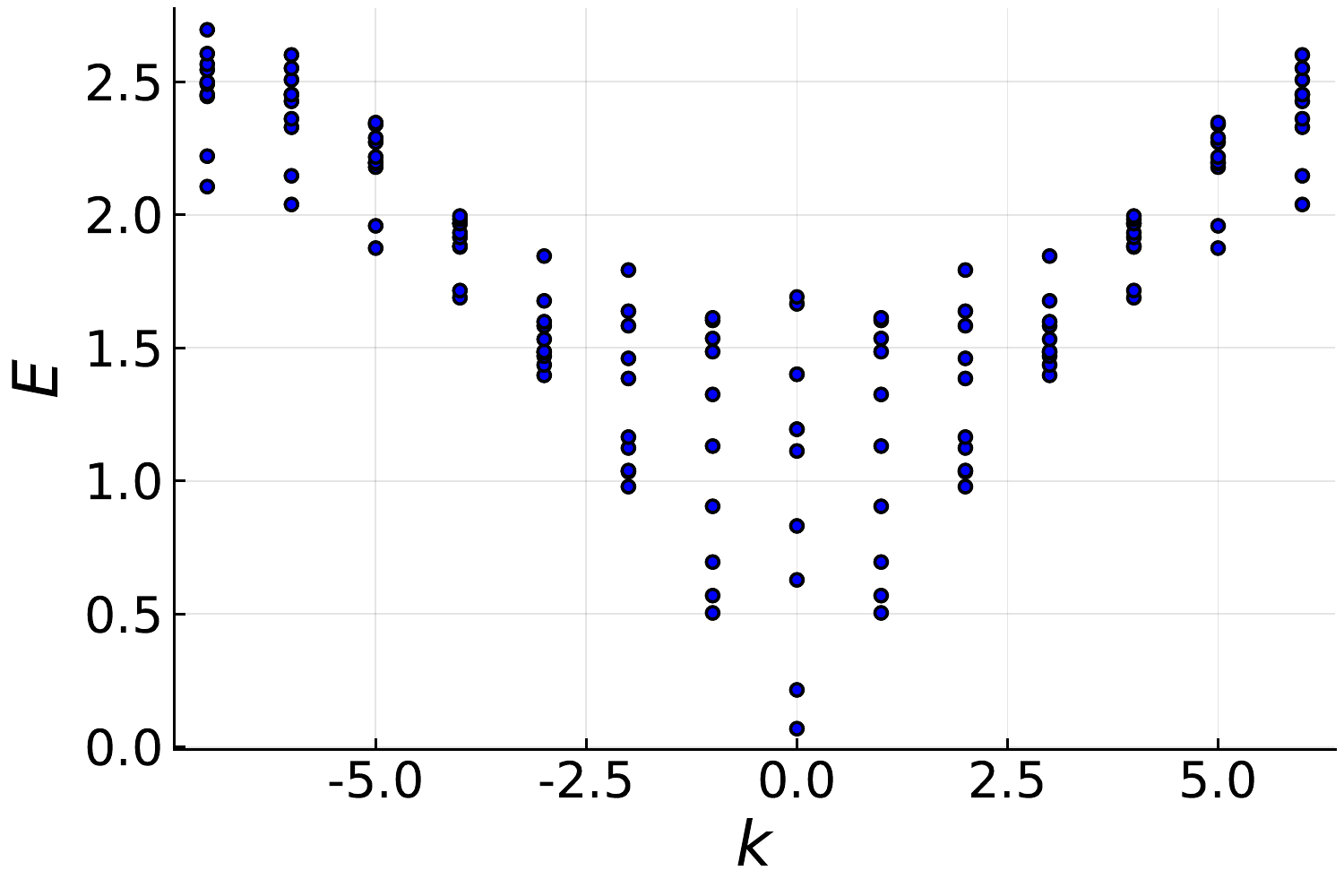}
\end{minipage}
\caption{Lowest 20 energies $E$ versus total momentum $k$ for the $L=14$ chain with anti-periodic boundary conditions (upper panel) and periodic boundary conditions (lower panel). The energy levels are shifted by an overall constant so that the ground state of the chain with anti-periodic boundary conditions has $E=0$. In the upper panel, we distinguish energy levels with momentum $k= \frac{2\pi}{L}(n+\frac{1}{2})$ and $k=\frac{2\pi}{L}n$ ($n$ is an integer) with different colors.}
\label{fig:spectrum}
\end{figure}

Figure~\ref{fig:spectrum} plots the many-body energy levels versus the total momentum $k$ for an $L  = 14$ system with anti-periodic (upper panel) and periodic (lower panel) boundary conditions.  
Eigenstates with global even fermion parity and odd fermion parity are shown.  
In both boundary conditions, the spectra are perfectly symmetric with respect to $k=0$---hence `right-movers' and `left-movers' are completely symmetric as expected from a $\mathcal{T}$-invariant quantum-spin-Hall edge. Furthermore, the dispersion of the lowest-energy states in each momentum near $k=0$ is crudely linear, indicating the Dirac-fermion nature of the excitations and allowing one to identify these lowest-lying levels as single-particle states. Other parts of the spectra, however, cannot be explained by a free Dirac fermion, indicating more complex Luttinger-liquid physics arising from interaction. Since the low-energy conformal field theory is unlikely to be a minimal model, there can be an infinite number of primary operators; hence, it is difficult to read off the full identity of the low-energy conformal field theory just from the spectrum.
 
We also point out some interesting degeneracies that appear in the spectra. For the periodic-boundary-condition spectrum from Fig.~\ref{fig:spectrum}, energy eigenstates \emph{at each momentum} are two-fold degenerate.  This degeneracy can be explained by some extra symmetries, apart from U$(1)$ and $\mathcal{T}$, that our Hamiltonian possesses.  Specifically, our Hamiltonian separately preserves the fermion parities $F_{1}$ and $F_{2}$ for layers 1 and 2, defined by
\begin{equation}
F_{a=1,2} = \prod_{i=1}^{n} i \gamma_{i,A}^{(a)}\gamma_{i,B}^{(a)}.
\end{equation}
Additionally, one can define a unitary modification of $\mathcal{T}$, which we denote $\mathcal{T}_{m}$, by removing complex conjugation operations and modifying the Majorana translation/layer-interchange from item 2 in the numbered list of the previous subsection as follows:
\begin{equation}
\begin{split}
\gamma_{i,A}^{(1)} \rightarrow \gamma_{i,B}^{(1)}, \quad \gamma_{i,A}^{(2)} \rightarrow \gamma_{i,B}^{(2)} \\
\gamma_{i,B}^{(1)} \rightarrow \gamma_{i+1,A}^{(1)}, \quad \gamma_{i,B}^{(2)} \rightarrow \gamma_{i+1,B}^{(2)}.
\end{split}
\end{equation} 
One can explicitly show that $\mathcal{T}_{m}$ also maps $H^{L}$ to $H^{R}$ and hence is a valid symmetry on the restricted subspace of interest. Since $\mathcal{T}_{m}$ commutes with a unit-cell translation but does not involve complex conjugation, it does not change momentum of an eigenstate. Finally, $\mathcal{T}_{m}$ changes the sign of both $F_{1}$ and $F_{2}$ due to the half-unit-cell Majorana translation. It follows that energy eigenstates at a given momentum always come in degenerate pairs, each one having opposite signs of $F_{1}$ and $F_{2}$.  When applied to the $k = 0$ ground state, we see that a zero-momentum, zero-energy mode is guaranteed to exist. 
 
For a chain with anti-periodic boundary conditions, $\mathcal{T}_{m}$ should be modified in the same manner as we modified $\mathcal{T}$ at the last part of the previous subsection. This modified symmetry action now preserves $F_{1}$ and $F_{2}$, and hence energy eigenstates at a given momentum are not always two-fold degenerate. However, any eigenstate with $F_{1} \neq F_{2}$ is still two-fold degenerate because of layer-interchange symmetry.  Such $F_{1} \neq F_{2}$ states [red circles in Fig.~\ref{fig:spectrum}, upper panel] exhibit odd total fermion parity, carry momentum $k=\frac{2\pi}{L}(n+\frac{1}{2})$ for integer $n$, and are indeed doubly degenerate in our numerics.  Even-fermion-parity states, by contrast, have $F_1 = F_2$ and momentum $k=\frac{2\pi n}{L}$; layer-interchange symmetry clearly does not imply degeneracy for those states; in particular, the ground state at $k=0$ is unique.
This degeneracy can also be understood as one state being the `particle-hole conjugate' of another.  While odd-fermion states always transform nontrivially under the particle-hole-like transformation, states with the integer momentum $k=\frac{2\pi n}{L}$ can be symmetric under that transformation and thus can exist without degeneracy, particularly when the states represent neutral excitations.
 
\begin{figure}
\includegraphics[width = 1.0\linewidth]{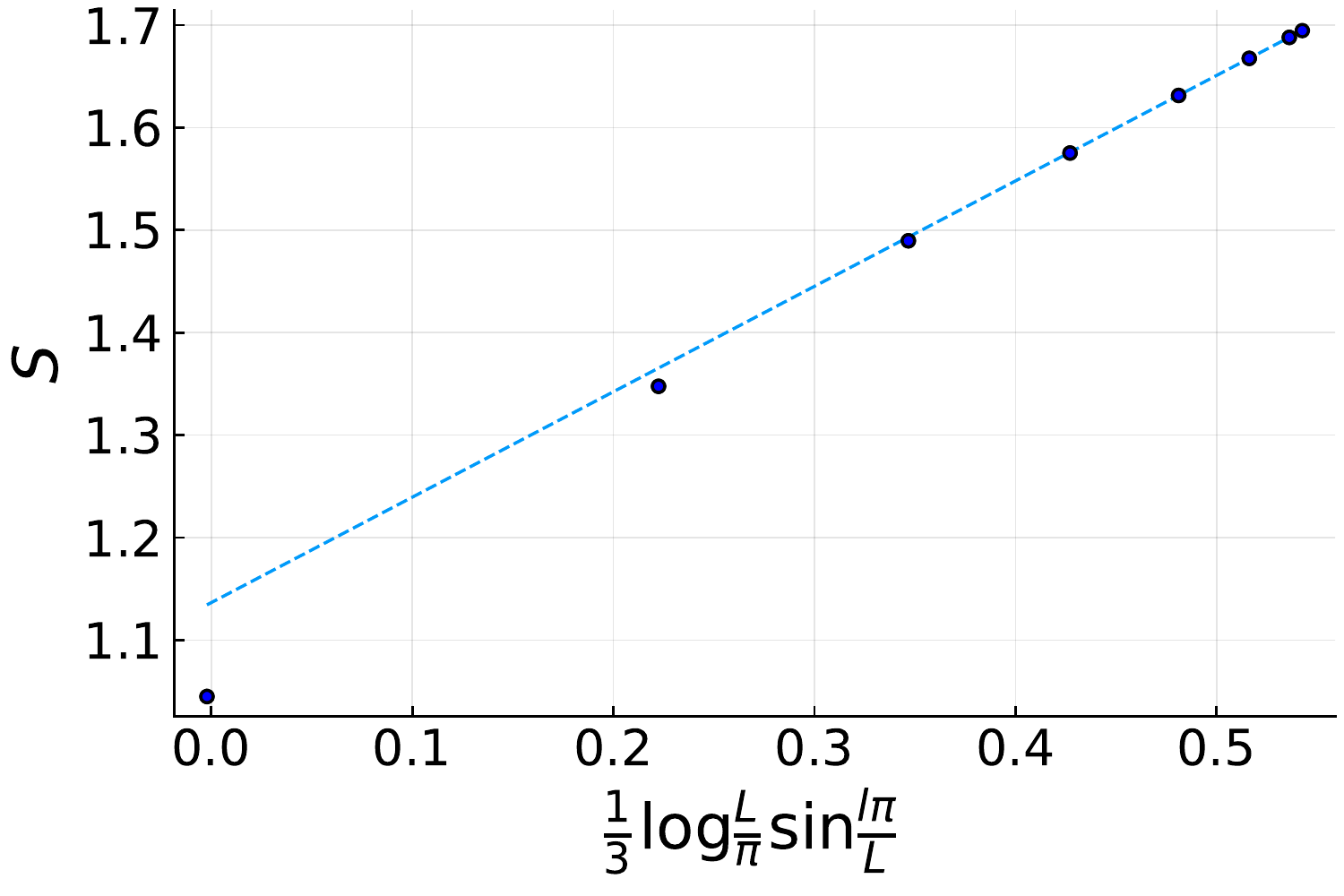}
\caption{Entanglement entropy $S$ versus $x(l) = \frac{1}{3} \ln{\frac{L}{\pi}\sin\left( \frac{l\pi}{L} \right)}$ for our strictly 1D boundary Hamiltonian, along with a linear regression line used to estimate the central charge $c$.  We find $c \approx 1$ as expected for a gapless, symmetric quantum-spin-Hall edge.}
\label{fig:EE}
\end{figure}

As an additional check, we extract the central charge $c$ by computing the entanglement entropy of the ground state with anti-periodic boundary conditions at $L=16$. It is known that if the system's low-energy physics is described by a conformal field theory, the entanglement entropy of a subsystem of size $l$ takes the form \citep{Calabrese2009}
\begin{equation}
S(l) = \frac{c}{3} \ln{\frac{L}{\pi}\sin\left( \frac{l\pi}{L} \right)}.
\end{equation}
Figure~\ref{fig:EE} shows the entanglement entropies versus $x(l) = \frac{1}{3} \ln{\frac{L}{\pi}\sin\left( \frac{l\pi}{L} \right)}$ for $l = 1,\ldots, 8$.  Performing a linear fit for the $l = 5,6, 7, 8$ data yields the relation $S(l) \approx  1.0285 x(l) + 1.1366$ (see dashed line in the figure), consistent with central charge $c = 1$ expected for a helical Luttinger liquid arising at a quantum-spin-Hall edge.  One can in principle perform other consistency checks such as extracting anomalous fermion scaling dimensions by computing fermion correlation function; we leave more detailed numerical studies of the 1D model for future work.


\section{Topological superconductor and insulator on a Klein bottle: construction and many-body invariants}
\label{sec:kleinCP}

The topological-superconductor construction presented in Ref.~\onlinecite{Wang2017}, along with our topological-insulator generalization, implicitly assume that the manifold on which the models are defined is orientable.  In this section, we first discuss how to adapt these models to a Klein bottle manifold in a $\mathcal{CP}$-symmetric manner. 
Then, we compute many-body topological invariants for $\mathcal{CP}$-symmetric topological phases \citep{Witten2016,Hsieh2014,Shiozaki2018} by changing the boundary condition along a non-contractible cycle and show that they acquire non-trivial values. We finalize the section with comments on how many-body invariants originally devised for $\mathcal{CP}$-symmetric topological phases may work for models with only $\mathcal{T}$ symmetry.
 
\subsection{Defining the lattice models on a Klein bottle}

\begin{figure}
\includegraphics[width=\linewidth]{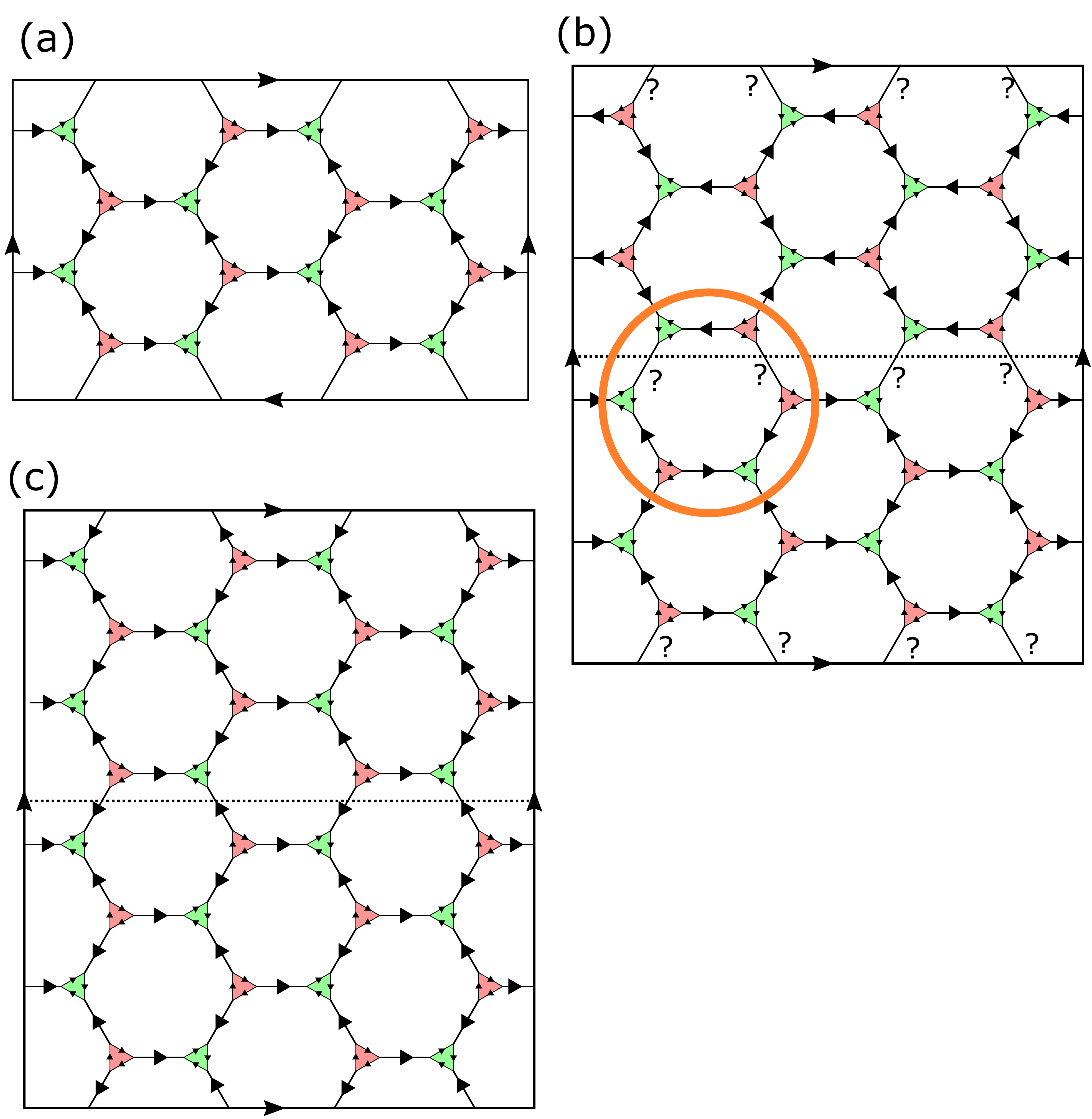}
\caption{(a) Sketch of the lattice defined on a Klein bottle.  Left and right edges of the rectangle are identified as indicated by the parallel arrows; upper and lower edges, however, are identified in a flipped manner as indicated by the antiparallel arrows.  (b) Double-cover representation of the Klein bottle; this representation defines a torus, but with the upper and lower rectangles representing reflected copies of each other.  The `branch cut' that separates the conjoined rectangles appears as a dotted line. As discussed in the main text, the encircled plaquette highlights a subtlety with choosing Kasteleyn arrows along edges crossing the branch cut.  (c) Kasteleyn orientation on the `twisted' double cover, which is identical to the orientation one would choose for an ordinary torus.}
\label{fig:cover}
\end{figure}

 Here, we illustrate some difficulties in naively defining our lattice models on a Klein bottle and how to overcome them. We focus on the topological superconductor; the generalization to the topological insulator is straightforward and will only be briefly discussed at the end. Imagine putting the honeycomb-lattice model on a rectangle as in Fig.~\ref{fig:cover}(a).  The two vertical edges of the rectangle are identified in the standard way, as indicated by the accompanying aligned arrows.  The two horizontal edges, however, are identified with opposite arrows.  Hence we have `twisted' the rectangle as if making a Mobius strip, indicating the non-orientable nature of the Klein bottle.
 
It is useful to examine the `double cover' representation of spin configurations and corresponding Majorana pairings on the Klein bottle; see Fig.~\ref{fig:cover}(b), which can be constructed by conjoining two rectangles with identical Majorana pairings and spins along horizontal edges, \emph{in a way that one rectangle is spatially reflected}. By identifying the horizontal edges and vertical edges of this `double rectangle', as illustrated by the arrows in Fig.~\ref{fig:cover}(b), one can construct a torus which is a double cover of the Klein bottle. For convenience we will retain the line between the conjoined rectangles, hereafter denoted as a branch cut. All $A_t$ and $B_{p}$ terms can be represented on this double cover as well.  Note that a single vertex or plaquette term corresponds to two different possible terms in the double cover due to its 2-to-1 nature.
  
 Consistently assigning Majorana pairings requires Kasteleyn orientations that satisfy the clockwise-odd rule.  It is relatively straightforward to choose Kasteleyn orientations within a single rectangle, utilizing the rules we discussed in Sec.~\ref{sec:model}. However, for edges that cross the branch cut, two problems arise:
 
\begin{itemize}
\item Long edges usually connect vertices on different sublattices. However, due to the peculiar identification scheme adopted in the double cover, long edges that cross the branch cut actually connect vertices on the same sublattice. Hence, from the rule we specified before, it is fundamentally ambiguous what arrows one chooses here. These problematic edges are indicated with question marks in Fig.~\ref{fig:cover}(b).
\item Even if one chooses the orientation along long edges in some way, it is now impossible to enforce the clockwise-odd rule consistently on each cycle. For example, consider the plaquette in Fig.~\ref{fig:cover}(b) circled in orange.  There, due to the Klein-bottle twist, arrows on small triangles above the dotted line are clockwise-even, while arrows on small triangles below the line are clockwise-odd. Hence, plaquette terms cannot be constructed in a way that preserves fermion parity.
\end{itemize}
 
 To rectify these issues, we attack the problem in reverse: We will construct a different double-cover representation, deduce how to correctly represent each Majorana pairing consistent with a given spin configuration, and then construct the Hamiltonian. Let us now assign sublattices and Kasteleyn orientations on our double rectangle as if the system was defined on an ordinary torus; see Fig.~\ref{fig:cover}(c).  The aforementioned pathology does not exist in this choice. However, all Kasteleyn orientations on the second rectangle are opposite of the original double cover, indicating that the second rectangle should be identified with the first in a different way. 
 
Specifically, we identify the second rectangle \emph{transformed by} $U_{\mathcal{CP},\text{TSC}}$ as the first, where $U_{\mathcal{CP},\text{TSC}}$ is defined as:
\begin{equation}
\label{eq:cpunitary}
\begin{split}
U_{\mathcal{CP},\text{TSC}} \text{ : } &\sigma_{p}^{z} \rightarrow -\sigma_{p}^{z}\\
&\gamma_{v_{A},\uparrow} \rightarrow \gamma_{v_{B}, \downarrow} \\
&\gamma_{v_{A},\downarrow} \rightarrow -\gamma_{v_{B}, \uparrow} \\
&\gamma_{v_{B},\uparrow} \rightarrow -\gamma_{v_{A}, \downarrow}\\
&\gamma_{v_{B},\downarrow} \rightarrow \gamma_{v_{A}, \uparrow}.
\end{split}
\end{equation}
Comparing with Eq.~\eqref{eq:cpact1}, we see that $U_{\mathcal{CP},\text{TSC}}$ can be understood as the `local' part of $\mathcal{CP}$ symmetry for the topological superconductor.  That is, $\mathcal{CP}_{\text{TSC}}$ is simply $U_{\mathcal{CP},\text{TSC}}$ followed by spatial inversion. This identification corresponds to what we will therefore call a $U_{\mathcal{CP},\text{TSC}}$-twisted double cover. The essential idea is that the above transformation explicitly maps any operators $i\gamma_{v,s}\gamma_{v',s'}$ that represent Majorana pairings within the second rectangle to $-i\gamma_{v,s}\gamma_{v',s'}$; hence, undoing this transformation effectively flips the Kasteleyn orientation, thus recovering the Kasteleyn orientation choice for the original double cover.  Note that we explicitly swap the sublattice index in the transformation $U_{\mathcal{CP},\text{TSC}}$; when comparing the sublattice distributions in Fig.~\ref{fig:cover}(b) and Fig.~\ref{fig:cover}(c), it is obvious that one should explicitly change the sublattice index of the upper rectangle to properly `untwist' the $U_{\mathcal{CP},\text{TSC}}$-twisted double cover.

\begin{figure}
\includegraphics[width=0.9\linewidth]{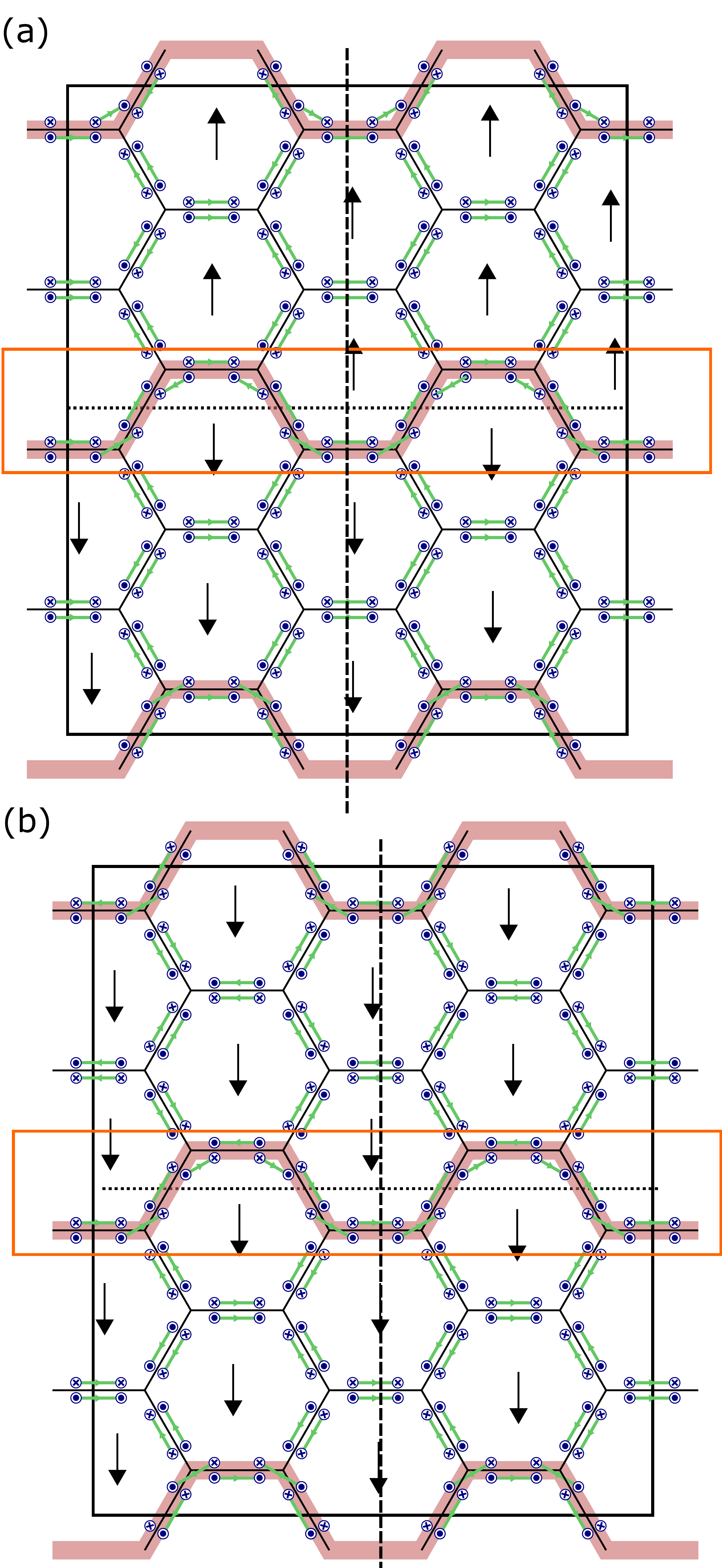}
\caption{(a) Twisted double-cover and (b) ordinary double-cover representation of all-down spin configurations and corresponding Majorana pairings on the Klein bottle. We added an arrow to the green lines to unambiguously specify the orientation of Majorana pairing.  Note the nontrivial Majorana pairings near the branch cut---which appear along a spin domain wall in (a).  }
\label{fig:doublerep}
\end{figure}

 Next, let us discuss how to represent spin configurations and Majorana pairings on the double cover. Starting from the $U_{\mathcal{CP},\text{TSC}}$-twisted double cover, we put bosonic spins on each plaquette, but considering the twist, bosonic spins on the second rectangle orient opposite those on the first. We pair Majorana fermions on the twisted double-cover following the usual vertex rule on the torus. To recover the original double cover from the twisted double cover, simply transform the upper rectangle with $U_{\mathcal{CP},\text{TSC}}$. We illustrate the double-cover representations of spin configurations and Majorana pairings in Fig.~\ref{fig:doublerep}(a) for the twisted double cover and Fig.~\ref{fig:doublerep}(b) for the untwisted one obtained by applying $U_{\mathcal{CP},\text{TSC}}$ to the upper rectangle. Note that within each rectangle Majorana pairings are fairly featureless.  The nontrivial pairing configurations reflecting the Klein-bottle manifold can be seen in the orange box, wherein Majorana pairings cross the branch cut.
 
\begin{figure}
\includegraphics[width=1.0\linewidth]{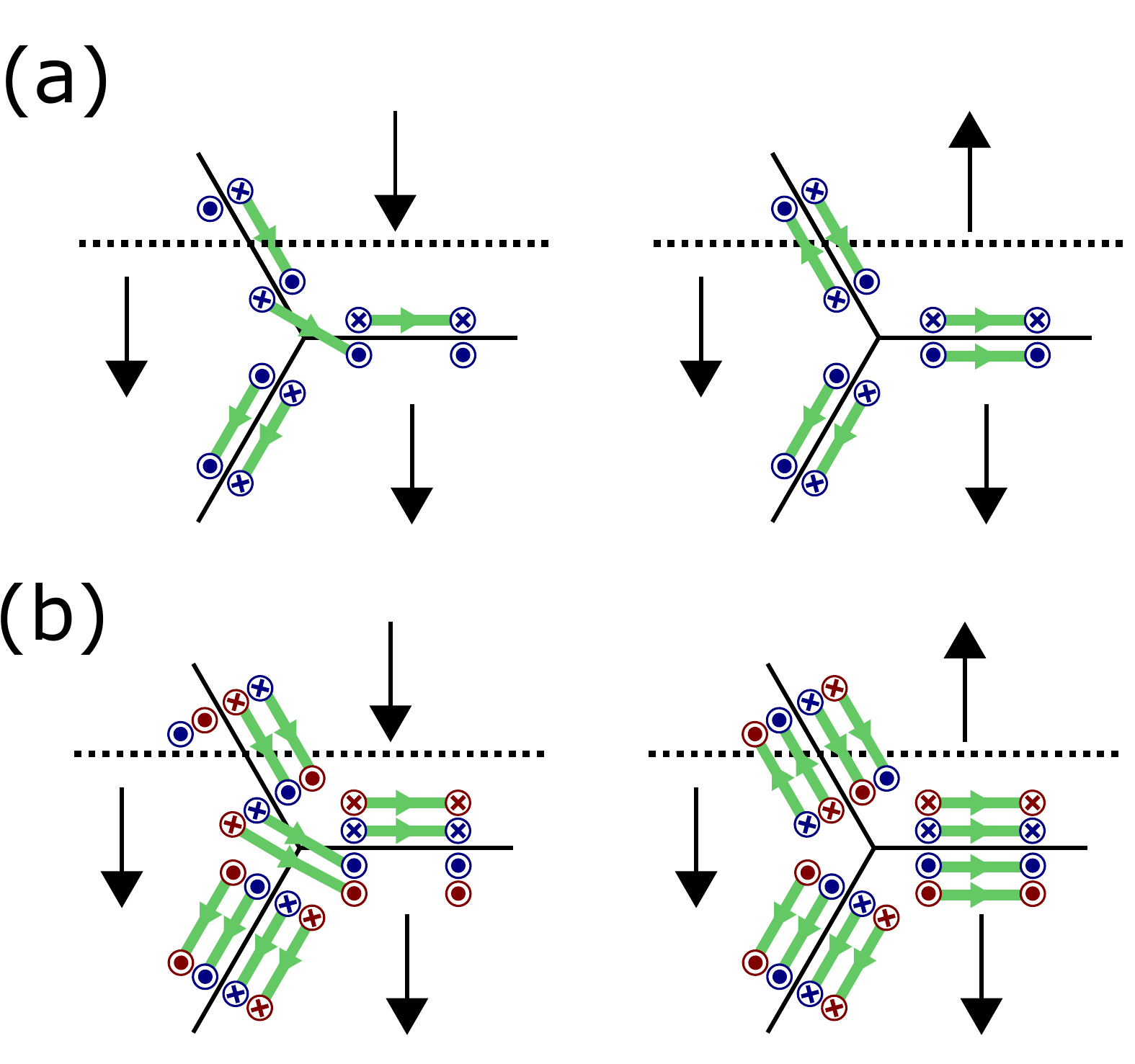}
\caption{Examples of vertex rules near the branch cut for (a) the topological-superconductor model and (b) the topological-insulator model. Here too we include arrows on the pairing lines to unambiguously specify the Majorana pairing orientations.  }
\label{fig:peculiarKlein}
\end{figure}

 We have thus constructed a double-cover representation with consistently assigned spin configurations and Majorana pairings that are free from the aforementioned pathologies. 
 From here, one can straightforwardly build $A_t$ and $B_{p}$ terms. The key modification of $A_{t}$ and $B_{p}$ comes from the following two features: First, due to the fact that untwisting the double cover flips bosonic spins, short-edge pairings no longer necessarily match with domain walls between Ising spins near the branch cut. Figure~\ref{fig:peculiarKlein}(a) illustrates some examples of anomalous vertex rules at the branch cut. Second, the untwisting procedure significantly modifies the nature of long-edge pairings across the branch cut, as is evident by explicitly constructing projectors corresponding to long-edge pairings of interest. We consider the following four projectors in the \emph{twisted} double-cover representation:
\begin{equation}
\begin{split}
&P_{v_{A}\uparrow,v_{B}'\uparrow} = \frac{1-i\gamma_{v_{A},\uparrow}\gamma_{v_{B}',\uparrow}}{2}, P_{v_{A}\downarrow,v_{B}'\downarrow} = \frac{1-i\gamma_{v_{A},\downarrow}\gamma_{v_{B}',\downarrow}}{2} \\ 
&P_{w_{B}\uparrow,w_{A}'\uparrow} = \frac{1+i\gamma_{w_{B},\uparrow}\gamma_{w_{A}',\uparrow}}{2}, P_{w_{B}\downarrow,w_{A}'\downarrow} = \frac{1+i\gamma_{w_{B},\downarrow}\gamma_{w_{A}',\downarrow}}{2}.
\end{split}
\end{equation}
To ensure the above projectors are associated with long-edge pairings across the branch cut, we will assume that $v_{A}$ and $w_{B}$ belongs to the upper rectangle in Fig.~\ref{fig:cover}(c); meanwhile, $v_{B}'$ and $w_{A}'$ belong to the lower rectangle. The untwisting procedure transforms the above four projector to
\begin{equation}
\label{eq:projklein}
\begin{split}
&P_{v_{B}\downarrow,v_{B}'\uparrow} = \frac{1-i\gamma_{v_{B},\downarrow}\gamma_{v_{B}',\uparrow}}{2}, P_{v_{B}\uparrow,v_{B}'\downarrow} = \frac{1+i\gamma_{v_{B},\uparrow}\gamma_{v_{B}',\downarrow}}{2} \\ 
&P_{w_{A}\downarrow,w_{A}'\uparrow} = \frac{1-i\gamma_{w_{A},\downarrow}\gamma_{w_{A}',\uparrow}}{2}, P_{w_{A}\uparrow,w_{A}'\downarrow} = \frac{1+i\gamma_{w_{A},\uparrow}\gamma_{w_{A}',\downarrow}}{2}.
\end{split}
\end{equation}
We note that Majorana fermions with opposite spins pair along long edges crossing the branch cut as seen in both panels of Fig.~\ref{fig:peculiarKlein}(a). Also, as seen in the right panel, two long-edge pairings that belong to the same edge across the branch cut should have opposite pairing signs.   
As for `arrows' along long edges crossing the branch cut, two pairings distinguished only by spins have opposite directions on this special type of long edge. 
 
One can locally (but not globally) convert $A_t$ and $B_{p}$ terms across the branch cut to usual vertex and plaquette terms in the `bulk' by applying $U_{\mathcal{CP},\text{TSC}}$ on one side.  Hence, this construction inherits all good local properties of the original model defined on an orientable manifold; most notably, our Hamiltonian is still a commuting-projector model, and $B_{p}$ acts as a nontrivial unitary operator on the subspace in which all vertex constraints are satisfied. Also, the `boundary-condition prescription' that we employed preserves $\mathcal{CP}$ symmetry---which arises fairly straightforwardly by observing that the untwisting operation $U_{\mathcal{CP},\text{TSC}}$ commutes with $\mathcal{CP}$ symmetry. (Commutation can be seen by how $\mathcal{CP}$ symmetry maps the projectors associated with long-edge pairings across the branch cut.) However, this prescription breaks $\mathcal{T}$-symmetry since under $\mathcal{T}$, $P_{v_{B}\downarrow,v_{B}'\uparrow}$ in Eq.~\eqref{eq:projklein} transforms to a \emph{complex conjugate} of $P_{v_{B}\uparrow,v_{B}'\downarrow}$, not $P_{v_{B}\uparrow,v_{B}'\downarrow}$ itself. The same is true for $P_{w_{A}\downarrow,w_{A}'\uparrow}$ as well. 
 
The above scheme can be straightforwardly extended to the topological-insulator model by adding one more layer. Recall that $\mathcal{CP}$ symmetry for the topological insulator involves layer interchange as well, so the preceding prescription enforces Majorana fermions to pair between different layers along the long edges crossing the branch cut; see Fig.~\ref{fig:peculiarKlein}(b), 
where the two layers are represented by different colors.
While this pairing breaks U$(1)$ symmetry, we will observe in a later subsection that this `layer twist' is a key reason why a U$(1)$-twisted boundary condition can be defined naturally on a Klein bottle.

\subsection{Many-body invariants for topological superconductors}

As an initial step towards defining a many-body invariant for the topological superconductor, we comment on a peculiar feature of the Klein-bottle manifold.  
Recall from Fig.~\ref{fig:2_vertexrule} that in our construction on orientable manifolds, Majoranas pair along short edges when there is an adjacent domain wall separating up and down bosonic spins. Hence, one can view our construction as essentially decorating topologically nontrivial Kitaev chains along spin domain walls.  
On closed, orientable manifolds that admit non-contractible cycles, domain walls and hence Kitaev chains can form along non-contractible cycles as well. Clearly there must always be an even number of such domain walls/Kitaev chains.  

But does the preceding statement continue to hold in the non-orientable Klein bottle manifold? To address this question we examine a Klein-bottle configuration with all bosonic spins down.  
As before, it is useful to consider the twisted double-cover representation, wherein such a bosonic spin configuration is represented by all-down spins in one rectangle and all-up spins in the other rectangle. Two non-contractible cycles thus form, one along each boundary between the two rectangles [see Fig~\ref{fig:doublerep}(a) for an illustration]. Nevertheless, because the double cover is a two-to-one representation, and despite all spins pointing down within the Klein bottle, \emph{the pairing rules across the branch cut imply that a Kitaev chain wraps around a single non-contractible cycle}.
Local transformations (i.e., $B_{p}$) that flip bosonic spins locally and reconfigure Majorana pairings accordingly may change the number of non-contractible cycles by an even number, \emph{but for all spin configurations and corresponding Majorana pairings the number of non-contractible Kitaev chains is always odd}.
 
 The above observations allow us to define a topological superconductor many-body invariant by changing boundary conditions. Consider the operator
\begin{equation}
F = \prod_{(vs,v's') \in \mathcal{P}_{\downarrow}} i\gamma_{v,s} \gamma_{v',s'},
\end{equation}
where $\mathcal{P}_{\downarrow}$ is a set of all $(vs,v's')$ vertex/spin labels that pair when all bosonic spins point down. Up to some factors of $i$, $F$ is a product of all Majorana operators that appear in the system; it thus defines the total fermion parity operator and indeed satisfies $F = \pm 1$. Now, we imagine flipping the boundary condition by flipping Kasteleyn orientations along edges that cross the vertical dashed line in Figs.~\ref{fig:doublerep}(a) and (b). If there is no domain wall along a given edge that crosses the vertical dashed line, then two Majorana pairings from that edge cross the dashed line; if there is a domain wall, due to presence of a non-trivial Kitaev chain along the domain wall, only one Majorana pairing crosses the dashed line. In the Klein bottle, \emph{there is always an odd number of domain walls that cross the vertical line and hence always an odd number of Majorana pairings across that line whose signs are explicitly inverted by the boundary condition change}.  The boundary-condition change consequently changes the sign of $F$.  


 It is reasonable to assert that on a closed manifold with a given boundary condition, the ground-state fermion parity can not change provided the bulk gap is finite so that the system remains in the same phase.  A trivial insulator can be smoothly deformed into a product state with all fermions strictly localized to the lattice sites; the ground state and hence the ground-state fermion parity are then clearly independent of boundary conditions. The feature in which a system defined on a Klein bottle has a different ground-state fermion parity compared to the torus thus defines a many-body topological invariant. Similar features arise in Majorana dimer models \citep{Ware2016} and $p+ip$ superconductors \citep{Green2000}, wherein putting the system on torus and changing boundary condition changes the fermion parity of the ground states.
 
\subsection{Many-body invariants for topological insulators}

Above we probed the nontrivial topological index of the topological superconductor by changing boundary conditions in a discrete way. In the case of the topological insulator, U$(1)$ symmetry allows one to continuously vary the boundary condition by twisting the pairings that cross the vertical dashed line in Figs.~\ref{fig:doublerep}(a) and (b) by a U$(1)$ variable $\theta$. Twisting is formally done as follows: For all projectors $P_{v_{1}s_{1},v_{2}s_{2}}^{(1)}P_{v_{1}s_{1},v_{2}s_{2}}^{(2)}$ associated with pairings that cross the vertical dashed line, apply the transformation 
\begin{equation}
\label{eq:projtwist}
\begin{split}
&P_{v_{1}s_{1},v_{2}s_{2}}^{(1)} P_{v_{1}s_{1},v_{2}s_{2}}^{(2)} \rightarrow P_{v_{1}s_{1},v_{2}s_{2}}^{(1)}(\theta)P_{v_{1}s_{1},v_{2}s_{2}}^{(2)}(\theta)
\\ &= \frac{1+ig_{v_{1}v_{2}}\gamma_{v_{1},s_{1},1}\gamma_{v_{2},s_{2},1}'}{2}\frac{1+ig_{v_{1}v_{2}}\gamma_{v_{1},s_{1},2}\gamma_{v_{2},s_{2},2}'}{2},
\end{split}
\end{equation}
where
\begin{equation}
\begin{pmatrix}
\gamma_{v_{2},s_{2},1}' \\
\gamma_{v_{2},s_{2},2}'
\end{pmatrix} = U(\theta) \begin{pmatrix}
\gamma_{v_{2},s_{2},1} \\
\gamma_{v_{2},s_{2},2}
\end{pmatrix}, \quad U(\theta) =\begin{pmatrix}
\cos{\theta} & \sin{\theta}  \\
-\sin{\theta} & \cos{\theta}
\end{pmatrix}.
\end{equation}
In the equations above we assume that $v_1$ and $v_2$ respectively reside on the right and left sides of the vertical dashed line.

\begin{figure}
\includegraphics[width=1.0\linewidth]{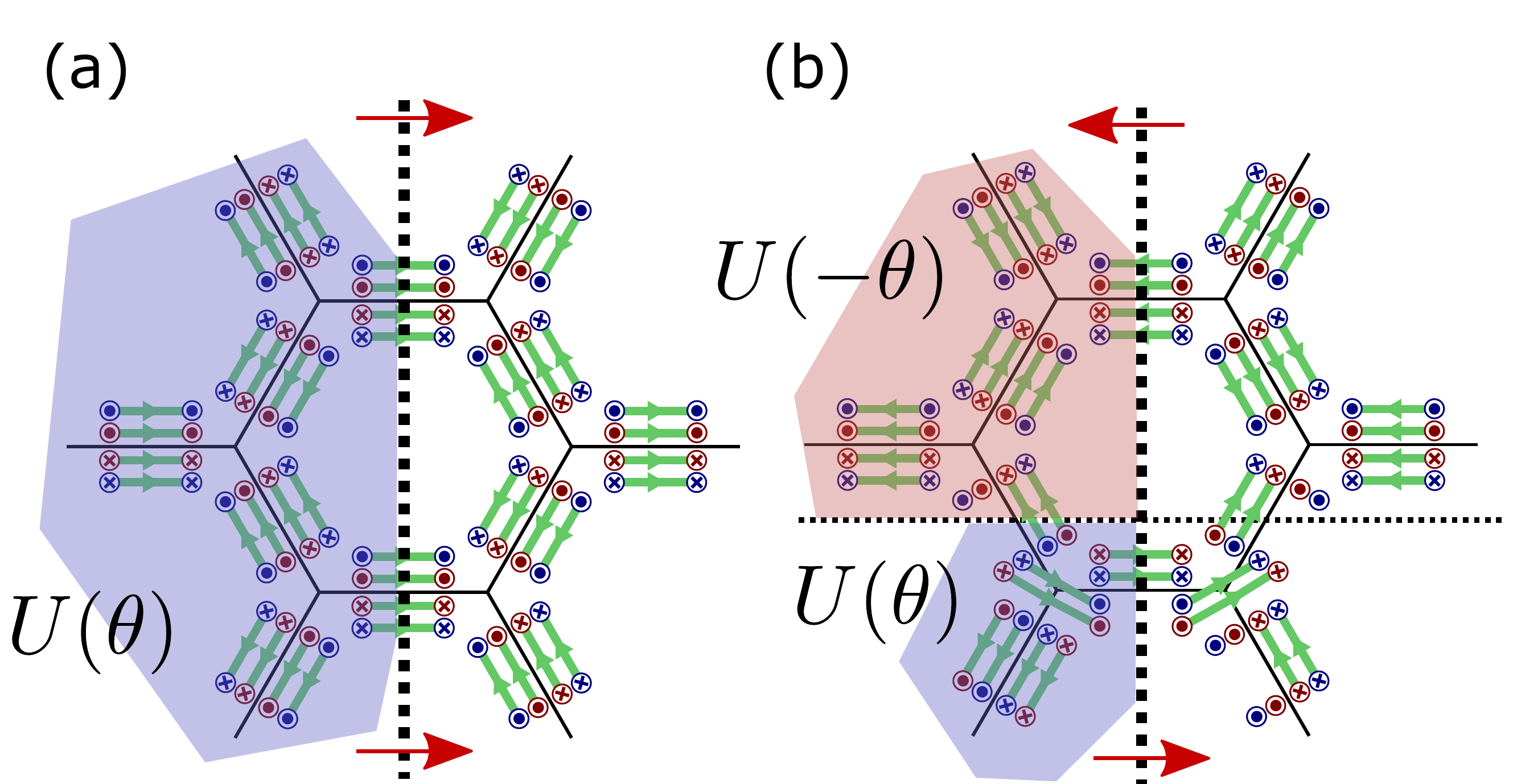}
\caption{Prescription for gauge-transforming Majorana fermions to locally `flatten' (a) the ordinary plaquette terms and (b) plaquette terms that extend across the branch cut (horizontal dashed line).  The vertical dashed line indicates the location of the boundary-condition twist. Red arrows on the vertical dashed line specify the direction in which the boundary condition is U$(1)$ twisted. Note that across the branch cut, the direction of the boundary-condition twist is reversed due to the structure of the Klein bottle.}
\label{fig:flattenBp}
\end{figure}

 This boundary condition \emph{preserves} $\mathcal{CP}$ symmetry, since although parity sends $\theta \rightarrow -\theta$, charge conjugation contributes another minus sign and returns $\theta$ to its original value. This boundary condition is `flat' in the sense that local terms $A_{t}$ and $B_{p}$ can be reverted to untwisted ones by local gauge transformations. For example, we see that the projector U$(1)$ twisted in Eq.~\eqref{eq:projtwist} reverts back to its original form by the transformation
\begin{equation}
\begin{pmatrix}
\gamma_{v_{1},s_{1},1} \\
\gamma_{v_{1},s_{1},2}
\end{pmatrix} \rightarrow 
U(\theta) \begin{pmatrix}
\gamma_{v_{1},s_{1},1} \\
\gamma_{v_{1},s_{1},2}
\end{pmatrix},
\end{equation}
which is an immediate consequence of a corollary of the lemma introduced in Appendix~\ref{app:lemma}. Using this transformation one can readily `flatten' $A_t$'s.  Similarly, one can flatten $B_{p}$'s that do not cross the branch cut by applying the same gauge transformation to Majoranas on one side of the vertical dashed line, e.g., in the blue area of Fig.~\ref{fig:flattenBp}(a). One should take a more careful approach for $B_{p}$'s that cross the branch cut since, due to the Klein-bottle structure, the boundary condition is twisted in opposite ways across the branch cut. Hence, to flatten $B_{p}$ terms locally, one should transform Majoranas above the branch cut [red area in Fig.~\ref{fig:flattenBp}(b)] with $U(-\theta)$ instead of $U(\theta)$, due to a different corollary presented in Appendix~\ref{app:lemma}. We recall that long-edge pairings across the branch cut pair Majoranas with different layer indices. Hence, the corresponding projector is preserved under the gauge transformation with $U(-\theta)$ on one side and $U(\theta)$ on the other.  If we had not twisted Majorana layers across the branch cut when defining the system on the torus in an earlier subsection, it would have been impossible for us to flatten those $B_{p}$'s. This once more emphasizes the importance of encoding layer interchange in the definition of $\mathcal{CP}$ symmetry. 

Because $A_t$ and $B_{p}$ can be locally flattened, all nice local properties are once more preserved in this boundary-condition twist. We emphasize, however, that no local gauge transformations map our system with specific $\theta$ into a system with a different U$(1)$ boundary-condition twist. 

Let us now compute the Berry phase $\gamma_{\text{GS}}$ of the ground state by adiabatically changing $\theta$ from 0 to $2\pi$. First, let us consider the Berry phase for a single snapshot in which all spin configurations are fixed to be down, with Majoranas paired accordingly. Denote this state as $\ket{\downarrow(\theta)}$. We will compute
\begin{equation}
\gamma_{\downarrow} = i\int_{0}^{2\pi} d\theta \bra{\downarrow(\theta)}\frac{\partial}{\partial \theta}\ket{\downarrow(\theta)}.
\end{equation}
As a first step we examine a 1D problem in which we compute the Berry phase for double Kitaev chains. 
One can easily compute that the Berry phase is precisely $\pi$. Since $\ket{\downarrow(\theta)}$ contains an odd number of such chains, the Berry phase $\gamma_\downarrow$ is $\pi$ mod $2\pi$. 
The full ground state of interest reads
\begin{equation}
\ket{\text{GS}(\theta)} = \frac{1}{2^{n/2}} \sum_{n_{p}=0,1}\prod_{p} \left( B_{p}(\theta) \right)^{n_{p}}\ket{\downarrow(\theta)},
\end{equation}
where $n$ is a total number of plaquettes. Since all terms in the summand have different bosonic spin configurations and hence are orthogonal to each other, one can write the Berry phase as
\begin{equation}
\begin{split}
&\gamma_{\text{GS}} = i\int_{0}^{2\pi} d\theta \bra{\text{GS}(\theta)}\frac{\partial}{\partial \theta}\ket{\text{GS}(\theta)} \\
&= \frac{1}{2^{n}} \sum_{n_{p}=0,1} i\int_{0}^{2\pi} d\theta \\
&\bra{\downarrow(\theta)} \left( \prod_{p} \left( B_{p}(\theta) \right)^{n_{p}}\right) \frac{\partial}{\partial \theta} \left( \prod_{p} \left( B_{p}(\theta) \right)^{n_{p}}\right) \ket{\downarrow(\theta)}.
\end{split}
\end{equation} 
The above equation implies that one can compute $\gamma_{\text{GS}}$ by simply averaging the Berry phase for $2^{n}$ different configurations. Remarkably, we will show in Appendix~\ref{app:Berry} that 
\begin{equation}
\label{eq:Berrykey}
\begin{split}
&\bra{\downarrow(\theta)} \left( \prod_{p} \left( B_{p}(\theta) \right)^{n_{p}}\right) \frac{\partial}{\partial \theta} \left( \prod_{p} \left( B_{p}(\theta) \right)^{n_{p}}\right) \ket{\downarrow(\theta)} \\
&= \bra{\downarrow(\theta)}\frac{\partial}{\partial \theta}\ket{\downarrow(\theta)} 
\end{split}
\end{equation} 
for any $\theta$ and any $n_{p}$---which guarantees that $\gamma_{\text{GS}} = \gamma_{\downarrow} = \pi ~(\text{mod } 2\pi)$. This $\pi$ Berry phase was computed from the boundary \citep{Hsieh2014} and bulk in band-theory frameworks \citep{Shiozaki2018} and found to be quantized in $\mathcal{CP}$-symmetric topological phases. Strikingly, this Berry phase---which is generically nontrivial to compute and usually evaluated with small-system numerics---can be analytically determined for any system size in our exactly solvable model! We note that while there are supporting evidences that the Berry phase is quantized to either $0$ or $\pi$, to our knowledge, a microscopic proof of robust quantization is not yet available.

\subsection{Towards bulk topological invariants of $\mathcal{T}$-symmetric topological phases without $\mathcal{CP}$ symmetry}

 Throughout the main text of this paper, we focused on honeycomb-lattice models with a specific choice of Kasteleyn orientation. We indeed eschewed fully general setups in favor of accessibility.  Appendix~\ref{app:moregeneral1}, however, shows how to define models on any trivalent lattice with any choice of Kasteleyn orientation. The particularly interesting setups relevant to this subsection occur when the trivalent lattice is chosen to violate parity symmetry; in these cases, there is no exact $\mathcal{CP}$ symmetry. It is natural to ask whether $(i)$ it is possible to define models on non-orientable manifolds when the lattice is not parity-symmetric and $(ii)$ if one can define the models on a Klein bottle, whether the many-body invariants we explored earlier in this section can be extended to these setups.
 
 In Appendix~\ref{app:moregeneral3}, we show that the answers to both questions are affirmative: The models can still be defined on a Klein bottle while maintaining commuting-projector properties, and the global-fermion-parity flip and $\pi$ Berry phase still characterize our exactly solvable model. While we relegate technical details to the appendix, we will briefly discuss here the role of time-reversal symmetry in defining models on a Klein bottle.
 
 Recall that in the earlier subsection we stated that putting systems on a Klein bottle \textit{breaks} $\mathcal{T}$-symmetry. The same is true for Klein-bottle constructions presented in Appendix~\ref{app:moregeneral3} as well. Hence, readers may naturally wonder in what sense $\mathcal{T}$ may protect many-body invariants of our interests. To see this within the context of the construction presented in this section, we emphasize that the unitary part of $\mathcal{T}$-symmetry encoded in Eq.~\eqref{eq:TTSCunitary} is actually identical to the local part of $\mathcal{CP}$ symmetry $U_{\mathcal{CP},\text{TSC}}$ in Eq.~\eqref{eq:cpunitary}, modulo some sublattice relabeling in the latter equation. The Klein-bottle construction presented in the appendix is a natural extension of this observation and \emph{employs the unitary part of the time-reversal symmetry}, without referring to any $\mathcal{CP}$ symmetry.
 
 While the previous paragraph gives us some confidence that many-body invariants on a Klein bottle may also characterize $\mathcal{T}$-symmetric topological phases, whether these invariants remain quantized or valid outside of the context of our special exactly solvable models requires further investigation. Also, we did not give any detailed prescription on how to construct these many-body invariants outside in the context of our  exactly solvable models; developing a systematic procedure would be an interesting future direction.

\section{Conclusions}
\label{sec:conc}

In this paper we constructed a commuting-projector model for an interacting 2D quantum-spin-Hall insulator.  While such phases are of course amenable to band-theory treatments in the non-interacting limit, our perspective illuminates complementary insights that are far from obvious from that more traditional approach.  Most strikingly, from our commuting-projector model we derived a strictly 1D lattice Hamiltonian that, \emph{within a restricted subspace}, faithfully reproduces the physics of a fully symmetric quantum-spin-Hall edge.  Additionally, we discovered that both the topological-superconductor model from Ref.~\onlinecite{Wang2017} and our topological-insulator model preserve $\mathcal{CP}$ symmetry when defined on parity-symmetric lattices, allowing us to explore bulk invariants by placing the systems on non-orientable manifolds. We further observed that computation of many-body invariants can be extended to models with $\mathcal{T}$ symmetry but without explicit $\mathcal{CP}$ symmetry.  The latter result raises the question of whether topological invariants of time-reversal-invariant topological phases can be also explored by defining systems on non-orientable manifolds.

To close we highlight a number of other interesting open questions connected to our study:

In our 1D lattice Hamiltonian that describes the quantum-spin-Hall edge, U$(1)$ symmetry is realized in the usual, local manner while $\mathcal{T}$ is implemented nonlocally.  Can one derive an alternative strictly 1D model in which the roles of these symmetries are reversed, i.e., unconventional realization of U$(1)$ but local $\mathcal{T}$ implementation?  It is also worthwhile to contemplate practical utility of such 1D Hamiltonians.  They may, for instance, enable efficient numerical studies of 2D quantum-spin-Hall edges with interactions, noise, etc.  As an alternative application, can one interface magnetic degrees of freedom and electrons to design experimental 1D setups that emulate the quantum-spin-Hall edge?  Architectures of this type could furnish new platforms for Majorana zero modes, fractional charges, anomalous pumping cycles, and more.  

On the more technical end, are the nontrivial $\pi$ Berry phase (for the topological-insulator model) and ground-state fermion-parity flip from changing fermionic boundary condition (for the topological-superconductor model) generic properties of \emph{time-reversal symmetric topological phases}? If so, is there an analytical proof available in the context of $(2+1)$-dimensional lattice models? 
How is our construction related to the relevant $\text{Pin}$-structure? Kasteleyn orientations are related to spin structures \citep{Cimasoni2007}, and there is generalization of Kasteleyn orientations to the $\text{Pin}^{-}$ structure as well \citep{Cimasoni2009}. The $\mathcal{T}^{2}=-1$ topological-superconductor model is known to be related to the $\text{Pin}^{+}$ structure \cite{Witten2016}, and indeed how the Kasteleyn orientation should be properly constructed for the model on a Klein bottle differs from the discrete $\text{Pin}^{-}$ structure given in Ref.~\onlinecite{Cimasoni2009}. 

Finally, are there higher-dimensional generalizations of the commuting-projector models that we studied here, e.g., for 3D topological insulators---potentially giving rise to strictly 2D lattice Hamiltonians that similarly capture the anomalous single-Dirac-cone surface states?

\acknowledgements
We thank Xie Chen, Gil Young Cho, Lukasz Fidkowski, Shinsei Ryu, Xiao-Qi Sun, and Zitao Wang for helpful discussions. This work was supported by the Army Research Office under Grant Award W911NF-17-1-0323; the NSF through grant DMR-1723367; the Caltech Institute for Quantum Information and Matter, an NSF Physics Frontiers Center with support of the Gordon and Betty Moore Foundation through Grant GBMF1250; and the Walter Burke Institute for Theoretical Physics at Caltech.

\appendix
\section{Models on general trivalent lattices with general Kasteleyn orientations}
\label{app:moregeneral}
 
 In this appendix, we briefly review the $\mathcal{T}$-symmetric topological-superconductor model on \emph{any trivalent lattice} defined on an orientable manifold, and for any choice of Kasteleyn orientation (as opposed to the specific choice of lattice and Kasteleyn orientation used elsewhere in this paper). Then, we show that on any parity-symmetric lattice and with any Kasteleyn orientation, the model preserves $\mathcal{CP}$ symmetry as well. Finally, we discuss how to define our models on a Klein bottle even when the lattice is not parity-symmetric and hence $\mathcal{CP}$ symmetry is absent. The latter discussion suggests that many-body invariants originally proposed for $\mathcal{CP}$-symmetric topological phases may also characterize $\mathcal{T}$-symmetric phases. We stress that although we primarily focus on the constructions of topological-superconductor models, the generalization to topological insulators proceeds straightforwardly by adding another layer of Majorana fermions.
  
\subsection{Topological-superconductor models }
\label{app:moregeneral1}

 The degrees of freedom are defined analogously to the honeycomb-lattice construction used earlier.  A single bosonic spin resides on each plaquette of the trivalent lattice, while four Majoranas appear at each edge---two at one end of the edge and two at the other. The pair of Majoranas at the same end of the edge will be distinguished by different spin indices. Alternatively, one can replace a vertex of the trivalent lattice with small triangles to make a triangle-decorated `pairing lattice' (i.e., the Fisher lattice in the honeycomb-lattice used in the main text); pairs of Majoranas with opposite spins can then be viewed as residing on each vertex of the pairing lattice.  As before, long edges correspond to edges of the pairing lattice from the original trivalent lattice and short edges to the edges arising from small-triangle decoration.
 
Given a Kasteleyn orientation on the pairing lattice (which can be chosen for any lattice on an orientable manifold \citep{Cimasoni2007}), we specify how Majoranas pair for each plaquette. From there the $A_{t}$ vertex terms and the $B_p$ plaquette terms can be constructed by using appropriate Majorana projectors dictated by the pairing rules below.

\begin{enumerate}
\item Determine which Majoranas should pair along each long edge first. Given a long edge $e$, vertices $v$ and $v'$ connected by $e$, and two neighboring plaquettes $p_{1}$ and $p_{2}$, these pairings are determined as follows: 
\begin{enumerate}
\item If Ising spins on $p_{1}$ and $p_{2}$ are identical, pair $(\gamma_{v,\uparrow},\gamma_{v',\uparrow})$ and $(\gamma_{v,\downarrow},\gamma_{v',\downarrow})$.
\item If those Ising spins are opposite, draw an arrow from the plaquette with up Ising spin to to the plaquette with down Ising spin, and cross product the Kastelyn arrow connecting $v$ and $v'$ with this arrow. If the cross product points out of the page, pair only $(\gamma_{v,\uparrow},\gamma_{v,\uparrow})$; otherwise, pair only $(\gamma_{v,\downarrow},\gamma_{v',\downarrow})$.
\end{enumerate}
\item If all three plaquette spins adjacent to a vertex $v$ are identical, the above rules pair all six Majoranas on the small triangle at vertex $t$. If not, exactly two among six Majoranas on the small triangle remain unpaired. Pair those two unpaired Majoranas.
\end{enumerate}
As an exercise, one can check that for the honeycomb lattice the pairing rule given in Fig.~\ref{fig:2_vertexrule} in the main text follows the above prescription.

Time-reversal symmetry $\mathcal{T}_{\text{TSC}}$ of the Hamiltonian is implemented by the anti-unitary transformation
\begin{equation}
\mathcal{T}_{\text{TSC}} \text{ : }  \sigma_{p}^{z} \rightarrow - \sigma_{p}^{z}, \quad \begin{pmatrix}
\gamma_{v,\uparrow} \\
\gamma_{v',\uparrow} \\
\gamma_{v,\downarrow} \\
\gamma_{v',\downarrow} 
\end{pmatrix} \rightarrow \begin{pmatrix}
0 & 0 & 1 & 0 \\
0 & 0 & 0 & -1 \\
-1 & 0 & 0 & 0 \\
0 & 1 & 0 & 0
\end{pmatrix} \begin{pmatrix}
\gamma_{v,\uparrow} \\
\gamma_{v',\uparrow} \\
\gamma_{v,\downarrow} \\
\gamma_{v',\downarrow} 
\end{pmatrix}.
\end{equation}
Here $v$ and $v'$ are vertices of the pairing lattice that are connected by a long edge, with the Kasteleyn arrow direction on that long edge pointing from $v'$ to $v$. One can prove that among two Majoranas $\gamma_{v_{1},s_{1}}$ and $\gamma_{v_{2},s_{2}}$ that can be paired in some spin configuration, the above transformation endows a minus sign to only one of them. Hence, combined with complex conjugation that sends $i \rightarrow -i$, $i\gamma_{v_{1},s_{1}}\gamma_{v_{2}s_{2}}$ maps to $i\gamma_{v_{1},\overline{s}_{1}}\gamma_{v_{2}\overline{s}_{2}}$, with the overlines denoting the opposite spin indices. As a consistency check, one can show that upon specializing to the honeycomb-lattice model with our previous Kasteleyn orientation choice, this transformation reduces to the operation specified in Eq.~\eqref{eq:TTSCunitary}.

 Finally, we discuss the relationship between two Hamiltonians with identical lattice but with different Kasteleyn orientations. We first recall that a `local transformation' at a vertex $v$ of the pairing lattice is defined by flipping arrows on the three edges that meet at $v$. It is known that a series of local transformations map all Kasteleyn orientations that represent the same spin structure \citep{Cimasoni2007}. For topological-superconductor models, flipping an arrow direction on a long edge changes how Majoranas pair for each spin configuration. Hence, one can think of the following local transformation at $v$:
\begin{equation}
L_{v} \text{ : } \begin{pmatrix} 
\gamma_{v,\uparrow} \\
\gamma_{v',\uparrow} \\
\gamma_{v,\downarrow} \\
\gamma_{v',\downarrow} 
\end{pmatrix} \rightarrow \begin{pmatrix}
0 & 0 & -1 & 0 \\
0 & 0 & 0 & 1 \\
-1 & 0 & 0 & 0 \\
0 & 1 & 0 & 0
\end{pmatrix} \begin{pmatrix}
\gamma_{v,\uparrow} \\
\gamma_{v',\uparrow} \\
\gamma_{v,\downarrow} \\
\gamma_{v',\downarrow} 
\end{pmatrix},
\end{equation}
where $v'$ is the vertex connected to $v$ via a long edge. Essentially, the above transformation adds minus signs to Majoranas at $v$ and flips the Majorana spins at both $v$ and $v'$. A series of $L_{v}$ transformations connect Hamiltonians defined with different Kasteleyn orientations in the same `topological sector'. We note that $L_{v}$ corresponds to $\mathbb{Z}_{2}$ gauge transformations on Majoranas combined with spin relabeling; if two Hamiltonians are related by a series of $L_{v}$'s, they should be regarded as identical.

\subsection{$\mathcal{CP}$ symmetry beyond honeycomb-lattice models}

 Here we present a proof that the topological-superconductor models defined on a torus or sphere and on parity-symmetric lattices always possess $\mathcal{CP}$ symmetry. While our proof is restricted to the case of two specific manifolds, it is natural to expect that a similar technique can be extended to any orientable manifold on which a reflection operation can be defined. 
 
\begin{figure}
\includegraphics[width=0.7\linewidth]{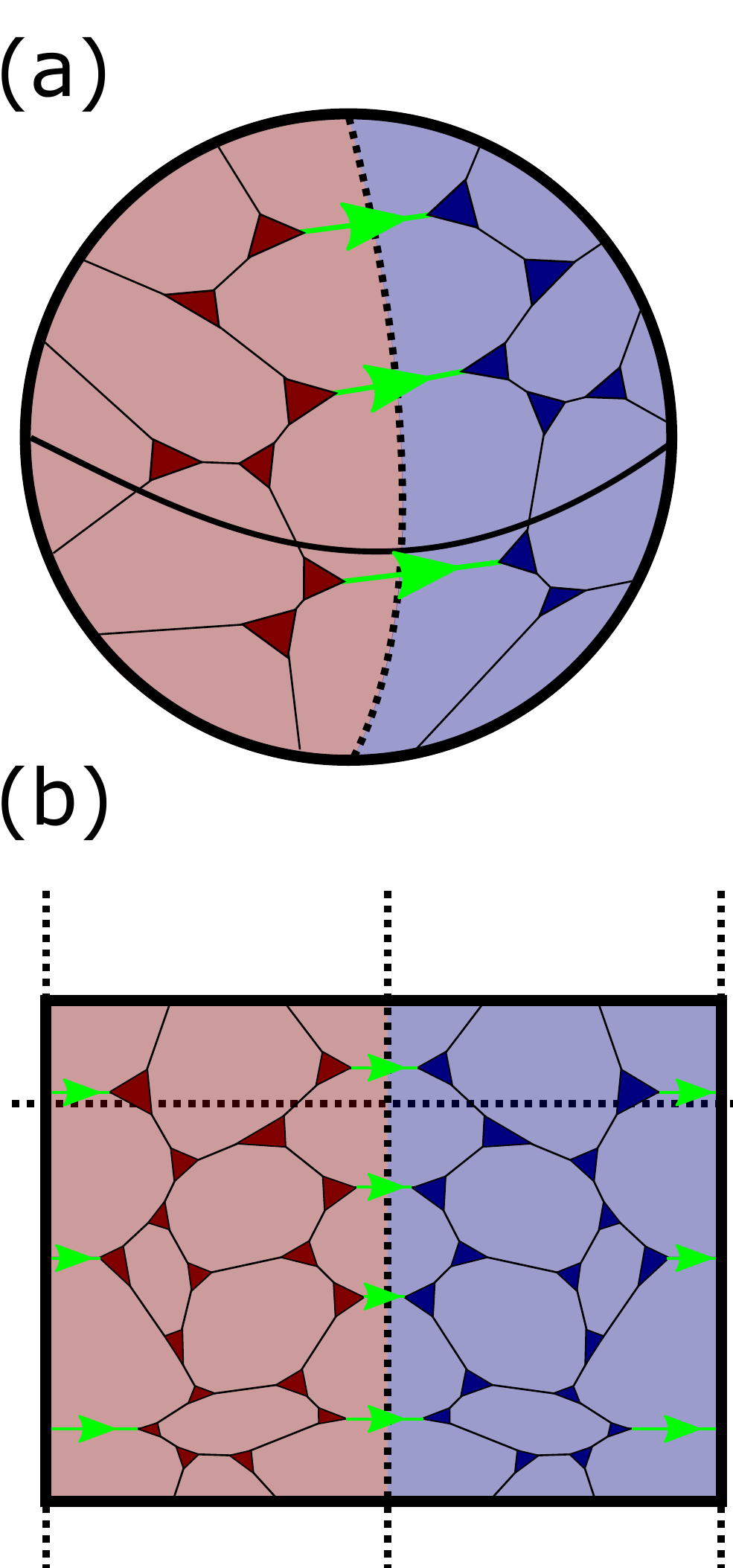}
\caption{Illustration of parity-symmetric trivalent lattices and how to assign Kasteleyn orientations on (a) the sphere and (b) the torus.}
\label{fig:canonicalCP}
\end{figure}
 
 First, we show that there are `canonical Kasteleyn orientations' in which the reflection operation flips all arrows. As shown in Fig.~\ref{fig:canonicalCP}(a), a line that is fixed under reflection [dotted vertical line in (a)] bisects the sphere into two parts marked blue and red in figure, one being a mirror image of the other. In the case of the torus, there are two lines that are fixed under reflection [three dotted vertical lines in Fig.~\ref{fig:canonicalCP}(b); the two at the extreme ends are actually identical].  These two lines similarly divide a torus into two parts, which are also mirror images of each other.  
 
 One can imagine first assigning arrows for edges that reside only on one part (e.g., the red region) such that all cycles within that part satisfy the clockwise-odd rule. Note that we are not assigning arrow directions to edges that cross the fixed lines and hence belong to both parts. Then, one can define arrow directions on the other part (e.g., the blue region) such that they are opposite of the mirror image of the arrows in the first part. Now, all edges except for the ones that lie across the fixed lines have arrow assignments, and all cycles isolated to each part satisfy the clockwise-odd rule. 
 
 Thanks to the two parts being mirror images of one another, the arrows for the edges that cross the fixed line are fairly straightforward to determine as well. Plaquettes that extend across the fixed lines are guaranteed to be clockwise even without arrows for the edges on a fixed line; indeed, due to reflection symmetry, the already-assigned arrows on the left and right have the same number of edges that are directed against the clockwise cycle. Hence, one can choose arrows along fixed lines such that all of them uniformly point from one area to the other, just like the bright green arrows in Figs.~\ref{fig:canonicalCP}(a) and (b). Plaquettes bisected by fixed lines then satisfy the clockwise-odd rule. 
 
 As a final step, we recall that there are four locally inequivalent choices of Kasteleyn orientations for the torus. The three other locally inequivalent canonical Kasteleyn orientations can be generated by flipping all arrows on any vertical or horizontal dotted line in Fig.~\ref{fig:canonicalCP}, or both. This global transformation does not alter the defining property that the reflection transformation on these Kasteleyn orientations are equivalent to flipping all arrows. 
 
One can write down the $\mathcal{T}$-symmetric topological-superconductor models based on these canonical Kasteleyn orientations. In addition to $\mathcal{T}_{\text{TSC}}$, these models enjoy $\mathcal{CP}_{\text{TSC}}$ symmetry defined by $U_{\mathcal{T},\text{TSC}}$ (unitary part of the time-reversal symmetry) followed by spatial reflection. As for the proof, one can manually check that our definition of $\mathcal{CP}_{\text{TSC}}$, when applied to a vertex term at the vertex $v$, properly gives its mirror partner---which suffices to prove $\mathcal{CP}_{\text{TSC}}$ invariance of the whole Hamiltonian.
 
 When the Hamiltonian is defined with some arbitrary Kasteleyn orientation without a mirror property, as we saw earlier, a series of local gauge transformations $L_{v}$ allow us to map the Hamiltonian to the one defined with canonical Kasteleyn orientations. Hence, we conclude that an emergence of $\mathcal{CP}$ symmetry on our exactly solvable model is a very general phenomenon whenever the underlying lattice is parity symmetric.
 
\subsection{Topological superconductor models on a Klein bottle without $\mathcal{CP}$ symmetry }
\label{app:moregeneral3}

We conclude this appendix by giving a general prescription of how to define topological-superconductor models on a Klein bottle, \emph{even when the underlying lattice is not parity symmetric and hence $\mathcal{CP}$ symmetry is absent}. We proceed similarly to the main text by building a twisted double-cover representation of Majorana pairings and spin configurations.
 
 The key ingredients in building a twisted double-cover representation is a choice of Kasteleyn orientation. Kasteleyn orientation on a twisted-double cover representation is essentially arrow assignments on a trivalent lattice on a torus, now with a special property: The lattice on the upper half is precisely the mirror image of the lower half, and the Kasteleyn orientation on the upper half is opposite of the mirror image of the lower half. We will show that such a Kasteleyn orientation can be generally chosen on a Klein bottle when each half contains an even number of vertices, using the result from Ref.~\onlinecite{Cimasoni2009}. 
 
 Before stating our proof, we set up some notation. There are two lines (referred to as `cuts') on the double-cover torus where the lower half and the upper half meet. Following Ref.~\onlinecite{Cimasoni2009} we will denote edges crossing these lines as `0-edges', while edges that do not cross the lines are dubbed `1-edges'. Reference~\onlinecite{Cimasoni2009} shows that as long as the number of vertices on either half is even, a Kasteleyn orientation on an identical lattice, but with modified relations between arrows on the lower half and upper half, is guaranteed to exist: In the modified relation, 0-edges follow the same rule described in the previous paragraph. However, arrows on 1-edges crossing one cut are just the mirror image of those on the other cut (instead of being opposite of the mirror image). All we need to do to construct the Kasteleyn orientations we want for our purpose is flip arrows on 1-edges that cross one of the cuts, so that now all arrows on the upper half---regardless of whether they are on 1-edges or 0-edges---are opposite of the mirror image of the lower half. This operation is benign in the sense that all clockwise-odd cycles remain clockwise-odd.

We emphasize that this proof is expected to only work on a Klein bottle; in fact, based on the fact that there is no Pin$^{+}$ structure on $\mathbb{RP}^{2}$, it is expected that there is no natural way to define our models consistently on some other non-orientable manifolds.
 
 The rest of the process is almost identical to the one presented in the main text: We build twisted double-cover representations by setting the bosonic spin configurations on the upper half as the opposite of the mirror image of the lower half, and pair Majoranas following the local rules as if the twisted double-cover torus was just an ordinary fully orientable torus. To build an ordinary double cover, we untwist the upper half with $U_{\mathcal{T},\text{TSC}}$. One can show that this operation generates Majorana pairings and spin configurations where the upper half is precisely the mirror image of the lower half, and we indeed thus construct the `double cover representation' of Majorana/spin configurations on a Klein bottle. From there, one can straightforwardly write down the Hamiltonian on a Klein bottle.
 
This method of putting the model on a Klein bottle explicitly breaks time-reversal symmetry. However, we note that $(i)$ this methods uses some form of $\mathcal{T}$ symmetry and indeed can be understood as twisting the boundary condition by the unitary part of time reversal and $(ii)$ many-body invariants presented in Sec.~\ref{sec:kleinCP} can be computed in this setup as well and can be shown to acquire the same values.

\section{Proof of a projector lemma and its corollaries} \label{app:lemma}

Here, we prove the following lemma: We are given four Majorana operators $\gamma_{v,1}$, $\gamma_{v,2}$, $\gamma_{v',1}$, $\gamma_{v',2}$ and $2 \times 2$ O$(2)$ matrices $M$ and $N$. We also define
\begin{equation}
\gamma'_{v,i}= M_{ij} \gamma_{v,j}, \quad \gamma'_{v',i}= N_{ij} \gamma_{v',j}.
\end{equation}
Here and in other parts of the proof repeated indices are implicitly summed. Then the following holds:
\begin{equation} \label{eq:lemmastep1}
\begin{split}
\frac{1+i\gamma'_{v,1}\gamma'_{v',1}}{2} & \frac{1+i\gamma'_{v,2}\gamma'_{v',2}}{2} \\
&= \frac{1}{4}  + \frac{i}{4} \gamma_{v,i} \left( M^{T} N \right)_{ij} \gamma_{v,j}  \\
&+ \frac{1}{4}\det \left( M^{T} N \right) \gamma_{v,1}\gamma_{v,2}\gamma_{v',1}\gamma_{v',2} .
\end{split}
\end{equation}
Notice that regardless of the detailed form of $M$ and $N$, as long as $M^{T}N$ remains the same, the right-hand side remains the same as well. 

The proof proceeds by brute force:
\begin{equation} \label{eq:lemmastep2}
\begin{split}
&\frac{1+i\gamma'_{v,1}\gamma'_{v',1}}{2}\frac{1+i\gamma'_{v,2}\gamma'_{v',2}}{2} \\
&= \frac{1+i M_{1i}\gamma_{v,i} N_{1j}\gamma_{v',j}}{2} \frac{1+i M_{2k}\gamma_{v,k} N_{2l}\gamma_{v',l}}{2} \\
&= \frac{1}{4} + \frac{i}{4} \gamma_{v,i} (M_{1i}N_{1j} + M_{2i}N_{2j})\gamma_{v,j} \\
&+ \frac{1}{4}M_{1i}\gamma_{v,i} M_{2k}\gamma_{v,k} N_{1j}\gamma_{v',j} N_{2l}\gamma_{v,l}
\end{split}
\end{equation}
By recognizing that $M_{1i}N_{1j} + M_{2i}N_{2j} = M_{ik}^{T}N_{kj}$, one can see that the second term on the right side of Eq.~\eqref{eq:lemmastep2} precisely matches the second term on the right side of Eq.~\eqref{eq:lemmastep1}. Observe next that
\begin{equation}
\begin{split}
M_{1i}\gamma_{v,i} M_{2k}\gamma_{v,k} &= (M_{11}M_{21} + M_{12}M_{22}) \\
&+ (M_{11}M_{22} - M_{12}M_{21})\gamma_{v,1}\gamma_{v,2}.
\end{split} 
\end{equation}
The first term above vanishes, due to the fact that the two row vectors of $M$ are orthogonal to each other. The coefficient in the second term is simply $\det{M}$---hence $M_{1i}\gamma_{v,i} M_{2k}\gamma_{v,k} = \det{M}\gamma_{v,1}\gamma_{v,2}$. One can see that the third term in the last line of Eq.~\eqref{eq:lemmastep2} is then precisely the last line of Eq.~\eqref{eq:lemmastep1}. The lemma is therefore proven.  

Below we list several useful corollaries.
\begin{enumerate}
\item Multiplication of two projectors
\begin{equation}
\frac{1+i\gamma_{v,1}\gamma_{v',1}}{2} \frac{1+i\gamma_{v,2}\gamma_{v',1}}{2}
\end{equation} 
is invariant under the transformation
\begin{equation}
\begin{split}
&\gamma_{v,i} \rightarrow U(\theta)_{ij} \gamma_{v,j}, \quad \gamma_{v',i} \rightarrow U(\theta)_{ij}  \gamma_{v',j} \\
&U(\theta) = \begin{pmatrix}
\cos{\theta} & \sin{\theta} \\ 
-\sin{\theta} & \cos{\theta}
\end{pmatrix}.
\end{split}
\end{equation}
 This corollary can be straightforwardly proven by noticing that the above transformation is equivalent to setting $M = N = U(\theta)$. 
\item The projector
\begin{equation}
\frac{1+i\gamma_{v,1}\gamma_{v',2}}{2} \frac{1+i\gamma_{v,2}\gamma_{v',2}}{2}
\end{equation}
appears in Sec.~\ref{sec:kleinCP}, where defining the model on a Klein bottle makes some Majoranas pair across the different layers. This projector is invariant under the U(1) transformation
\begin{equation}
\gamma_{v,i} \rightarrow U(\theta)_{ij} \gamma_{v,j}, \quad \gamma_{v',i} \rightarrow U(-\theta)_{ij}  \gamma_{v',j}.
\end{equation}
This property can be proven as follows: The above projector can be obtained by choosing $M = I$, $N = \begin{pmatrix} 0 & 1 \\ 1 &0 \end{pmatrix}$, and the transformation is equivalent to choosing $M' = U(\theta)$ and $N' = \begin{pmatrix} 0 & 1 \\ 1 &0 \end{pmatrix} U_{-\theta}$. It is easy to show that $M^{T}N = M'^{T}N'$.
\item Let us consider
\begin{equation}
P(\theta) = \frac{1+i\gamma_{v,1}\gamma_{v',1}'(\theta)}{2} \frac{1+i\gamma_{v,2}\gamma_{v',2}'(\theta)}{2},
\end{equation}
where $\gamma_{v',1}'(\theta)$ and $\gamma_{v',2}'(\theta)$ are defined as
\begin{equation}
\begin{pmatrix}
\gamma_{v',1}'(\theta) \\
\gamma_{v',2}'(\theta)
\end{pmatrix} = \begin{pmatrix}
\cos{\theta} & \sin{\theta} \\
-\sin{\theta} & \cos{\theta}
\end{pmatrix} \begin{pmatrix}
\gamma_{v',1} \\
\gamma_{v',2}
\end{pmatrix}.
\end{equation}
Then one can show:
\begin{equation}
P(\theta) \frac{\partial P(\theta)}{\partial \theta} P(\theta) =0
\end{equation}
This can be easily proven by noticing that
\begin{equation}
\begin{split}
&\frac{\partial P(\theta)}{\partial \theta} = \frac{i}{4} \begin{pmatrix} \gamma_{v,1} & \gamma_{v,2}
\end{pmatrix} \begin{pmatrix} 
-\sin{\theta} & \cos{\theta}\\
-\cos{\theta} & -\sin{\theta}
\end{pmatrix} \begin{pmatrix} 
\gamma_{v',1}\\
\gamma_{v',2}
\end{pmatrix} \\
&=\frac{i}{4}(\gamma_{v,1}\gamma_{v',2}'(\theta) + \gamma_{v,2}\gamma_{v',1}'(\theta)).
\end{split}
\end{equation}
Then, one can deduce
\begin{equation}
\begin{split}
&P(\theta) \frac{\partial P(\theta)}{\partial \theta} P(\theta) = P(\theta) \frac{i}{4}(\gamma_{v,1}\gamma_{v',2}'(\theta) + \gamma_{v,2}\gamma_{v',1}'(\theta)) P(\theta)\\
&=\frac{i}{4}(\gamma_{v,1}\gamma_{v',2}'(\theta) + \gamma_{v,2}\gamma_{v',1}'(\theta))\\
&\frac{1-i\gamma_{v,1}\gamma_{v',1}'(\theta)}{2} \frac{1-i\gamma_{v,2}\gamma_{v',2}'(\theta)}{2}P(\theta).
\end{split}
\end{equation}
The product of the projectors on the third line is explicitly 0. This property will be used in Appendix~\ref{app:Berry}.

\end{enumerate}

\section{More on the gapless edge Hamiltonian}
\label{app:edge}

This appendix fills in technical gaps that we left open in Sec.~\ref{sec:edgegapless}.

\subsection{Strategy for proving $[B_{p},C_{I}]=0$}

 Here, we introduce a trick that allows $C_{I}$ to be modified such that it can be treated more like an ordinary plaquette flip term $B_{p}$. This modification allows one to prove $[B_{p},C_{I}]=0$ straightforwardly.  
First we summarize the key elements in the proof of $[B_{p},B_{p'}]=0$ (following similar techniques used, e.g., in Ref.~\onlinecite{Ware2016}, \onlinecite{jones2019}, and \onlinecite{Son2018}). If $p$ and $p'$ are non-neighboring, this relation is somewhat obvious. The non-trivial feature of the proof comes from the fact that when $B_{p}$ and $B_{p'}$ are neighboring, there are Majorana projectors in the expression for $B_{p}$ and $B_{p'}$ that do not commute with each other and should be handled carefully. We emphasize that $(i)$ the precise details of proof are only dependent on projectors involving Majoranas around two vertices and $(ii)$ there is a step in the proof that relies on the fact that in $B_{p}$, projectors that project onto the Majorana pairings consistent with the original spin configuration and projectors that project onto the Majorana pairings consistent with the new spin configurations together form a loop around the plaquette $p$. 
 
Let us investigate how $(i)$ and $(ii)$ are retained or violated in $C_{I}$. Think of a bulk plaquette $p$ neighboring to $I$, and two vertices at which $I$ and $p$ meet.
The operator $C_{I}$ actually contains all Majorana projectors involving degrees freedom around those two vertices as if it was a normal `complete' plaquette. Hence, one may naively expect that projectors that should be treated carefully are actually just identical to those involved in the proof of $[B_{p},B_{p'}]=0$. However, $C_{I}$ is explicitly incomplete and hence Majorana projectors \emph{do not} form a loop anymore, violating $(ii)$. Hence, if we devise an equivalent modification of $C_{I}$ in which Majorana projectors now form a closed loop without affecting Majorana projectors around the two vertices of interest, then we can apply the same technique for proving $[B_{p},B_{p'}]=0$.

\begin{figure}
\includegraphics[width=0.8\linewidth]{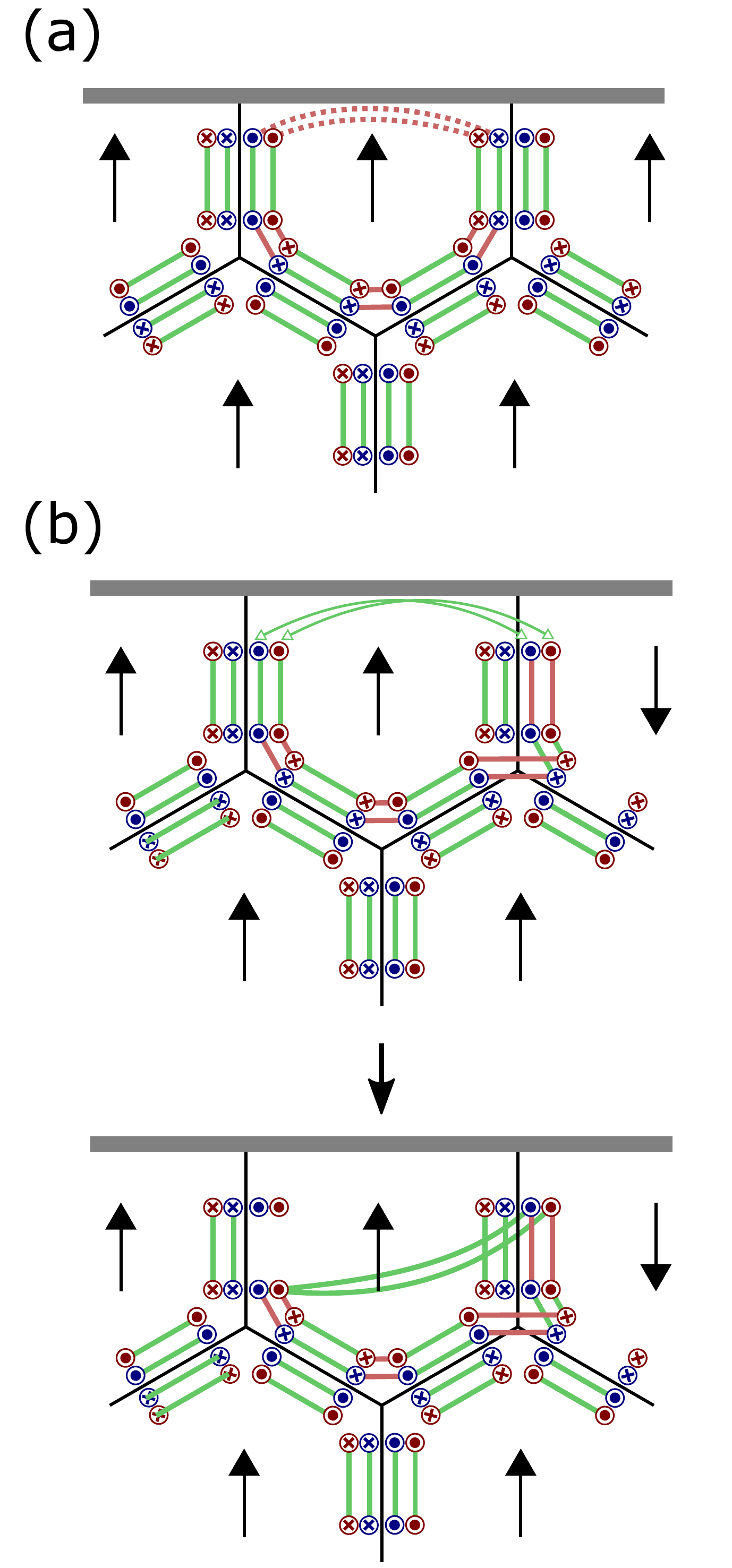}
\caption{Illustration of the modification of $\mathcal{C}_{u_{I}}$ when the two spins on incomplete plaquettes surrounding $I$ are (a) identical and (b) opposite. In the case of (a), the modification inserts a projector associated with Majorana pairings indicated by red dashed lines. For (b), the modification introduces a local unitary transformation that can be interpreted as changing the locations of Majoranas, thus allowing the projectors involved in $\mathcal{C}_{u_{I}}$ to form a loop.}
\label{fig:bpcimod}
\end{figure}

We divide into cases in which the two incomplete plaquette spins surrounding $I$ are $(i)$ the same or $(ii)$ opposite. For case $(i)$, assume that the two incomplete plaquette spins are up, and that the spin at plaquette $I$ is up as well. Then, in the expression for $\mathcal{C}_{u_{I}}$ in Eq.~\eqref{eq:cI}, the projectors in the first parenthesis explicitly keep four Majoranas at the edge unpaired, whereas the projectors in the second parenthesis keep these previously unpaired Majoranas paired with Majoranas in the bulk. Denote those four Majoranas as $\gamma_{e,s,1}$ $\gamma_{e,s,2}$, $\gamma_{e',s',1}$, and $\gamma_{e',s',2}$. One can insert the projectors
\begin{equation}
\label{eq:projaddition}
P_{es,e's'}^{\text{TI}} = \frac{1 \pm i\gamma_{e,s,1}\gamma_{e's',1}}{2} \frac{1 \pm i\gamma_{e,s,2}\gamma_{e's',2}}{2}
\end{equation} 
to get an equivalent expression, where the signs should be determined by the clockwise-odd rule around the incomplete plaquette $I$. This modification makes $\mathcal{C}_{u_{I}}$ satisfy $(ii)$ for the $u_{I}$ we considered so far. To see why this works, let us label Majoranas (on the first layer) in a way that the second parenthesis of $\mathcal{C}_{u_{I}}$ enforces Majorana pairings between $(\gamma_{a_{1},1},\gamma_{a_{2},1})$, $(\gamma_{a_{3},1},\gamma_{a_{4},1})$,$\cdots$, $(\gamma_{a_{2n-1},1},\gamma_{a_{2n},1})$, and the first parenthesis projects onto the state with Majorana pairings $(\gamma_{a_{2},1},\gamma_{a_{3},1}),\cdots,(\gamma_{a_{2n-2},1},\gamma_{a_{2n-1},1})$; we chose to combine spin and vertex indices of Majoranas into a single index $a_{i}$ for simplicity. Note that $\gamma_{e,s,1}$ and $\gamma_{e',s',1}$ respectively corresponds to $\gamma_{a_{1},1}$ and $\gamma_{a_{2n},1}$. By definition, states that are not projected out by the second parenthesis have definite eigenvalue $\pm 1$ of the following operator:
\begin{equation}
\prod_{i=1}^{n} i \gamma_{a_{2i-1},1} \gamma_{a_{2i},1} = -(i\gamma_{a_{2n},1}\gamma_{a_{1},1})\prod_{i=1}^{n-1} i \gamma_{a_{2i},1} \gamma_{a_{2i+1},1}.
\end{equation}
The product of operators on the right side is also fixed by the first parenthesis---implying that states obtained after applying the original $\mathcal{C}_{u_{I}}$ have definite $i\gamma_{a_{2n},1}\gamma_{a_{1},1}$ eigenvalues despite lacking projectors involving the latter Majoranas. Hence, adding the projector in Eq.~\eqref{eq:projaddition} that projects onto the certain eigenstate of $i\gamma_{a_{2n},1}\gamma_{a_{1},1}$ is harmless and does not change how $\mathcal{C}_{u_{I}}$ acts. The same logic applies for Majoranas in layer 2. See Fig.~\ref{fig:bpcimod}(a) for an illustration.

One can insert the same projector in the second parenthesis if the configuration $u_{I}$ has spin down at plaquette $I$ (which is the `Hermitian conjugate' of the first case we considered). One can also apply a similar modification when two surrounding incomplete plaquette spins point down instead of up.
 
 Next we introduce a modification when the two incomplete plaquette spins around $I$ are opposite---fixed for concreteness to up on the left and down on the right---and with the spin at $I$ pointing up. The spin configuration $u_{I}$ then enforces two Majoranas $\gamma_{a_{1},1}$ and $\gamma_{a_{1},2}$ to be unpaired.  After action of $\mathcal{C}_{u_{I}}$, the aforementioned Majoranas pair with $\gamma_{a_{2},1}$ and $\gamma_{a_{2},2}$, and a different set of Majoranas $\gamma_{a_{3},1}$ and $\gamma_{a_{3},2}$ will now be unpaired; see Fig.~\ref{fig:bpcimod}(b). We introduce the unitary transformation
\begin{equation}
U = \frac{1 \pm \gamma_{a_{1},1}\gamma_{a_{3},1}}{\sqrt{2}}\frac{1 \pm \gamma_{a_{1},2}\gamma_{a_{3},2}}{\sqrt{2}} \frac{1+\sigma_{I}^{z}}{2} + \frac{1-\sigma_{I}^{z}}{2},
\end{equation}
which exchanges $\gamma_{a_{1},1}$ and $\gamma_{a_{3},1}$ and $\gamma_{a_{1},2}$ and $\gamma_{a_{3},2}$ \emph{only when $\sigma_{I}^{z} = +1$}.   (Whether we use the $+$ or $-$ sign above is unimportant.)  This unitary transformations commutes with any bulk term, hence proving $[B_{p}, U^{\dagger} C_{I} U] = 0$ implies $[B_{p},C_{I}] = 0$. The intuitive effect of this transformation is clear: As illustrated in Figs.~\ref{fig:bpcimod}(b) and (c), this transformation fixes Majoranas that remain unpaired, allowing $\mathcal{C}_{u_{I}}$ to be treated as closed loops for the $u_{I}$ we are considering. This can be explicitly confirmed via projector algebra, which we leave as an exercise for readers who want to gauge their understanding of this paper (the proof can be done in three lines). As before, the above modification can be straightforwardly generalized to the other spin configurations falling into case $(ii)$.  

\subsection{More detailed justification for polarizing bulk spins to derive the edge Hamiltonian}

Next we provide a more detailed and technical justification for deriving the edge Hamiltonian by fixing bulk spins and stripping out degrees of freedom that are frozen due to lack of quantum fluctuations of bulk spins in the low-energy space. To justify this procedure, we investigate the low-energy physics and the spectrum of the Hamiltonian $H = -\sum A_{t} -\sum C_{I}$. We will work in the subspace $\mathcal{S}_{A_{t}}$ in which $A_{t} =1$ is enforced. Also, we define $\mathcal{S}_{u_{b}}$, the subset of $\mathcal{S}_{A_{t}}$ in which bulk spin configurations are fixed to be $u_{b}$. By definition, $\cup_{u_{b}}\mathcal{S}_{u_{b}} = \mathcal{S}_{A_{t}}$.  

\begin{figure}
\includegraphics[width=0.8\linewidth]{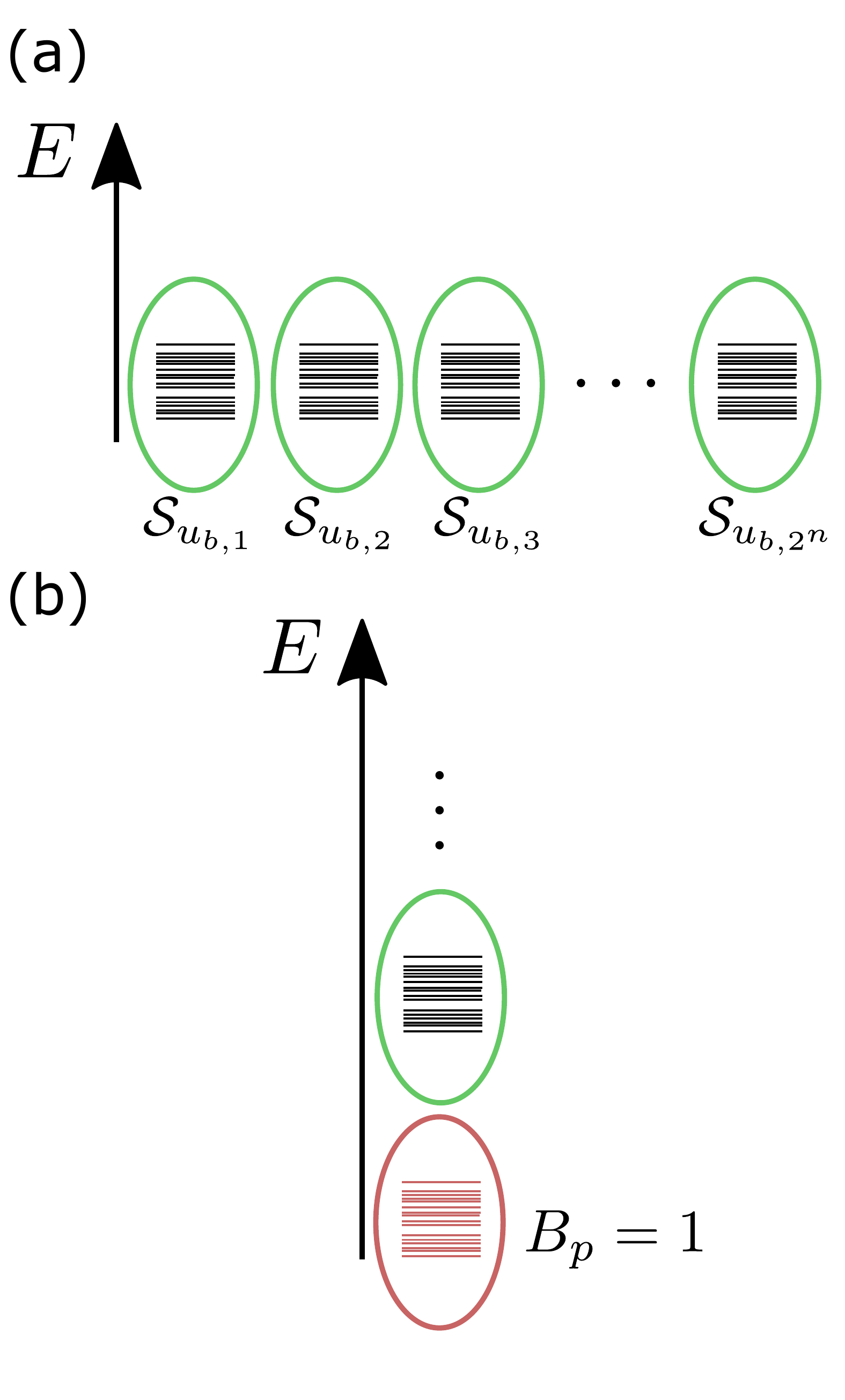}
\caption{Schematic energy spectrum of the sector that satisfies $A_{t} =1$ for all $t$'s, for $(a)$ $H = -\sum A_{t} -\sum C_{I}$ and $(b)$ $H = -\sum A_{t} -\sum B_{p} - \sum C_{I}$. Each green/red circle contains an identical sub-spectrum resulting from nontrivial action of $C_{I}$ within states in the circles. The red circle in (b) corresponds to the true low-energy sector of the edge termination for our topological-insulator model, while any circle contains identical spectral information.}
\label{fig:leveldraw}
\end{figure}

One can construct a canonical isomorphism from $S_{u_{b}}$ to $S_{u_{b}'}$, the bulk spin configuration $u_{b}'$ obtained by flipping spins at a set of plaquettes $ \{ p \}$ from $u_{b}$, defined as:
\begin{equation}
\ket{\psi_{u_{b}}} \rightarrow \prod_{ \{p \} }B_{p}\ket{\psi_{u_{b}}}.
\end{equation} 
Moreover, $C_{I}$ commutes with $B_{p}$. These two facts imply that the spectra of $H = -\sum A_{t} -\sum C_{I}$ within \emph{all} $\mathcal{S}_{u_{b}}$ sectors are identical. When the full Hilbert space is considered, the low-energy spectrum of $H = -\sum A_{t} -\sum C_{I}$ is $2^{n}$-fold degenerate ($n$ is the number of bulk plaquette spins), each degenerate state in the spectrum coming from a different subspace $S_{u_{b}}$. See Fig.~\ref{fig:leveldraw}(a) for a schematic.

 Adding the term $-\sum_{p} B_{p}$ lifts this massive degeneracy, splitting the $2^{n}$ degenerate levels into different levels that can be labeled by $B_p$ eigenvalues; this is possible because $B_{p}$ commutes with $C_{I}$ and $A_{t}$. Within the levels that possess the same set of $B_p$ eigenvalues, their relative energies are purely given by $C_{I}$ and hence are \emph{identical} to those given by the Hamiltonian $H= -\sum_{I} C_{I}$ defined on the restricted Hilbert space $\mathcal{S}_{u_{b}}$. See Fig.~\ref{fig:leveldraw}(b) for an illustration. The low-energy subspace of interest satisfies $B_{p}=1$ for all $p$, and therefore one can fix bulk spin configurations to study the low-energy physics of the Hamiltonian $H = -\sum A_{t} -\sum C_{I} - \sum B_{p}$. The cost is loss of an explicit on-site time-reversal symmetry.  However, we saw that the effective 1D model we derived from this procedure retains some non-trivial incarnation of $\mathcal{T}$-symmetry in the low-energy space.
 
\subsection{Simplifying the 1D model with unitary transformations and cutting out Majoranas}

\begin{figure}
\includegraphics[width=0.8\linewidth]{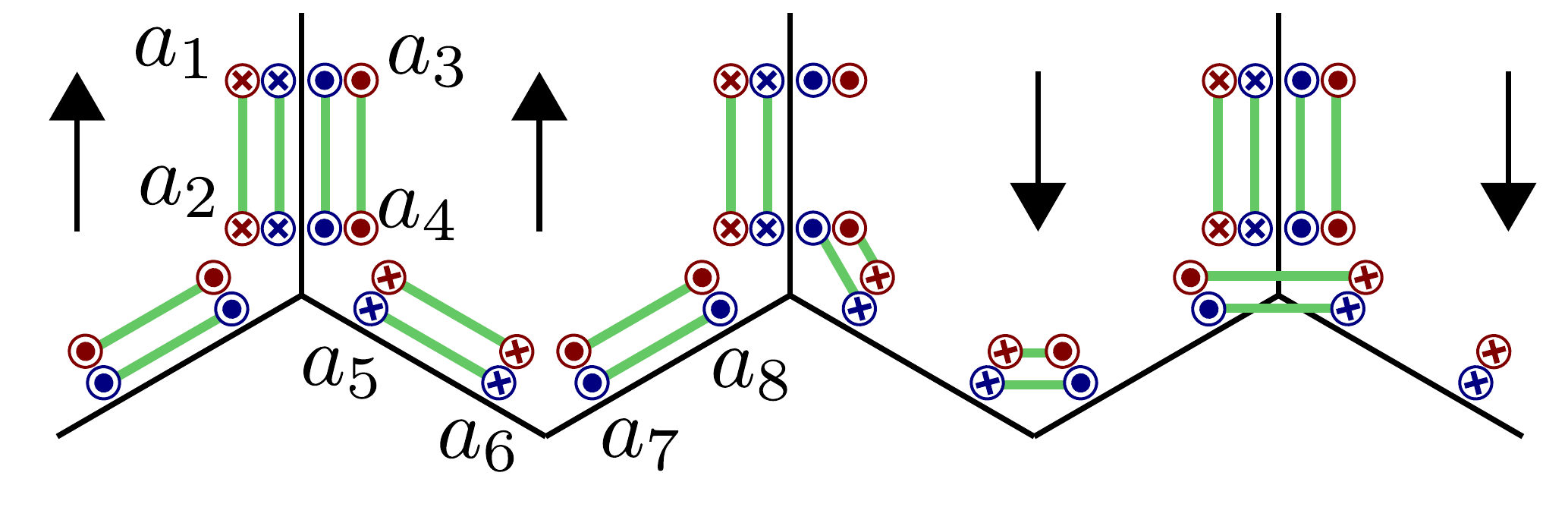}
\caption{The `site plus spin indices' $a_{1},a_{2},\cdots a_{8}$ for the original 1D edge Hamiltonian.}
\label{fig:aindex}
\end{figure}

 After polarizing bulk spins and removing Majoranas that are frozen in the low-energy space, we obtain a 1D model with a bosonic spin in each plaquette (i.e., unit cell) and sixteen Majoranas per unit cell, eight per layer. We will label the latter as $\gamma_{i,a_{1},1}, \gamma_{i,a_{2},1}, \cdots \gamma_{i, a_{8},1}$ in layer 1 and similarly for layer 2; here $i$ denotes a unit cell index and $a_{1},\cdots ,a_{8}$ are internal indices within a unit cell (see Fig.~\ref{fig:aindex} for the definition).  We now introduce a three-step local unitary transformation that reduces the number of Majorana degrees of freedom that fluctuate at low energies to four per unit cell, two for each layer: 

\begin{enumerate}
\item Whenever $\sigma_{i-1}^{z} = -1$ and $\sigma_{i}^{z}=+1$ apply the following unitary transformation 
\begin{equation}
\begin{split}
X_{i} &= \frac{1-\gamma_{i,a_{1},1}\gamma_{i,a_{3},1}}{\sqrt{2}} \frac{1-\gamma_{i,a_{2},1}\gamma_{i,a_{4},1}}{\sqrt{2}} \\
& \frac{1-\gamma_{i,a_{1},2}\gamma_{i,a_{3},2}}{\sqrt{2}} \frac{1-5\gamma_{i,a_{2},2}\gamma_{i,a_{4},2}}{\sqrt{2}}
\end{split}
\end{equation}
to each site.
The above transformation exchanges $\gamma_{a_{1},1}$ and  $\gamma_{a_{3},1}$, and $\gamma_{a_{2},1}$ and $\gamma_{a_{4},1}$ in layer 1, with the same exchanges occurring also in layer 2. In the original model, Majoranas with both spins can fluctuate; by applying the following changes, we will freeze Majoranas with a certain spin (in the language of this section, $\gamma_{i,a_{1},1}$, $\gamma_{i,a_{2},1}$, $\gamma_{i,a_{1},2}$, and $\gamma_{i,a_{2},2}$) and discard them.
\item Whenever $\sigma_{i}^{z} = +1$, apply the transformation
\begin{equation}
Y_{i} = \frac{1+\gamma_{i,a_{5},1}\gamma_{i,a_{7},1}}{\sqrt{2}} \frac{1+\gamma_{i,a_{5},2}\gamma_{i,a_{7},2}}{\sqrt{2}},
\end{equation}
which changes $\gamma_{i,a_{5},1}$ and $\gamma_{i,a_{7},1}$ in layer 1 only when the spin at $i$ points down (identical changes occur in layer 2). As a result of this transformation, in the low-energy subspace, $\gamma_{i,a_{6},1}$ and $\gamma_{i,a_{7},1}$ always remain paired and can be removed for the sake of studying the low-energy physics. The same goes for the analogous Majoranas in layer 2.
\item Whenever $\sigma_{i-1}^{z} = +1$ and $\sigma_{i}^{z}=-1$ apply the unitary transformation
\begin{equation}
Z_{i,+-} = \frac{1+\gamma_{i,a_{3},1}\gamma_{i,a_{5},1}}{\sqrt{2}} \frac{1+\gamma_{i,a_{3},2}\gamma_{i,a_{5},2}}{\sqrt{2}}.
\end{equation}
Meanwhile, when $\sigma_{i-1}^{z} = -1$ and $\sigma_{i}^{z}=+1$, apply 
\begin{equation}
Z_{i,-+} = \frac{1+\gamma_{i-1,a_{8},1}\gamma_{i,a_{3},1}}{\sqrt{2}} \frac{1+\gamma_{i-1,a_{8},2}\gamma_{i,a_{3},2}}{\sqrt{2}}.
\end{equation}
Both transformations are applied when there are unpaired Majoranas between sites $i-1$ and $i$, and move the unpaired Majoranas (which due to the very first transformation we applied are always $\gamma_{i,a_{3},1}$ and $\gamma_{i,a_{3},2}$) to either $a_{8}$ or $a_{5}$. Also, this transformation causes $\gamma_{i,a_{3},1}$ and $\gamma_{i,a_{4},1}$ to always pair in the low-energy subspace---hence they can be discarded. Likewise, $\gamma_{i,a_{3},2}$ and $\gamma_{i,a_{4},2}$ can be thrown away.
\end{enumerate}

\begin{figure*}
\includegraphics[width=0.6\linewidth]{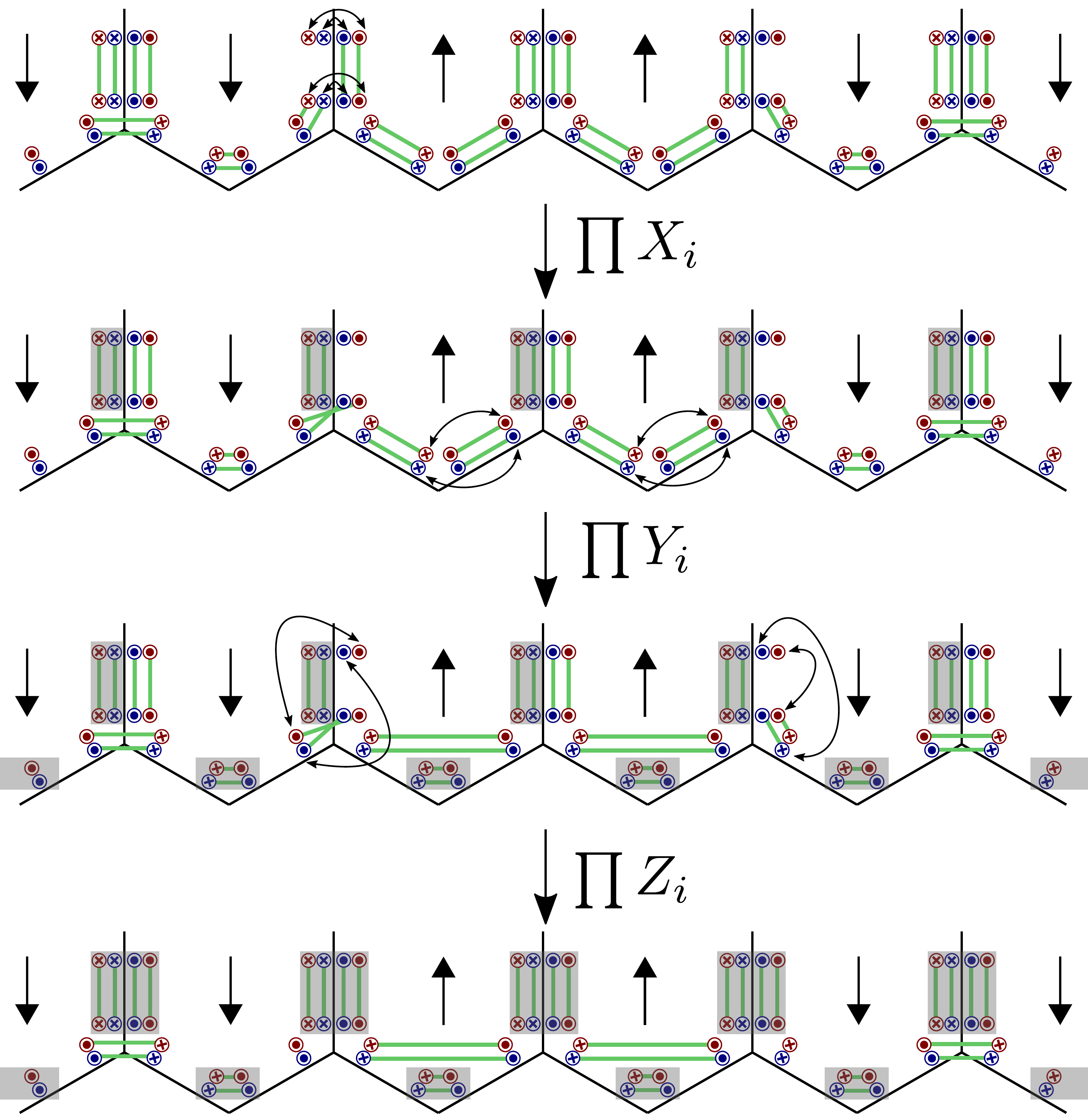}
\caption{An example of the three-step unitary transformation applied to a state that satisfies ${A}_{t}=1$ for all $t$'s. The bidirectional arrows indicate which Majoranas are `exchanged' at each step; Majoranas in the grey square are frozen in the low-energy space after the unitary transformation. We see that at the end only four Majoranas per unit cell fluctuate.}
\label{fig:unitaryexample}
\end{figure*}

Figure~\ref{fig:unitaryexample} illustrates the effect of the above three transformations. Applying these unitary transformation and discarding Majoranas that are frozen in the low-energy space yields a 1D Hamiltonian $H^{M}$ that is almost identical to $H^{L}$ (up to some $\mathbb{Z}_{2}$ Majorana gauge transformation), subject to the following difference: In $H^{M}$, there are two types of vertex terms, one originating from the fully trivalent `complete vertices' in Fig.~\ref{fig:aindex} and another originating from `incomplete vertices' where only two edges meet. Terms orignating from incomplete vertices are notably absent in $H^{L}$. However, removing these terms does not affect the subspace with $A_{t}=1$.  Hence, we conclude that the low-energy physics of $H^{L}$ is completely identical to the 1D edge of our exactly-solvable topological insulator model. 

\section{More on the Berry-phase calculation}
\label{app:Berry}
 
The goal of this appendix is to prove Eq.~\eqref{eq:Berrykey}. We start by proving a related lemma.
 
\subsection{Lemma} 
Let $\ket{u(\theta)}$ denote a state with bosonic spin configuration $u$ and Majorana pairings consistent $u$. Then, for any $u$ and $p$, the following holds:
\begin{equation}
\label{eq:lemmaBerry}
\bra{u(\theta)} B_{p}(\theta) \frac{\partial B_{p}(\theta)}{\partial \theta} \ket{u(\theta)} = 0.
\end{equation}
Since $B_{p}(\theta)$ acts on some fixed spin configuration, we can write
\begin{equation}
\bra{u(\theta)} B_{p}(\theta) \frac{\partial B_{p}(\theta)}{\partial \theta} \ket{u(\theta)} = \bra{u(\theta)} \mathcal{B}_{u_{p}}(\theta)^{\dagger} \frac{\partial \mathcal{B}_{u_{p}}(\theta)}{\partial \theta} \ket{u(\theta)},
\end{equation}
where $\mathcal{B}_{u_{p}}(\theta)$ only contains Majorana projectors (the spin parts of the two $B_p$ operators must undo one another to get a nontrivial result).  Let us denote $P_{v_{1}s_{1},v_{2}s_{2}}^{\text{TI}}(\theta)$ as any pair of projector that appears in the expression for $\mathcal{B}_{u_{p}}(\theta)$, and define $\mathcal{B}_{u_{p},v_{1}s_{1},v_{2}s_{2}}(\theta)$ as the version of $\mathcal{B}_{u_{p}}(\theta)$ with $P_{v_{1}s_{1},v_{2}s_{2}}^{\text{TI}}(\theta)$ erased. We then have
\begin{equation}
\begin{split}
&\bra{u(\theta)} \mathcal{B}_{u_{p}}(\theta)^{\dagger} \frac{\partial \mathcal{B}_{u_{p}}(\theta)}{\partial \theta} \ket{u(\theta)} = \bra{u(\theta)} \mathcal{B}_{u_{p}}(\theta)^{\dagger}\\
& \sum_{(v_{1}s_{1},v_{2}s_{2}) \in \overline{\mathcal{P}}_{u_{p}}}  \frac{\partial P_{v_{1}s_{1},v_{2}s_{2}}^{\text{TI}}(\theta)}{\partial \theta} \mathcal{B}_{u_{p},v_{1}s_{1},v_{2}s_{2}}(\theta) \ket{u(\theta)} \\
&+ \sum_{(v_{1}s_{1},v_{2}s_{2}) \in \mathcal{P}_{u_{p}}}  \mathcal{B}_{u_{p},v_{1}s_{1},v_{2}s_{2}}(\theta) \frac{\partial P_{v_{1}s_{1},v_{2}s_{2}}^{\text{TI}}(\theta)}{\partial \theta}  \ket{u(\theta)}.
\end{split}
\end{equation}
 While this looks complicated, it is just a straightforward application of the product rule of derivatives. Recall that following the notation of Eq.~\eqref{eq:tibp}, $(v_{1}s_{1},v_{2}s_{2}) \in \overline{\mathcal{P}}_{u_{p}}$ corresponds to projectors that reconfigure Majorana pairings and are in the first parenthesis of Eq.~\eqref{eq:tibp}. In this case, following similar logic that local-fermion-parity conservation fixes one of the pairings without explicit projection, $\mathcal{B}_{u_{p},v_{1}s_{1},v_{2}s_{2}}(\theta) = \mathcal{B}_{u_{p}}(\theta)$! The case $(v_{1}s_{1},v_{2}s_{2}) \in \mathcal{P}_{u_{p}}$ corresponds to projectors in the second parenthesis of Eq.~\eqref{eq:tibp} that project onto Majorana pairings consistent with a spin configuration $u_p$ around plaquette $p$. If Majorana pairings between $(v_{1}s_{1},v_{2}s_{2})$ are reconfigured by the first parenthesis, similar logic as before guarantees that $\mathcal{B}_{u_{p},v_{1}s_{1},v_{2}s_{2}}(\theta) = \mathcal{B}_{u_{p}}(\theta)$. If they are not reconfigured, we use the fact that $P_{v_{1}s_{1},v_{2}s_{2}}^{\text{TI}}(\theta)$ commutes with all projectors $\mathcal{B}_{u_{p}}(\theta)$, which allows us to convert $\mathcal{B}_{u_{p},v_{1}s_{1},v_{2}s_{2}}(\theta)$ back to to $\mathcal{B}_{u_{p}}(\theta)$ as well. We thus deduce
\begin{equation}
\begin{split}
&\bra{u(\theta)} \mathcal{B}_{u_{p}}(\theta)^{\dagger} \frac{\partial \mathcal{B}_{u_{p}}(\theta)}{\partial \theta} \ket{u(\theta)} \\
&=  \sum_{(v_{1}s_{1},v_{2}s_{2}) \in \overline{\mathcal{P}}_{u_{p}}} \bra{u(\theta)} \mathcal{B}_{u_{p}}(\theta)^{\dagger}\frac{\partial P_{v_{1}s_{1},v_{2}s_{2}}^{\text{TI}}(\theta)}{\partial \theta} \mathcal{B}_{u_{p}}(\theta) \ket{u(\theta)} \\
&+ \sum_{(v_{1}s_{1},v_{2}s_{2}) \in \mathcal{P}_{u_{p}}}  \bra{u(\theta)} \mathcal{B}_{u_{p}}(\theta)^{\dagger} \mathcal{B}_{u_{p}}(\theta) \frac{\partial P_{v_{1}s_{1},v_{2}s_{2}}^{\text{TI}}(\theta)}{\partial \theta}  \ket{u(\theta)}.
\end{split}
\end{equation}
Finally, using the fact that on the third line $P_{v_{1}s_{1},v_{2}s_{2}}^{\text{TI}}\ket{u(\theta)} = \ket{u(\theta)}$ and that $P^2 = P$, one can equivalently substitute the partial derivatives in the above expression via
\begin{equation}
\frac{\partial P_{v_{1}s_{1},v_{2}s_{2}}^{\text{TI}}(\theta)}{\partial \theta} \rightarrow P_{v_{1}s_{1},v_{2}s_{2}}^{\text{TI}}\frac{\partial P_{v_{1}s_{1},v_{2}s_{2}}^{\text{TI}}(\theta)}{\partial \theta} P_{v_{1}s_{1},v_{2}s_{2}}^{\text{TI}}.
\label{projector_replacement}
\end{equation}
Appendix~\ref{app:lemma} proved that the expression on the right side of Eq.~\eqref{projector_replacement} vanishes. This proves Eq.~\eqref{eq:lemmaBerry}.

\subsection{Proof of Eq.~\eqref{eq:Berrykey} using the lemma}

Next, we note that
\begin{equation}
\begin{split}
&\bra{\downarrow(\theta)} \left( \prod_{p} \left( B_{p}(\theta) \right)^{n_{p}}\right) \left( \prod_{p} \left( B_{p}(\theta) \right)^{n_{p}}\right)  \frac{\partial}{\partial \theta}\ket{\downarrow(\theta)} \\
&= \bra{\downarrow(\theta)}\frac{\partial}{\partial \theta}\ket{\downarrow(\theta)},
\end{split}
\end{equation}
where the definition of each symbol is identical to the ones in Eq.~\eqref{eq:Berrykey}. The proof thus really boils to down to showing
\begin{equation}
\bra{\downarrow(\theta)} \left( \prod_{p} \left( B_{p}(\theta) \right)^{n_{p}}\right) \left( \frac{\partial}{\partial \theta} \prod_{p} \left( B_{p}(\theta) \right)^{n_{p}}\right)  \ket{\downarrow(\theta)} = 0.
\label{EqToProve}
\end{equation}
In Eq.~\eqref{EqToProve} the derivative acts only on $B_p$ operators in the second product.  

To proceed, let us set an ordering for plaquettes $p_{1}$, $p_{2}$, $\cdots$ $p_{n}$; the precise details of the ordering do not matter. Let us further order $B_{p}(\theta)$ terms in the second parenthesis according to the chosen ordering, but order terms in the first parenthesis with the \emph{inverse} ordering (which can be understood as the `Hermitian conjugate' of ordering in the second parenthesis). Then, the product rule of partial derivatives allows one to show that 
\begin{equation}
\label{eq:midstepBerry}
\begin{split}
&\bra{\downarrow(\theta)} \left( \prod_{p} \left( B_{p}(\theta) \right)^{n_{p}}\right) \left( \frac{\partial}{\partial \theta} \prod_{p} \left( B_{p}(\theta) \right)^{n_{p}}\right)  \ket{\downarrow(\theta)} \\
&= \sum_{j=1}^{n} \bra{\downarrow(\theta)} \\
&\left( \prod_{i=1}^{j-1} \left( B_{p_{i}}(\theta) \right)^{n_{p_{i}}}\right) (B_{p_{j}}(\theta))^{n_{j}} \left( \prod_{i=j+1}^{n} \left( B_{p_{i}}(\theta) \right)^{n_{p_{i}}}\right) \\
&\left( \prod_{i=j+1}^{n} \left( B_{p_{i}}(\theta) \right)^{n_{p_{i}}}\right) \frac{\partial(B_{p_{j}}(\theta))^{n_{j}}}{\partial \theta} \left( \prod_{i=1}^{j-1} \left( B_{p_{i}}(\theta) \right)^{n_{p_{i}}}\right) \ket{\downarrow(\theta)} .
\end{split}
\end{equation}
The identical product of projectors in the second and the third parenthesis multiply to 1. Also,
\begin{equation}
\left( \prod_{i=1}^{j-1} \left( B_{p_{i}}(\theta) \right)^{n_{p_{i}}}\right) \ket{\downarrow(\theta)} = \ket{u_{j}(\theta)},
\end{equation}
where $\ket{u_{j}(\theta)}$ is a state with a spin configuration obtained by flipping spins of $p_{1}, p_{2}, \cdots p_{j}$ from the all-down spin configuration, and with Majorana pairings consistent with that new spin configuration. Hence, we can greatly simplify Eq.~\eqref{eq:midstepBerry} to be
\begin{equation}
\begin{split}
&\bra{\downarrow(\theta)} \left( \prod_{p} \left( B_{p}(\theta) \right)^{n_{p}}\right) \left( \frac{\partial}{\partial \theta} \prod_{p} \left( B_{p}(\theta) \right)^{n_{p}}\right)  \ket{\downarrow(\theta)} \\
&= \sum_{j=1}^{n} \bra{u_{j}(\theta)} (B_{p_{j}}(\theta))^{n_{j}} \frac{\partial(B_{p_{j}}(\theta))^{n_{j}}}{\partial \theta} \ket{u_{j}(\theta)} .
\end{split}
\end{equation} 
If $n_{j} =0$ then the summand trivially vanishes. If $n_{j} =1$ we can apply the lemma of the previous subsection to show that it also vanishes. Hence, we proved Eq.~\eqref{eq:Berrykey}, and the Berry phase of our topological-insulator model on a Klein bottle may be computed by computing the Berry phase at one spin configurations with corresponding Majorana pairings.

\bibliography{ref}

\end{document}